\documentclass{pasa}%

\usepackage{natbib}
\usepackage{color}

\title[Dawes Review 4: Spiral Structures in Disc Galaxies]{Dawes Review 4: Spiral Structures in Disc Galaxies}
\author[C. Dobbs \& J. Baba]{
	Clare Dobbs$^1$, \and 
	Junichi Baba$^{2}$
	\\
\affil{$^1$School of Physics and Astronomy, University of Exeter, Stocker Road, Exeter, EX4 4QL, UK}%
\affil{$^2$Earth-Life Science Institute,
	Tokyo Institute of Technology
	2-12-1-I2-44 Ookayama, Meguro, Tokyo 152--8551, Japan}
}%
\jid{PASA}
\doi{10.1017/pas.\the\year.xxx}
\jyear{\the\year}

\begin{document}%

\begin{abstract}

The majority of astrophysics involves the study of spiral galaxies, and stars and planets within them, but how spiral arms in galaxies form and evolve is still a fundamental problem. Major progress in this field was made primarily in the 1960s, and early 1970s, but since then there has been no comprehensive update on the state of the field. In this review, we discuss the progress in theory, and in particular numerical calculations, which unlike in the 1960s and 1970s, are now commonplace, as well as recent observational developments. We set out the current status for different scenarios for spiral arm formation, the nature of the spiral arms they induce, and the consequences for gas dynamics and star formation in different types of spiral galaxies. We argue that, with the possible exception of barred galaxies, spiral arms are transient, recurrent and 
initiated by swing amplified instabilities in the disc. We suppose that unbarred $m=2$ spiral patterns are induced by tidal interactions, 
and slowly wind up over time. However the mechanism for generating spiral structure does not appear to 
have significant consequences for star formation in galaxies.
\end{abstract}
\begin{keywords}
keyword1 -- keyword2 -- keyword3 -- keyword4 -- keyword5
\end{keywords}
\maketitle%

The Dawes Reviews are substantial reviews of topical areas in astronomy, published by authors of international standing at the invitation of the PASA Editorial Board. The reviews recognise William Dawes (1762-1836), second lieutenant in the Royal Marines and the astronomer on the First Fleet. Dawes was not only an accomplished astronomer, but spoke five languages, had a keen interest in botany, mineralogy, engineering, cartography and music, compiled the first Aboriginal-English dictionary, and was an outspoken opponent of slavery.

\section{Introduction}
Spirals galaxies represent some of the most beautiful, and fascinating objects in the Universe. According to the Galaxy Zoo project, spiral galaxies make up about two thirds of all massive galaxies, whilst around one third are ellipticals, and a few per cent merging galaxies \citep{Lintott2011,Willett2013}. Star formation overwhelmingly occurs in spiral galaxies, and in particular is associated with spiral arms. Thus understanding the nature of spiral arms is essential both for understanding star formation, and galaxy evolution.

Spiral galaxies are generally classified into different types according to the presence of a bar (S and SB for unbarred and barred galaxies, and sometimes SAB for weakly barred galaxies) and the degree of winding (or pitch angle) of the spiral arms {\citep{Hubble1926,Reynolds1927,deVaucouleurs1959}}.  The latter is scaled from Sa-Sd or SBa to SBd with the `d' classification representing the most open arms, and the `a' classification the most tightly wound. The sequence also represents a decrease in the size and luminosity of the bulge from Sa (or SBa) galaxies to Sd (or SBd), and an increase in gas content from Sa to Sd galaxies.

A second classification scheme was proposed by \citet{ElmegreenElmegreen1982} and \citet{Elmegreen1987} to classify spiral galaxies  into 12 types according to the number and length of spiral arms. Thus galaxies with many fragmented short arms are different types to those with two long arms. Galaxies could also be denoted as having two inner arms, and multiple outer arms. A simpler, but similar division of spiral galaxies (see e.g. \citealt{Elmegreen1990}) is into 3 types: flocculent spiral galaxies (with many short arms, such as NGC2841), multi-armed spirals (e.g. M33) and grand design galaxies (with two main spiral arms, e.g. M51). All of these types may or may not exhibit bars. Around 60 \%, of galaxies exhibit some grand design structure, either in the inner or entire part of the disc \citep{ElmegreenElmegreen1982,Grosbol+2004}.

The Hubble classifications are usually associated with the long-term evolution of galaxies, whereas the classification by Elmegreen is instead associated with their current properties and environment. 
Historically, Sa galaxies, and ellipticals were termed early type galaxies, whilst Sc and Sd galaxies were termed late type galaxies, though this is opposite to the evolutionary sequence which has since been established. Instead trends in star formation rate, bulge-to-disc ratio, and age of disc stars now indicate an evolutionary sequence from Sc to Sa types \citep{Sandage1986,Kennicutt1998}.
Sa type galaxies are thought to have already used up much of their gas and exhibit lower star formation rates compared to Sc and Sd types, although mergers and galaxy interactions will also influence the properties of the galaxies (e.g. \citealt{Elmegreen1990}. We also note correlations with Hubble type are only a general trend -- \citet{Kennicutt1981} indicates that the pitch angle correlates only in an average sense with galaxy type, and there is quite substantial spread.

The flocculent, or grand design nature of spiral galaxies, is directly linked to the mechanism which generates the spiral arms. There are three main mechanisms hypothesised to produce spiral arms, i) (quasi-stationary) density wave theory, ii) local instabilities, perturbations, or noise which are swing amplified into spiral arms, and iii) tidal interactions. Bars may also play a role in inducing spiral arms. Note that these mechanisms are not necessarily mutually exclusive, for example a tidal interaction could theoretically induce a wave which obeys density wave theory. Typically though, local instabilities are associated with flocculent or multi-armed galaxies, whereas grand design galaxies are presumed to have undergone a tidal interaction, have a bar driving arms, and/or obey steady state density wave theory. In addition to the various classifications of spiral galaxies, and spiral arm formation mechanisms, there are also three kinematic types of spiral arm i) material arms, which obey the kinematics of the disc, ii) kinematic spiral arms, which rotate slower than the angular velocity of the disc, and iii)  stationary spiral arms, which rotate rigidly and do not wind up. In the following sections we discuss these (and a few other) supposed mechanisms, and which type of spiral arms, and spiral galaxies are produced.

There are also several simple properties of spiral arms that we can observe that can give insight on the nature of spiral arms (we go into much more depth on observational tests for spiral arms in Section 4), i) the number of spiral arms, ii) the pitch angle, iii) amplitude, iv) arm shape and v) lifetime. How many spiral arms a galaxy exhibits is one of the most fundamental questions regarding the theory of spiral arms. In the absence of a bar, or perturber, this will most simply depend on the relative disc and halo masses, and their dimensions. A galaxy will only form spiral arms at all if the disc is sufficiently gravitationally dominated. To a rough approximation (the susceptibility of the disc to asymmetric perturbations) this is governed by the Toomre parameter $Q$ for stars and/or gas (see Section 2.1.1).  If the disc is unstable, an estimate of the expected number of spiral arms can be made by considering the stability of different wavenumbers in the appropriate dispersion relation (i.e. the value of $k$, the wavenumber, such that $e^{-i\omega(k) t}$ grows fastest). Alternatively, and more appropriately for perturbations growing from local instabilities or noise, the number of arms can be estimated by swing amplification theory, as described in Section 2.2.1, where again the number of spiral arms corresponds to the value which produces the greatest amplification. Tidally interacting galaxies naturally produce two-armed spiral galaxies.

Other observable properties of spiral arms were investigated by Kennicutt 1981, and many other works since (e.g. \citealt{ConsidereAthanassoula1988, Block+1994, PuerariDottori1992, SeigarJames1998, Ma2002, Seigar+2006, Elmegreen+2011, Kendall+2011}). Although the pitch angle is historically used to classify galaxies according to the Hubble sequence, the differences in spiral arm shape, i.e. the pitch angle of the spiral arms appears to be most dependent on the maximum rotation velocity, and thus the local shear in the disc, rather than the global mass distribution \citep{Kennicutt1981, KennicuttHodge1982, Garcia1993, SeigarJames1998, Seigar+2006}.
For example Figures 8 and 10 of \citet{Kennicutt1981} show that the pitch angle correlates much better with the maximum rotational velocity than the properties of the bulge. However there is still considerable scatter (see Figure~7 of \citealt{Kennicutt1981}) in the correlation with rotation velocity that there is scope for tidal interactions, or density wave theory to introduce some spread (see also \citealt{Grand+2013}). There is also no correlation with pitch angle and arm class, i.e. the Elmegreen classification scheme of whether the galaxy is flocculent or grand design \citep{PuerariDottori1992}. \citet{Kennicutt1981} also examined the shapes of spiral arms, finding that they did not fit exactly into the category of either density wave theory (logarithmic) or tidally induced (hyperbolic spirals).
The lifetimes of spiral arms are obviously much more difficult to test observationally (see \citealt{Sellwood2011}). Here we have relied more on computer simulations, and theory to predict the lifetimes of spiral arms for different  scenarios. Generally though, arms in flocculent galaxies are expected to be fairly short lived (few 100 Myrs) and arms in grand design spirals somewhat longer lived ($\sim$1 Gyr).  

Although so far we have discussed spiral galaxies as either flocculent or grand design, observations in the 1990s showed that galaxies could exhibit characteristics of both flocculent and grand design structure, typically with grand design arms seen in the infrared (old stars) and a more flocculent structure seen in the optical (gas and young stars) \citep{BlockWainscoat1991,Thornley1996,ThornleyMundy1997a}.  Some galaxies also appear to exhibit a 3 armed structure in the optical and 2 armed in the IR \citep{Block+1994}. The existence of such galaxies poses a further challenge for theories of spiral structure.

The main previous review on spiral structure is \citet{Toomre1977}, though there have also been a couple of shorter reviews by Sellwood in recent years  \citep{Sellwood2010review,Sellwood2011}. A historical review of spiral arm theory in the 1960s and 70s is also given by \citet{Pasha2004a,Pasha2004b}. A review specific to the Milky Way is currently being written by Benjamin (Benjamin 2014, in preparation).  In this review, we aim to bring together the different aspects of studies of spiral structure including simulations and observational tests, as well as the theory.
The outline of this review is as follows. In Section 1.1, we discuss the historical context of spiral galaxies, and the origin of different theories for spiral structure. In Section 2 we go into much more detail on the possible mechanisms for generating spiral structure, including density wave theory, swing amplification, bars, tidal interactions, stochastic star formation and dark matter halos. We also include discussion of computer simulations to test these theories. In Section~3 we examine the gas response to spiral arms, including again density wave theory, local instabilities and tidal perturbations. 
In Section~4 we discuss possible observational tests to distinguish between the various scenarios of spiral structure. 
Finally in Section 5, we present our conclusions.

\subsection{Historical overview}
For a comprehensive review of the history of spiral structure, we recommend \citet{Pasha2004a,Pasha2004b}, who gives a very detailed, and personal description of the developments in spiral structure, particularly in the 1960s. Here we given a brief overview up to about the time of the \citet{Toomre1977} review, although much of the background theory is also considered in much more detail in Section~2.

\begin{figure}[h]
\begin{center}
\includegraphics[scale=0.35]{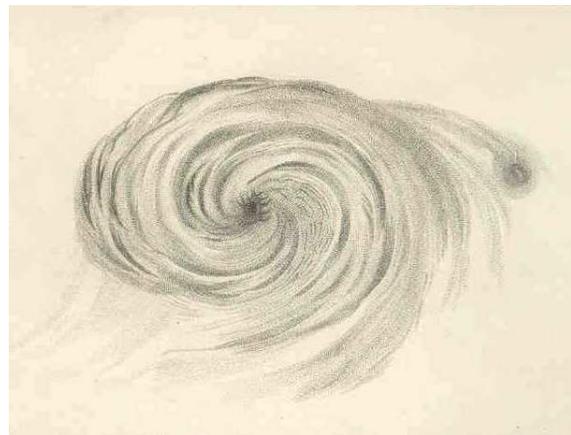}
\caption{A sketch of M51 by Lord Rosse \citep{Rosse1850}.}\label{fig:rosse}
\end{center}
\end{figure}

Spiral galaxies have been observed for over 150 years, although until the 1920s, they were classed as `spiral nebulae', and assumed to lie within our own Galaxy. The spiral structure of M51 was identified by Lord Rosse in 1850 \citep{Rosse1850} as the first spiral nebulae (Figure~\ref{fig:rosse}). Rosse also identified point sources within these nebulae, hence establishing that they were not simply clouds of gas. The Curtis-Shapley `Great Debate' then later ensued about whether these nebulae were extragalactic. This matter was clarified by Hubble, who confirmed that the spiral nebulae were indeed external to the Milky Way, and thus spiral galaxies, by determining the distance first to M33, and then M31, using Cepheid variables \citep{Hubble1926b,Hubble1929}. The distances to M31 and M33 demonstrated that these objects were far too distant to lie within the Milky Way. 

Following the establishment of the nature of spiral nebulae, astronomers considered the nature of the spiral arms themselves. The first main proponent of this work was Lindblad, who first considered spiral arms in terms of Maclaurin ellipsoids (flattened spheroids rotating in an equilibrium state) \citep{Lindblad1927}, following previous work by Jeans and Poincare. He considered an instability occurring at  the edge of an ellipsoid, which induces high eccentricity in the orbits at the outer edges, pertaining to circular orbits nearer the centre. \citet{Lindblad1935} later derived a condition for gravitational instability, and thereby spiral arms, in a series of rotating spheroids. Lindblad wrote that spiral arms are analogous to a harmonic wave in an unstable Maclaurin spheroid \citep{Lindblad1927, Lindblad1940}. He considered spiral arms in terms of individual stellar orbits (and indeed, \citealt{Kalnajs1973} later showed that a spiral perturbation can be represented by a series of unaligned elliptical orbits) rather than a collective process. The idea of spiral arms as a wave was not actively considered until the 1960s.

The 1960s in fact saw the next major development in spiral arm theory, when indeed spiral arms started to be considered as collective processes governed by the gravity of the galactic disc. The pioneering work of \citet{Toomre1964} and \citet{LinShu1964} (following also the stability analysis of \citealt{Safranov1960} for discs) studied gravitational instabilities in the context of an infinitesimally thin, rotating, stellar disc. 
Both papers started with the linearised equations of motion, and Poisson's equation for a stellar disc, 
and established solutions which have the Fourier decomposition
\citep{Shu1992b,BinneyTremaine2008}:
\begin{eqnarray}
 {\psi(R,\phi,t) = {\rm Re} [A(R) e^{i(\omega t - m\phi)}], \label{eqn:spiralarm}} 
\end{eqnarray}
where $\Omega_{\rm p}=\omega/m$ and $\Omega_{\rm p}$ is the angular velocity of the perturbation, or pattern speed.
Equation \ref{eqn:spiralarm} assumes that the complex function $A(R)$, 
which determines the amplitude and radial phase of the perturbations, 
varies slowly with $R$ (the tight winding approximation, see Section 2.1.1). 
Thus these solutions represent waves with crests at periodic displacements. In addition to the form of the wave, these results also established the dispersion relations for fluid and stellar discs (with \citealt{LinShu1966}, and \citealt{Kalnajs1965}), and the stability criteria for discs subject to axisymmetric perturbations (see Section~2). At this point however, there is complete flexibility regarding the value of $m$ (the number of arms), the superposition of waves of different $m$, what range of $R$ or $\phi$ the solution covers, the sign of $\Omega_{\rm p}$, 
and thus whether the arms are leading or trailing, or the length of time the perturbation exists.

\citet{LinShu1964} proposed that in fact there is a preference for lower values of $m$, and that such waves are relatively stable with time. \citet{LinShu1964, LinShu1966} also proposed a global solution for the disc, rather than the local perturbations assumed by \citet{Toomre1964}. Such global stable waves would be standing waves in the disc, and hence they were called `quasi-stationary', a term first introduced by \citet{Lindblad1963}.  The motivation for supposing the stability of these waves, in particular for $m=2$ was largely observational. 
Most galaxies were observed to be spirals at that time \citep{Hubble1943}, 
so either the spiral arms are long lived, or they are continually replenished. Furthermore, fixed spiral arms would remove the so called `winding problem'. In addition, disproportionately many galaxies have 2 spiral arms, so a tendency for systems to exhibit $m=2$ would explain this predominance.

\citet{GoldreichLynden-Bell1965a}, consider the action of gravitational instabilities, first in a uniformly rotating gas disc, then under differential rotation \citep{GoldreichLynden-Bell1965}. They supposed that spiral arms are a superposition of many unstable wavelengths in the gas. In their picture, it is the instabilities in the gas which form gaseous spiral arms, which in turn form stars and lead to stellar spiral arms. This is somewhat different from the picture of a stellar dominated disc, where instabilities are thought to arise in the stars, leading to a gravitational potential well for the gas to fall into, shock and form molecular clouds (see Section 3.7). Unlike the simpler analysis of discs subject to axisymmetric perturbations (see Section 2), these studies investigate asymmetric perturbations in a shearing disc. \citet{GoldreichLynden-Bell1965}, and \citet{JulianToomre1966}, demonstrated the significance of a differentially rotating disc. Gravity is enhanced as a region undergoes shear.  Hence it is easier for perturbations to grow via the disc self gravity. This effect was later coined swing amplification, discussed further in Sections~2.1.3 and 2.2. 

Meanwhile there were some important observational developments following the theoretical work of Lindblad. One was the finding that spiral arms tended to be trailing in character \citep{Hubble1943}. A second was that, rather than uniform rotation, galaxies were indeed observed to rotate differentially (e.g. \citealt{Burbidge1964, Rubin1970}).

In the late 60s, and 70s, authors started to consider the response of gas to the stellar disc. Assuming a static spiral potential of the form proposed by \citet{LinShu1964}, the solution for the gas response can be obtained \citep{Fujimoto1968,Roberts1969}. In particular the gas is found to undergo a shock caused by the stellar spiral spiral arms. The detection of dark dust lanes alongside spiral arms \citep{Sandage1961, Lynds1970} gave strong observational evidence that the gas undergoes a spiral shock, the dense shock being seen as dark clouds in the dust lanes that go on to form stars (Roberts 1969). In fact it is now evident that regardless of how spiral arms are generated, spiral structure is only very weak in the old stars, whereas the spiral structure we see by eye is dominated by the gas and young stars (e.g. \citealt{Elmegreen+2011}).

At the same time however, results were starting to query whether steady spiral modes could be sustained in galaxies. \citet{Lynden-BellOstriker1967} showed, in the `anti-spiral theorem' that stable spiral modes do not exist in a steady state, although it is possible to obtain a solution with asymmetric spirals, i.e. one trailing and one\citet{Toomre1969} also showed that the waves will not remain in a fixed position within the disc, rather the pattern will propagate inwards to outwards with the group velocity on a timescale of a few galactic rotations -- suggesting that density waves need to be constantly replenished. Consequently, a mechanism to maintain density waves was required. 
\citet{Mark1974, Mark1976b} suggested that it could be possible to maintain spiral density waves by means of reflection between two radii of the disc -- setting up a standing wave. \citet{Toomre1969} instead proposed tidally interacting galaxies were the primary means of generating $m=2$ spiral structure. 

Since the 1970s, the debate about stationary versus transient spirals has continued. In addition to theoretical arguments, numerical simulations have become much more widespread to test theories of spiral structure. Observations are also starting to provide some information on the dynamics of spiral galaxies.

\section{Generation of spiral structure}

In this section we describe the different mechanisms for generating spiral structure, 
namely quasi-stationary density wave theory (Section 2.1), 
recurrent transient spiral instabilities (Section 2.2), 
bars (Section 2.3), tidal interactions (Section 2.4), stochastic star formation (Section 2.5), 
and exotic mechanisms such as perturbations from dark matter halos (Section 2.6).
 
\subsection{Quasi-stationary density wave theory}
In this section we present the theory that global spiral arms are slowly evolving patterns 
that rotate with fixed pattern speeds in the disc, quasi-stationary density wave theory. 
Much of this material is theoretical, as we discuss in Section~2.1.5, 
this theory has not yet been demonstrated in the context of numerical simulations.

Inspired by the idea of kinematic density waves suggested by \citet{Lindblad1960,Lindblad1963}, 
\citet{LinShu1964} proposed a self-consistent density wave theory. 
Unlike Lindblad's approach \cite[for reviews]{Toomre1977,Pasha2004a,Pasha2004b}, 
Lin \& Shu treated the galactic disc as a `continuum' consisting of either stars or gas,
and derived the dispersion relation of the density waves for a rotating disc. 
Qualitatively, they assumed that the spiral arms are not material in nature, 
but instead made up of areas of greater density, with the stars and gas moving through the spiral arms.
The difference in the dynamics is often compared to cars moving along a traffic jam. 
Material arms are analogous to a queue of cars moving at a speed $v_0$ identical to 
all other vehicles on the road. For density waves, instead suppose a queue of cars moving 
at $v_1<v_0$, whereas other cars on the road will slow down to $v_1$ 
as they enter the queue and speed up as they leave.

After the derivation of the dispersion relation for a fluid disc by \citet{LinShu1964}, 
the dispersion relation for a stellar disc was derived by \citet{LinShu1966} 
and \citet{Kalnajs1965}. 
The dispersion relations for fluid and stellar discs are called the Lin-Shu dispersion relation 
and Lin-Shu-Kalnajs dispersion relation, respectively.
We first outline the physical meaning of the dispersion relations 
based on the linear tight-winding density wave theory of fluid and stellar discs (Section 2.1.1). 
Then, we explain the global mode theory of galactic discs (Section 2.1.4).
The behaviour of gas in quasi-stationary density waves will be described in Section 3.5. 

\subsubsection{Dispersion relations of tight-winding density waves}

Lin and Shu derived dispersion relations of fluid and stellar discs 
under the following approximations and assumptions:

\begin{enumerate}
\item Linear perturbations: 
They linearized the equation of continuity, the equations of motion (Euler equation), 
the equation of state, and Poisson equation, and then studied the behaviour of
the linear perturbations. 
In this case, it is assumed that the unperturbed disc is axisymmetric and has no radial motions.

\item 
Tight-winding approximation (short wavelength or WKB\footnote{
Named after the Wentzel-Kramers-Brillouin approximation of quantum mechanics.
} approximation): 
Deriving the dispersion relation for a general spiral wave is extremely complicated 
because of the long-range force nature of gravity (see Section 2.1.4). 
They assumed the spiral arm has a small pitch angle in order that 
distant density perturbations can be neglected. 
In other words, the response of the matter to the gravity perturbations becomes local. 
In this approximation, the dispersion relation of density waves can 
be written down in terms of local quantities. 
If we write the radial dependence of any perturbation quantity 
(Equation \ref{eqn:spiralarm}) in terms of an amplitude and phase as 
\begin{equation}
 A(R) = \varPhi(R) e^{if(R)}, 
 \label{eqn:shapefunction}
\end{equation}
the tight-winding approximation corresponds to the assumption 
that the phase $f(R)$ varies rapidly in comparison
with amplitude $\varPhi(R)$, i.e., 
\begin{equation}
 \left| \frac{df}{dR}\right| \gg \left| \frac{1}{\varPhi}\frac{d\varPhi}{dR}\right|.
\end{equation}

\item Quasi-stationary spiral structure hypothesis (QSSS hypothesis): 
They hypothesized that `global' spiral arms hardly change their shape 
during many galaxy rotational periods (`quasi-stationarity') 
based on empirical grounds \citep{LinShu1964,BertinLin1996,Bertin2000}. 
This is equivalent to assuming that the rotation of global spiral arms is 
rigid-body rotation with a specific angular velocity and pitch angle.  
This fixed angular speed is called the pattern speed. 
In the inner parts of galaxies, stars and gas rotate faster than the pattern speed, 
and overtake the spiral arm. 
In the outer parts of galaxies, stars and gas rotate slower than the pattern speed, 
and the spiral arms overtake the stars and gas. 
\end{enumerate}

Taking into account these approximations and assumptions, 
the Lin-Shu theory is often called 
a linear tight-winding, or quasi-stationary density-wave theory. 

Since the stars occupy most of the mass of the galactic disc, 
the dispersion relation of the density wave of a stellar disc is 
important in understanding the spiral arms. 
However, the dispersion relation for a fluid disc is simpler, so we introduce this first, 
before discussing the dispersion relation for a stellar disc. 
We refer the reader to \citet{BinneyTremaine2008} and \citet{Shu1992b}
for the mathematical details on derivation of dispersion relations,
as well as \citet{BertinLin1996} and \citet{Bertin2000} 
for further discussion on the concept of quasi-stationary density wave theory.

The dispersion relation for linear tight-winding perturbations in the razor-thin fluid disc 
(Lin-Shu dispersion relation; LS dispersion relation) is given by
\begin{eqnarray}
 (\omega - m\Omega)^2 = c_s^2k^2 + \kappa^2 -  2\pi {\rm G} \Sigma_{\rm 0} |k|,
 \label{eq:LinShu}
\end{eqnarray}
\citep{LinShu1964}. Here $\Omega$, $\Sigma_0$, $c_s$ and $\kappa$ are 
the angular frequency, surface density, sound speed and 
epicyclic frequency 
\begin{equation}
 \kappa=\sqrt{R \frac{\partial\Omega^2}{\partial R}+4\Omega^2}
\end{equation}
of the fluid disc, respectively. 
These quantities depend on the galacto-centric radius $R$
and define the axisymmetric basis state.
$\omega$, $k$, and $m$ are the angular frequency, radial wave-number, and the number of 
spiral arms, respectively. 
The radial wave-number $k$ is related to the phase of 
the radial dependence of the perturbations $f(R)$ (Equation \ref{eqn:shapefunction}) via
\begin{equation}
 k(R) \equiv \frac{df(R)}{dR}. 
\end{equation}
We define $k>0$ for a trailing spiral arm and $k<0$ for a leading spiral arm. 
Note that the so-called pattern speed $\Omega_{\rm p}$ is defined as $\omega/m$.

In the inertial frame, $\omega$ on the left hand side of Equation (\ref{eq:LinShu}) is 
the angular frequency of the density wave. 
In the rotating frame at some radius $R$ in the disc, 
$(\omega-m\Omega)$ is the angular frequency of the density wave experienced by a star at $R$.
A perturbation to the disc will be of the form $\exp[-i(\omega-m\Omega)t]$. 
Positive $(\omega-m\Omega)^2$ means that the perturbations to the disc will be stable. 
However negative $(\omega-m\Omega)^2$ means that the perturbations will be of the form 
$\exp[\pm|\omega-m\Omega|t]$ and there is a perturbation whose amplitude grows exponentially, 
thus the disc is unstable.
Therefore, the right-hand side of equation (\ref{eq:LinShu}) indicates the stability of the density waves. 
Here, we can introduce a dimensionless parameter
\begin{equation}
 Q \equiv \frac{\kappa c_s}{\pi G \Sigma_0},
 \label{eqn:gasQ}
\end{equation}
known as the Toomre $Q$ parameter, 
such that if $Q>1$, $(\omega-m\Omega)^2>0$ for all radial wave-numbers $k$, while if $Q<1$, 
$(\omega-m\Omega)^2$ becomes negative for a range of radial wave-numbers.
Therefore $Q$ gives us a criterion whether the disc is unstable or not to tight-winding perturbations\footnote{
A physical interpretation of $Q$ arises from comparing the timescale 
for gravitational collapse $ \sim (\lambda/G \Sigma_0)^{1/2}$ to those for shear, 
$\sim 1/\kappa$ and pressure ($\sim \lambda/c_s$). 
Requiring that a region size $\lambda$ collapse on a timescale shorter than 
the time for shear or pressure to react leads also to $Q$ as for 
Equation~\ref{eqn:gasQ}, but without the numerical denominator \citep{Pringle2007}.}.  
We can also define a critical unstable wavelength $\lambda_{\rm crit}=2 \pi /k_{\rm crit}$,
where $(\omega-m\Omega)^2 = 0$ for a cold fluid disc (i.e., $c_s=0$).
In this case, all perturbations with wavenumber $|k|<k_{\rm crit}$ or wavelength 
$\lambda > \lambda_{\rm crit}$ are unstable, where
$k_{\rm crit} = \kappa^2/(2\pi G \Sigma_0)$ or $\lambda_{\rm crit} = 4\pi^2 G \Sigma_0/\kappa^2$.
Note $k_{\rm crit} = k_{\rm min}/2$, where $k_{\rm min}$ is defined such that 
$(\omega(k_{\rm min})-m\Omega)^2 = 0$ for a neutrally stable fluid disc ($Q=1$).

\begin{figure*}[htbp]
\begin{center}
\includegraphics[width=0.4\textwidth]{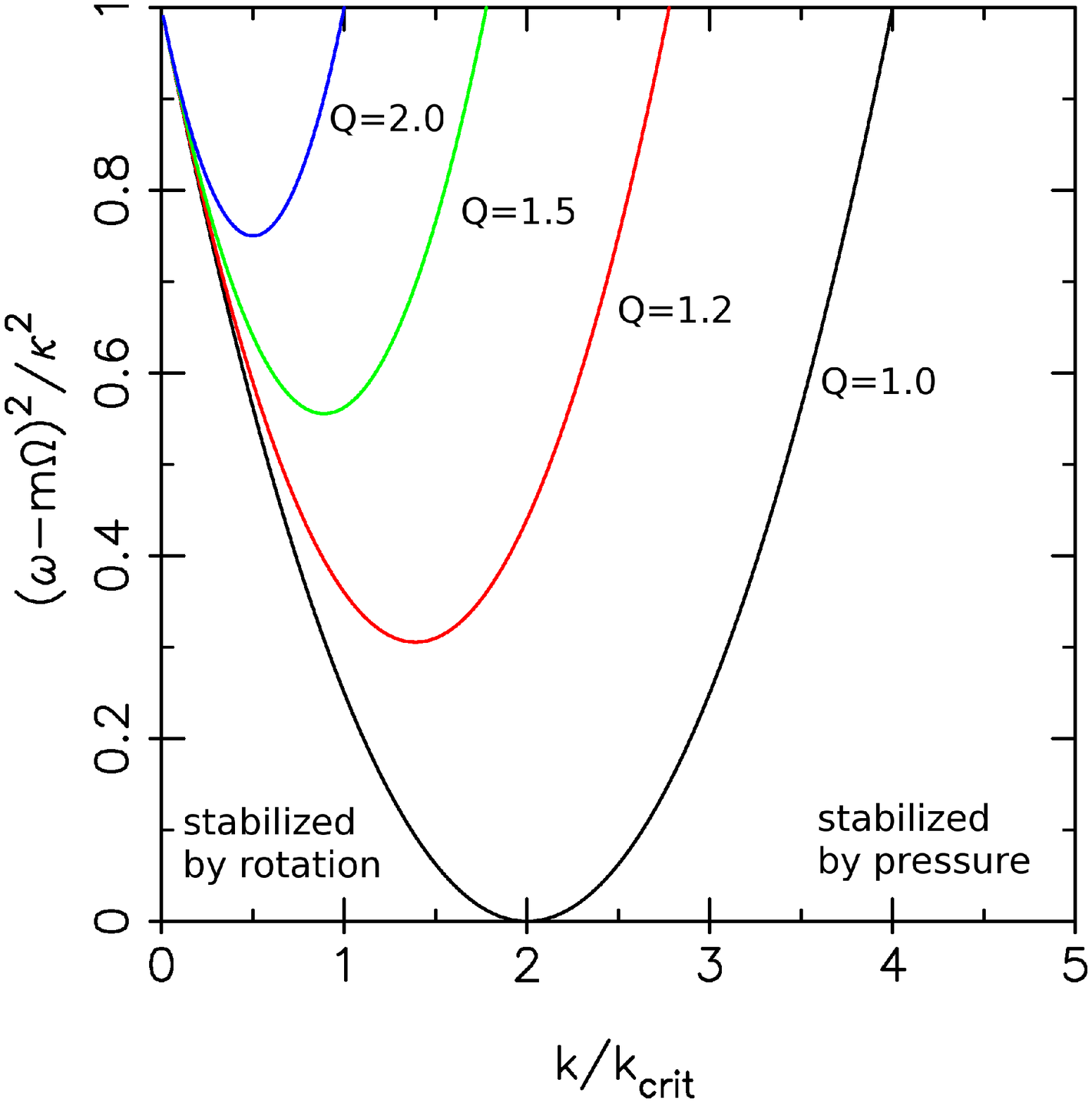}
\includegraphics[width=0.4\textwidth]{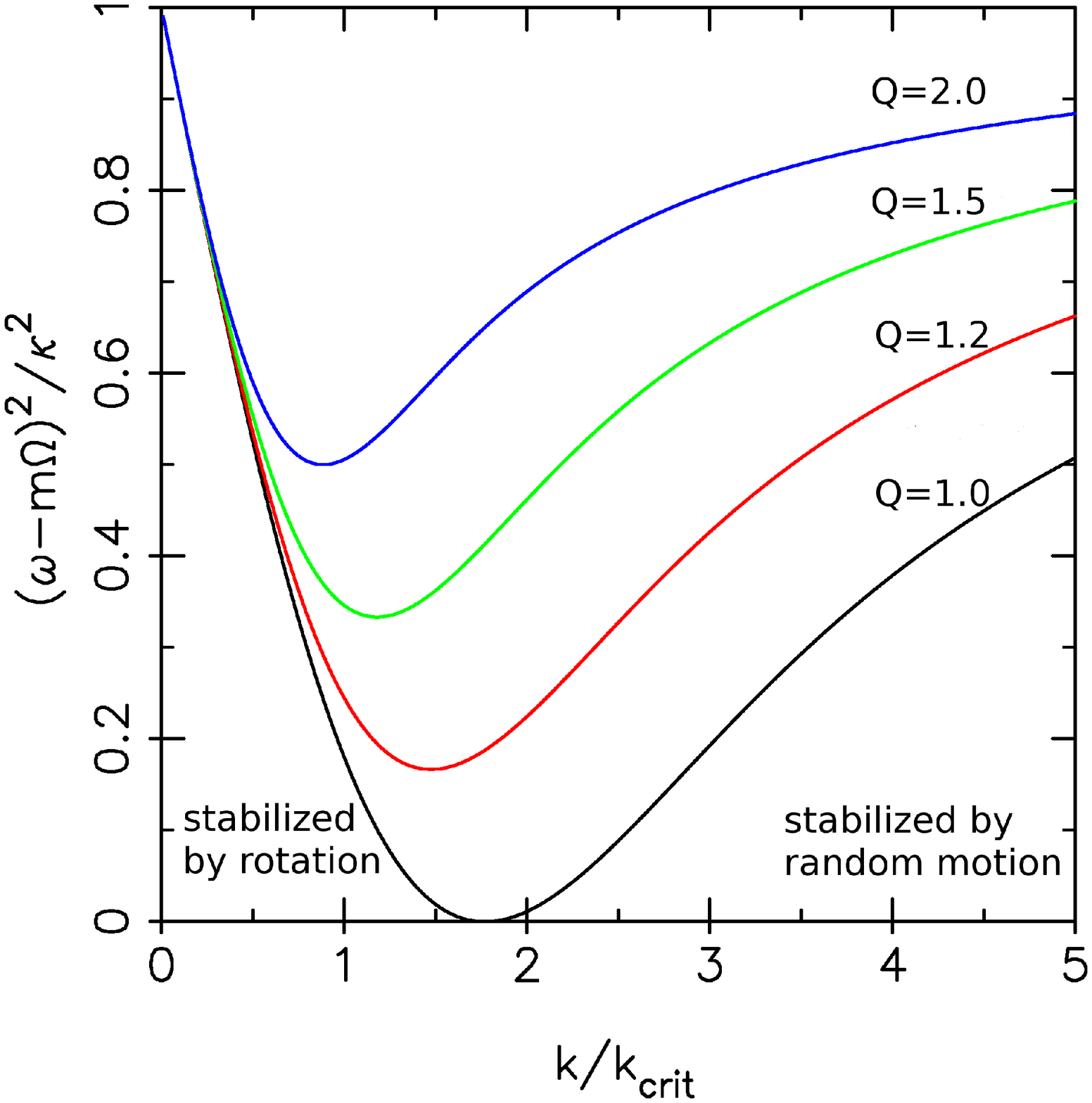}
\caption{
	Dispersion relations for tight-winding density waves in a fluid disc (left) and stellar disc (right).
	Waves of a wavenumber smaller than that at the minimum frequency ($|k| \ll k_{\rm crit}$)	
	are called long waves, while those with $|k| \gg k_{\rm crit}$ are called short waves. 
	The critical wavenumber $k_{\rm crit}$ is defined as $\kappa^2/(2\pi {\rm G} \Sigma_0)$.
	}
\label{fig:DispersionRelation}
\end{center}
\end{figure*}

Figure \ref{fig:DispersionRelation}a shows the Lin-Shu dispersion relations for different $Q$ values. 
Figure~\ref{fig:DispersionRelation}a shows that high $Q$ values (stability) occur for density waves 
with large wavelengths and angular frequencies.
The physical meaning of each term of the right-hand side of equation (\ref{eq:LinShu}) is as follows. 
The first term, $c_s^2k^2$, expresses the effect of pressure, which being positive stabilizes 
the fluid against perturbations. This  is the same as the dispersion relation of sound waves. 
The second term, $\kappa^2$ represents rotation, which again stabilizes the disc. 
The third term, which incorporates the self gravity of the disc, promotes the growth of instabilities.
When the effect of self-gravity exceeds the limit where $(\omega-m\Omega)^2$ is non-negative, 
a real root does not exist and it is impossible for a stable density wave to exist. 
Thus, the LS dispersion relation shows that the (gaseous) spiral density wave can be 
considered to be a sort of acoustic wave taking into account the effects of rotation and self-gravity. 

Consider now a stellar disc. The dispersion relation for linear tight-winding perturbations 
a in razor-thin stellar disc with a modified Schwarzschild
distribution (Lin-Shu-Kalnajs dispersion relation; LSK dispersion relation) is given by 
\begin{eqnarray}
 \label{eq:LinShuKalnajs}
 &&(\omega - m\Omega)^2 = \kappa^2 -  2\pi G \Sigma_0 |k| 
 	\mathcal{F}\left(\frac{\omega-m\Omega}{\kappa},\frac{\sigma_{\rm R}^2k^2}{\kappa^2}\right)\\
 && \mathcal{F}(s,\chi) 
 	 \equiv \frac{2}{\chi}(1-s^2)e^{-\chi}\sum_{n=1}^{\infty}\frac{I_{\rm n}(\chi)}{1-s^2/n^2},
\end{eqnarray}
\citep{LinShu1966,Kalnajs1965}. 
Here, $\sigma_{\rm R}$ is the radial velocity dispersion 
of the stellar disc, and $I_{\rm n}$ is a modified Bessel function.
Figure \ref{fig:DispersionRelation}b shows the LSK dispersion relations for 
different $Q$ values defined by
\begin{equation}
 Q \equiv \frac{\kappa \sigma_R}{3.36 G \Sigma_0}.
  \label{eqn:starQ}
\end{equation}
The behaviour of the dispersion relation is similar to the LS relation
for smaller radial wave-number (larger wavelength), 
but in the  the larger radial wave-number (shorter wavelength) regime, 
the behaviour of the two is decidedly different. 
For the short-wave regime, the LSK dispersion relation approaches 
$(\omega - m\Omega)^2/\kappa^2=1$ asymptotically, 
but the LS dispersion relation extends to $(\omega - m\Omega)^2/\kappa^2>1$.
This difference originates in the essential difference between the pressure for a fluid disc, 
and the velocity dispersion for a stellar disc: 
In the case of fluid discs, pressure will become large at small wavelengths.
In contrast, since the stellar disc is collisionless, 
there is no such repelling force.
Instead the frequencies of perturbations cannot become larger than 
the epicyclic frequency $\kappa$.

\subsubsection{Propagation of tight-winding density waves}
Although we have discussed waves as being quasi-stationary in the previous section, in reality if a wave is induced in the disc, it will propagate radially with some group velocity, rather than being stationary. In this and the next sections, we will discuss the group velocity and describe the further developments in density wave theory (the setting up of suitable boundary conditions) which allow the possibility, at least theoretically, of setting up a standing wave.
The propagation of the tight-winding density waves is reviewed throughly in \citet{Grosbol1994}
and \citet{BinneyTremaine2008}.

The original density wave theory \citep{LinShu1964} is based on the QSSS hypothesis which 
assumes that the amplitude and shape of the spiral arm are independent of time.
However, since the angular frequency $\omega$ depends on radial wave-number $k$ 
via the LS or LSK dispersion relations (eqs. \ref{eq:LinShu} and \ref{eq:LinShuKalnajs}), the
energy and angular momentum of the density waves propagate radially as wave packets \citep{Toomre1969}. 
This propagation velocity is the group velocity, given by $ v_g = \partial \omega(k,R)/\partial k$. 
If we consider propagation of density waves in a fluid disc,
following the LS dispersion relation (Equation \ref{eq:LinShu}), the group velocity of a wave packet is
\begin{eqnarray}
 v_g &&= \frac{\partial \omega(k,R)}{\partial k} 
 = \pm \frac{|k|c_s^2 - \pi G \Sigma_0}{\omega-m\Omega}, 
\end{eqnarray}
where positive and negative signs indicate trailing ($k>0$) and leading ($k<0$) spiral waves, respectively.
 
\begin{figure}[htbp]
\begin{center}
\includegraphics[width=0.4\textwidth]{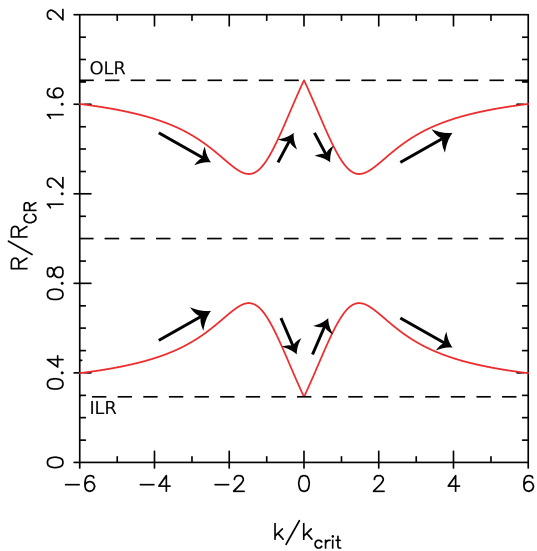}
\caption{
	Propagation diagram for tight-winding stellar density waves following the LSK dispersion relation (Equation \ref{eq:LinShuKalnajs}).
	The disc is assumed to have a flat rotation curve and constant Toomre's $Q=1.2$.
	The horizontal dashed lines are the OLR radius (upper), CR radius (middle), and ILR radius (lower), respectively.
	The arrows indicate the directions of group velocities.
	Long waves ($|k/k_{\rm crit}| \ll1 $) are reflected at the Lindblad resonances, while
	short waves ($|k/k_{\rm crit}| \gg 1 $) are absorbed there due to Landau damping.
	}
\label{fig:DensityWavePropagation}
\end{center}
\end{figure}

The sign of the numerator of this equation is negative for short waves and positive for long waves,
and the sign of the denominator is negative and positive for $R<R_{\rm CR}$ and $R>R_{\rm CR}$, respectively.
Thus, short trailing and long leading spiral waves will propagate away from the coronation (CR) radius,
while the short leading and long trailing spiral waves will approach the CR radius 
(propagation directions are indicated by arrows in Figure \ref{fig:DensityWavePropagation}).
Note that if the disc has a large $Q$, a forbidden region emerges in the vicinity of the CR, 
where due to the pressure or random motions the density waves diminish.

The behavior is essentially same for stellar density waves
except for regions around the inner and outer Lindblad resonances (ILR and OLR). 
The difference around the ILR/OLR originates in
the difference in dynamical behavior between stars and fluid (Section 2.1.1).
The propagation digram for stellar density waves following the LSK dispersion relation (Equation \ref{eq:LinShuKalnajs})
is shown in Figure \ref{fig:DensityWavePropagation}. 
Long stellar density waves ($|k/k_{\rm crit}| \ll 1$) are reflected 
at the Lindblad resonances \citep{GoldreichTremaine1978,GoldreichTremaine1979}
while short waves ($|k/k_{\rm crit}| \gg 1$) are absorbed there
due to Landau damping \citep{Lynden-BellKalnajs1972}.
Thus, both long and short stellar density waves cannot pass through the Lindblad resonances. 
Therefore, the permitted region for stellar density waves is restricted between the ILR and OLR radii
(with the exception again of the forbidden region).
However, this does not necessarily imply that stationary density waves will exist here.

If we apply the group velocity formula to the solar neighborhood, 
$v_{\rm g} \sim 12~\rm km~s^{-1}$ the stellar density wave takes $\sim 400$ Myr to propagate $5$ kpc radially. 
This timescale is comparable to the rotation period of the Galaxy.
Therefore, the stellar density waves will have a short lifetime of order $< 1~\rm Gyr$ \citep{Toomre1969}.

This problem can be solved if the density waves are reflected in the central region 
before reaching the ILR, and amplified by some mechanism. 
An absorption of the short stellar density waves at the ILR can be avoided if the Toomre's $Q$ parameter 
increases significantly (forming a so-called $Q$-barrier) refracting the density wave outside the ILR.
Short trailing stellar density waves can be excited near the CR from long trailing stellar density waves 
by `the wave amplification by stimulated emission of radiation' (WASER) in lighter discs \citep{Mark1974,Mark1976b},
or from short leading density waves  by the swing amplification mechanism in heavier discs 
\citep{GoldreichLynden-Bell1965,JulianToomre1966,GoldreichTremaine1978,Toomre1981}.
With these assumptions, `standing-wave' patterns \footnote{
Here `standing' means that the density waves do not propagate radially but do propagate azimuthally with a
pattern speed. }
can exist between a reflecting radius in the inner part of the galaxy and CR radius, 
where the waves can be amplified \citep{Bertin+1989a,Bertin+1989b}.
The spiral density waves should be located between, but not reaching the ILR and OLR.

\subsubsection{Swing amplification}

The quasi-stationarity of spiral arms requires wave amplification mechanisms such as WASER \citep{Mark1974,Mark1976b} 
or swing amplification \citep{GoldreichLynden-Bell1965,JulianToomre1966,GoldreichTremaine1978,Toomre1981}.  
In the WASER (swing amplification) mechanism, a trailing (leading) wave is turned into a trailing wave 
when crossing CR and is greatly amplified in the process. 
It is noted that there is no conflict between the swing amplification and WASER 
\citep{LinThurstans1984,Bertin+1984,Bertin+1989b}. 
The reason they have been considered differently is largely historical, reflecting opposing views at the time.
However, if discs have non-negligible self-gravity at CR, 
the swing amplification mechanism can greatly dominate amplification by the WASER mechanism. 
On the other hand, in a system where the disc mass is only a small fraction that supports the rotation curve, 
the WASER mechanism can underline the growth of the most important spiral mode, 
as long as $Q \simeq 1$ at CR \citep{Shu1992b,BertinLin1996,Bertin2000}.

We focus on the swing amplification mechanism
as a wave amplification mechanism for sustaining quasi-stationary density waves 
between the ILR and OLR.
The so-called swing amplification works
when short leading waves are reflected to short trailing waves at the CR radius, 
or when a density enhancement formed by self-gravity is stretched out by differential rotation.
The dynamical response takes the form of wavelets in the surrounding medium, 
each amplified by its own self-gravity through the swinging of 
leading features into trailing ones due to shear. 

The swing amplification operates through a combination of three ingredients: 
the shearing flow, epicyclic motions, and the disc self-gravity.
 \citet{Toomre1981} interpreted the swing amplification mechanism 
in terms of the wave-particle interaction between spiral arms and stars. 
Since the direction of epicyclic motion of a star is the same as the direction 
which the spiral arm is sheared by differential rotation, stabilisation by rotation is reduced, 
and the perturbation can grow via the usual Jeans instability  
\citep{GoldreichLynden-Bell1965,JulianToomre1966,GoldreichTremaine1978}. 
The timescale of epicyclic motion ($\kappa^{-1}$) is comparable to the timescale of 
involvement with the spiral arm ($A^{-1}$ where $A$ is Oort's constant), and unless $Q \gg 1$, 
the structure can grow in a short time comparable to $\kappa^{-1}$.
The resulting spiral structure from this process is generally expected to be chaotic 
\citep{Sellwood2011} rather than lead to a symmetric spiral pattern.

Consider a local region of a galactic disc away from the galaxy center. 
Since the galactic rotation is parallel to this local region (curvature can be ignored), 
we set an $x$ and $y$-axis aligned with the radial and rotational directions of the galaxy respectively.
In this case, the equations of motion of the stars are given by
\begin{eqnarray}
 \ddot{x} - 2\Omega_{\rm 0} \dot{y} - 4\Omega_{\rm 0}A_{\rm 0} x = f_{\perp}\sin \gamma,\\
 \ddot{y} + 2\Omega_{\rm 0} \dot{x} = f_{\perp} \cos \gamma,
\end{eqnarray}
(using the Hill approximation). Here, $x = R- R_{\rm 0}$, $y = R_{\rm 0} (\phi - \Omega_{\rm 0}t)$, 
and $\Omega_{\rm 0}$ and $A_{\rm 0} \equiv -\frac{1}{2} R_{\rm 0}(d\Omega/dR)_{\rm 0}$ 
are the angular velocity and Oort's constant at $R_{\rm 0}$, respectively.
$f_{\perp}$ indicates the gravitational force perpendicular to the spiral arm. 
$\gamma$ is an angle between the spiral arm and radial direction of the galaxy: 
$\gamma=90^\circ$, $\gamma<0$, and $\gamma>0$ correspond to 
a ring, leading, and trailing structures, respectively.

\begin{figure}[htbp]
\begin{center}
\includegraphics[width=.45\textwidth]{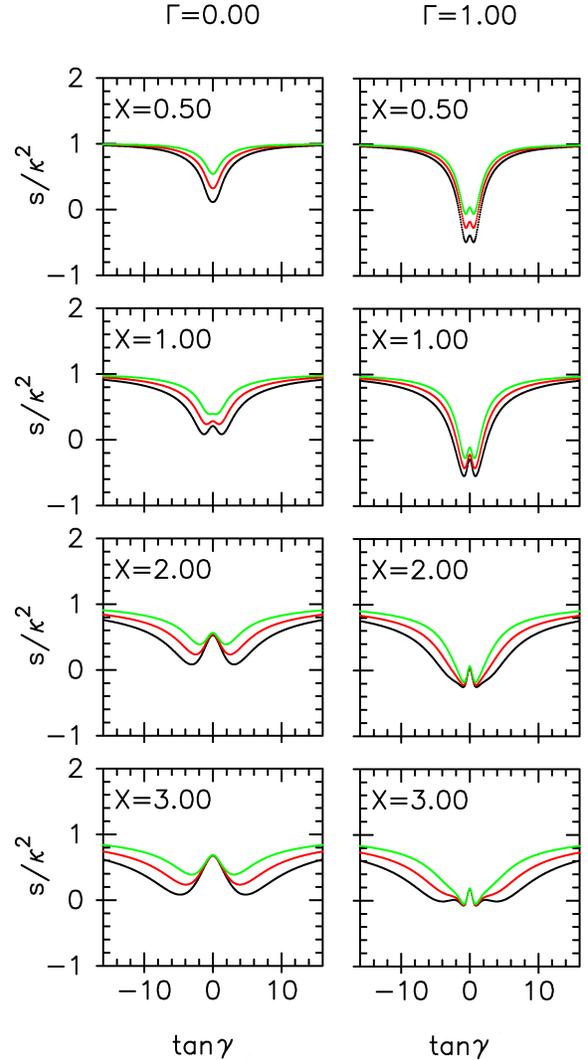}
\caption{
	Squared spring rate $S(\gamma)$ as a function of the angle $\gamma$ 
	between the spiral arm and radial direction of the galaxy
	for $\Gamma = 0.0$ (rigid rotation) and $\Gamma=1.0$ (flat rotation).
	Different lines indicate $Q=1.0$ (black), $1.2$ (red), and $1.5$ (green), respectively.
	Spring rates are calculated based on the equations of motion in 
	\citet{Toomre1981} and \citet{Athanassoula1984}.
	The squared spring rate is always positive in the case of $\Gamma = 0.0$, 
	but it can be negative in the case of $\Gamma=1.0$. Thus, the normal displacement of the stars 
	around the spiral arm $\xi$ can grow exponentially as the spiral arm is sheared by differential rotation. 
}
\label{fig:SpringRate}
\end{center}
\end{figure}

Defining the normal displacement of the star perpendicular to the spiral arm,
\begin{eqnarray}
 \xi = x \sin \gamma + y\cos \gamma,
\end{eqnarray}
as a new variable, the equations of motion reduce to an equation 
\begin{eqnarray}
 \ddot{\xi} + S(\gamma)\xi = 0,
\end{eqnarray}
where the squared spring rate is given by
\begin{eqnarray}
 S(\gamma) &=& \kappa^2 - 8\Omega_{\rm 0} A_{\rm 0} \cos^2\gamma 
 		+ 12 A_{\rm 0}^2\cos^4\gamma - 2\pi G \Sigma_{\rm 0} k \mathcal{F} \\
		&=& 
		\left( 1 - \frac{2\Gamma}{2-\Gamma}\cos^2\gamma 
			+ \frac{3}{2}\frac{\Gamma^2}{2-\Gamma}\cos^4\gamma
			-\frac{\mathcal{F}}{X}\sec \gamma
		\right) \kappa^2, 
\end{eqnarray}
and 
\begin{eqnarray}
 && \Gamma = -\frac{d\ln \Omega}{d\ln R}, \\
 && X = \frac{k_{\rm crit} R}{m} = \frac{\kappa^2 R}{2\pi G \Sigma_0 m} \label{eq:X}.
\end{eqnarray}
Figure \ref{fig:SpringRate} shows dependence of the spring rate on galaxy parameters of $(\Gamma,Q,X)$.
In the case of $\Gamma=0$ (i.e, the galaxy has rigid-body rotation), the spring rate is always positive.
Thus, the stars cannot be trapped by the spiral arm, and then the spiral arm does not amplify.
In other words, the swing amplification cannot work without differential rotation.

\begin{figure*}[htbp]
\begin{center}
\includegraphics[width=.85\textwidth]{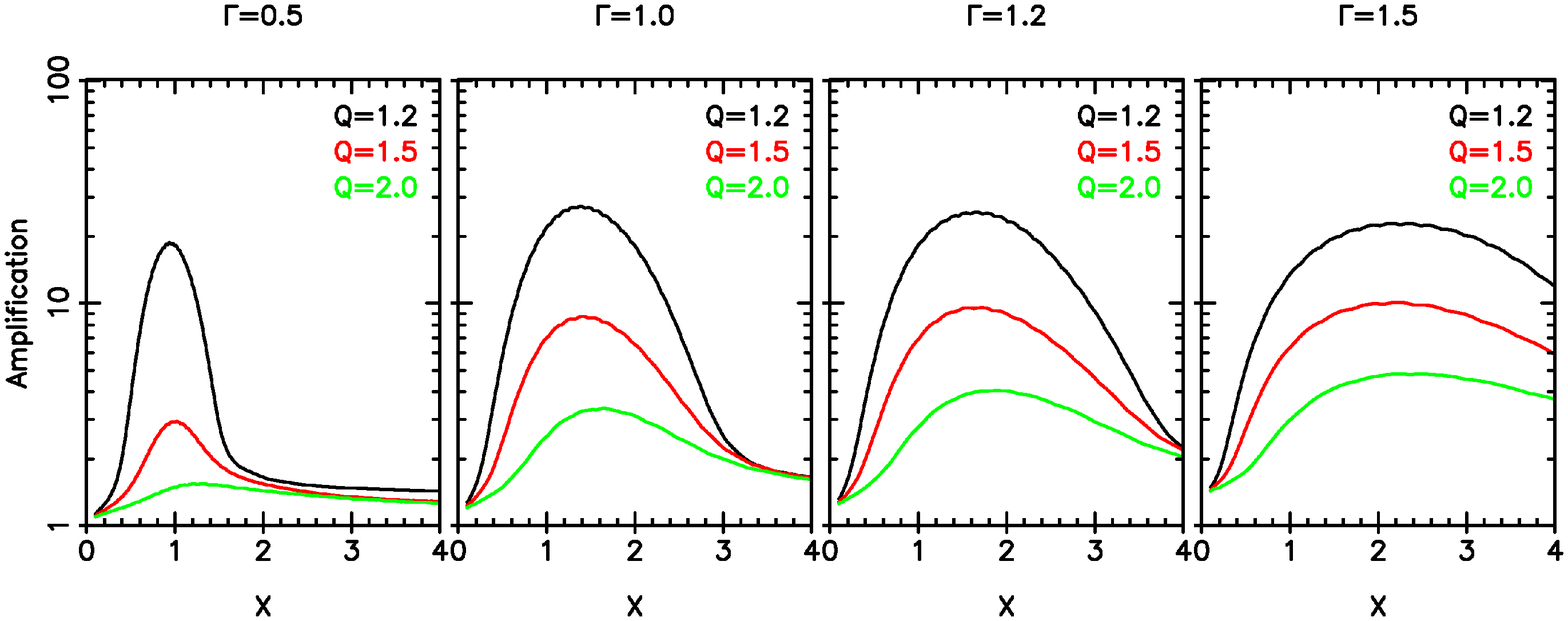}
\caption{
	The maximum amplification factor is shown as a function of the $X$, $\Gamma$ and $Q$ parameters.
	The amplification factor is calculated based on the equations of motion given in 
	\citet{Toomre1981} and \citet{Athanassoula1984}.
}
\label{fig:SWAfactor}
\end{center}
\end{figure*}

By the transformation of variables to $\gamma$, instead of time $t$, Equation 17 becomes
\begin{eqnarray}
 \frac{d^2\xi}{d\tan \gamma^2} + \frac{2(2-\Gamma)}{\Gamma^2}\frac{S(\gamma)}{\kappa^2}\xi = 0.
\end{eqnarray}
Numerical integration of this differential equation gives the dependence of the swing amplification 
factor on the galaxy parameters $(\Gamma,Q,X)$ shown in Figure \ref{fig:SWAfactor}. 
The effect of self-gravity and the winding of the spiral arm work in synergy, 
so that  a star comes to stay at the spiral arm for a long time, and spiral arms are amplified temporarily. 
Note that the above argument is based on the linear analysis by \citet{Toomre1981} 
and \citet{Athanassoula1984}. 
\citet{Fuchs2001a} solved linearised collisionless Boltzmann and Poisson equations 
self-consistently and showed that the result is essentially the same.
Further, non-linear effects are studied in $N$-body simulations of local regions of stellar discs 
\citep{Toomre1990,ToomreKalnajs1991,Fuchs+2005}, as well as $N$-body simulations of 
global stellar discs \citep{SellwoodCarlberg1984,CarlbergFreedman1985,Bottema2003,
Fujii+2011,Baba+2013,D'Onghia+2013}.
\citet{D'Onghia+2013} carefully demonstrated the growth of spiral arm 
features by swing-amplification and found a 
nonlinear evolution that is not fully consistent 
with the classic swing-amplification picture of \citet{JulianToomre1966} and
 lasted longer than predicted by swing amplification \citep{ToomreKalnajs1991} (see also Section 2.1.5).

In order for the swing amplification mechanism to work continuously \citep{ToomreKalnajs1991}, 
we need to understand how leading waves are generated. 
One possibility is the case where there is no ILR. 
A trailing wave does not suffer from Landau damping at the ILR, 
instead the wave turns into a leading wave as it crosses the galaxy center.
This is the so-called feedback loop proposed by \citet{Toomre1981}.

\subsubsection{Global mode theory}

Although linear density wave theory was successful in demonstrating the existence 
of a tight-winding spiral wave, the tight-winding density wave theory has room for improvement. 
Firstly, since they utilized the WKB approximation, 
this theory cannot be applied to very long waves (or open spiral arms) strictly. 
Secondly, the presence of {\it neutral} spiral density waves itself is theoretically questionable. 
It is critically problematic that a density wave propagates 
through a galactic disc radially in a few galactic rotations, 
and eventually disappears by absorption at the inner/outer Lindblad resonances (ILR/OLR).
Thus, the `quasi-stationarity' hypothesis is not ensured \citep[][Section 2.1.2]{Toomre1969}. 
Finally, the tight-winding theory cannot predict the number of spiral arms $m$
and sign of the wave-number $k$ (i.e., trailing or leading). In other words, the theory cannot explain why 
actual spiral galaxies prefer to have trailing two-armed spirals ($k>0$ and $m=2$)
and what determines the angular frequency of the spiral density wave.
In response to these criticisms, the tight-winding density wave theory developed into a global mode theory
\citep[e.g.,][]{Lau+1976,Bertin+1977,Aoki+1979,Iye1978,Bertin1983,Bertin+1984,
Bertin+1989a,Bertin+1989b,BertinLin1996}.

A key nontrivial step at the basis of the derivation of the dispersion relation is the 
reduction of the long-range gravity law to a WKB dispersion relation between the perturbed 
potential and the perturbed density. Numerical integration of the basic
perturbed equations is required.
Since the first global mode analysis was applied to rotating fluid discs 
by \citet{Hunter1965}, 
there have been many studies, mainly in the 1970s-80s
(e.g. \citealt{Bardeen1975,Aoki+1979,Iye1978,Takahara1978}),
as well as extending the analysis to rotating stellar discs \citep{Kalnajs1972}.
In order to analyze the eigen-value problem of a stellar system, 
it is necessary to solve the density perturbations and the responsive orbital perturbations
in satisfying the linearized collisionless Boltzmann equation and the Poisson equation, self-consistently.
Although \citet{Kalnajs1972} solved the eigenvalue problem of the Maclaurin disc 
using the so-called `matrix method', 
numerical integration is required to solve the eigen-value problems of stellar discs 
(e.g. \citealt{AthanassoulaSellwood1986,SellwoodAthanassoula1986,Sellwood1989,
EarnSellwood1995,VauterinDejonghe1996,PichonCannon1997,
Polyachenko2004,Polyachenko2005,JalaliHunter2005}).
However, these studies are somewhat 
limited due to the mathematical complexity.

Global mode analysis is based on a point of view that 
the spiral arms are manifestations of the gravitationally `unstable' 
global eigen-oscillations of disc galaxies \footnote{
\citet{Lynden-BellOstriker1967} have proved the so-call
anti-spiral theorem which argues that there is no neutral spiral mode 
unless there exists degeneracy of modes or dissipation mechanism. 
}.
This eigen-oscillation problem of the galactic disc resembles 
the problem of oscillating patterns of the skin when a drum is struck. 
Similar to the way oscillation patterns are controlled by how to stretch and how to strike the skin, 
oscillation of a galactic disc is controlled by the density and velocity-dispersion 
distributions of a galactic disc. 
However, there are two differences between eigen-oscillation problems of the drum and galactic disc.
First, changes of the gravity from the oscillation should be taken into account to
solve the eigen-oscillation problem of the galactic disc. This makes the problem very complicated. 
In the case of a drum, there is only a traverse wave, 
but for the oscillation of the galactic disc, there is also a longitudinal wave as well as a transverse wave.
The transverse and longitudinal waves in the galactic disc are equivalent to the
bending (warp) of a galactic disc and spiral arms, respectively.

\begin{figure}[htbp]
\begin{center}
\includegraphics[width=0.5\textwidth]{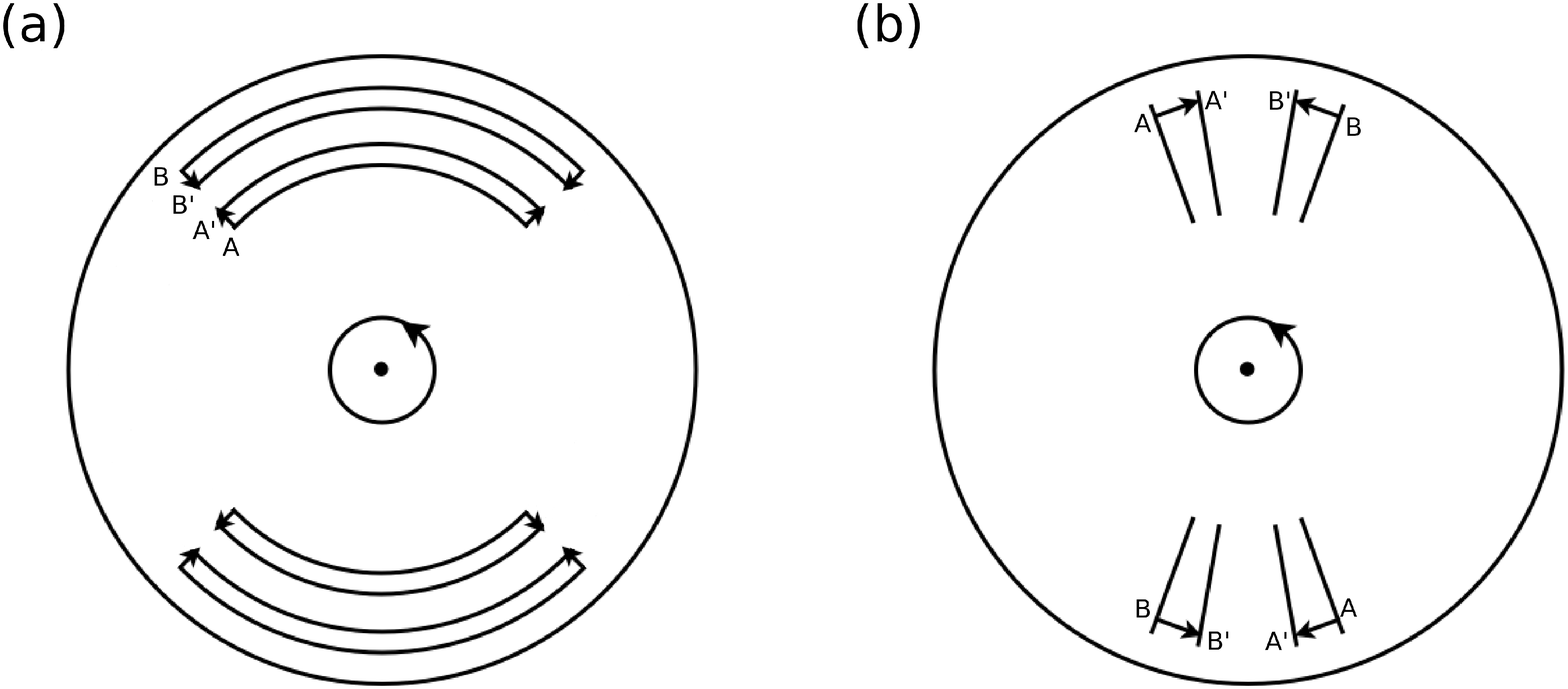}
\caption{
	Axisymmetric perturbations (a) and bar-like perturbations (b) on an axisymmetric disc.
	The disc rotates anti-clock wise.
	Directions of the perturbations are indicated by small arrows.
	}
\label{fig:discInstability}
\end{center}
\end{figure}

\begin{figure*}[htbp]
\begin{center}
\includegraphics[width=0.4\textwidth]{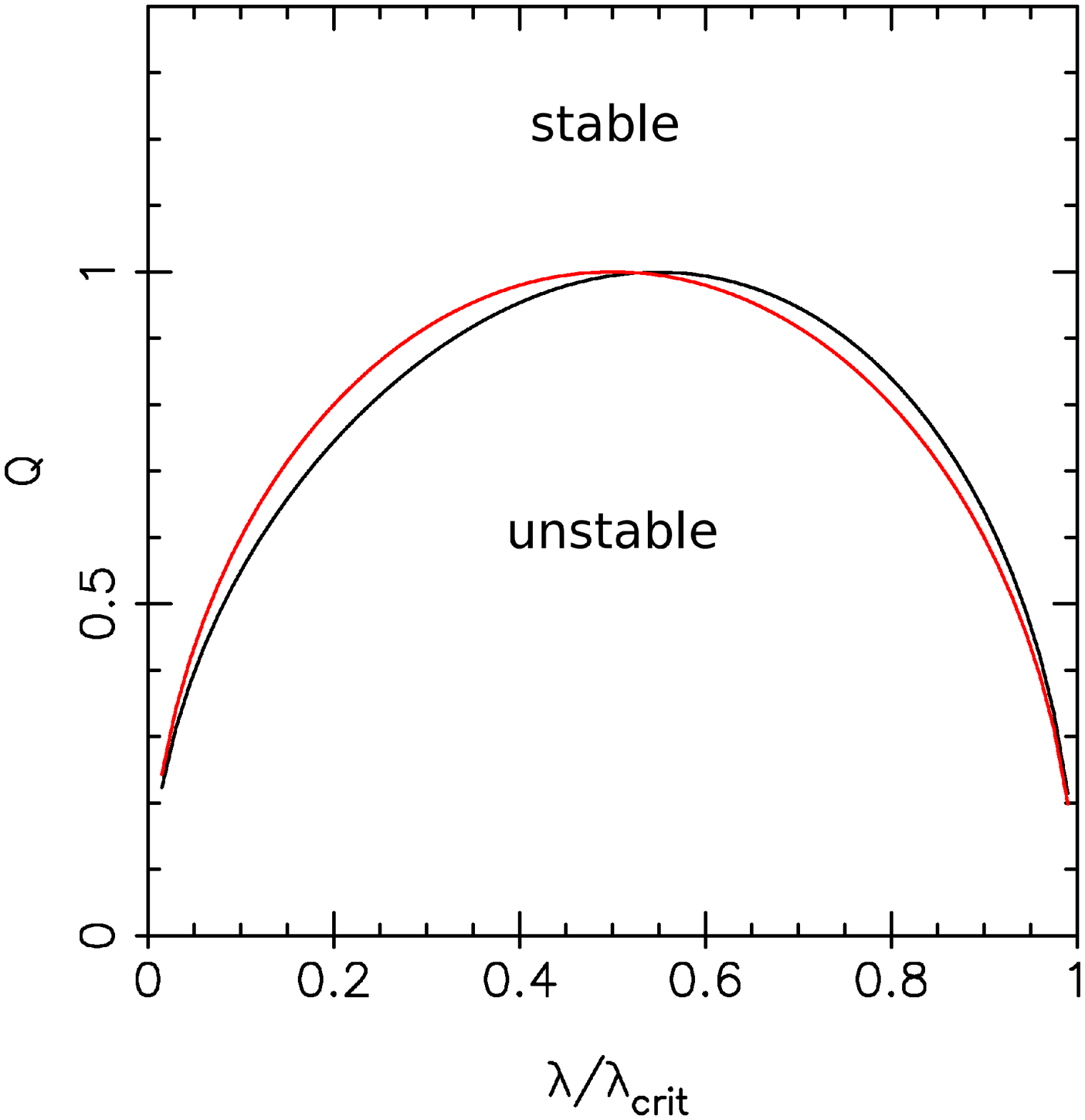}
\includegraphics[width=0.4\textwidth]{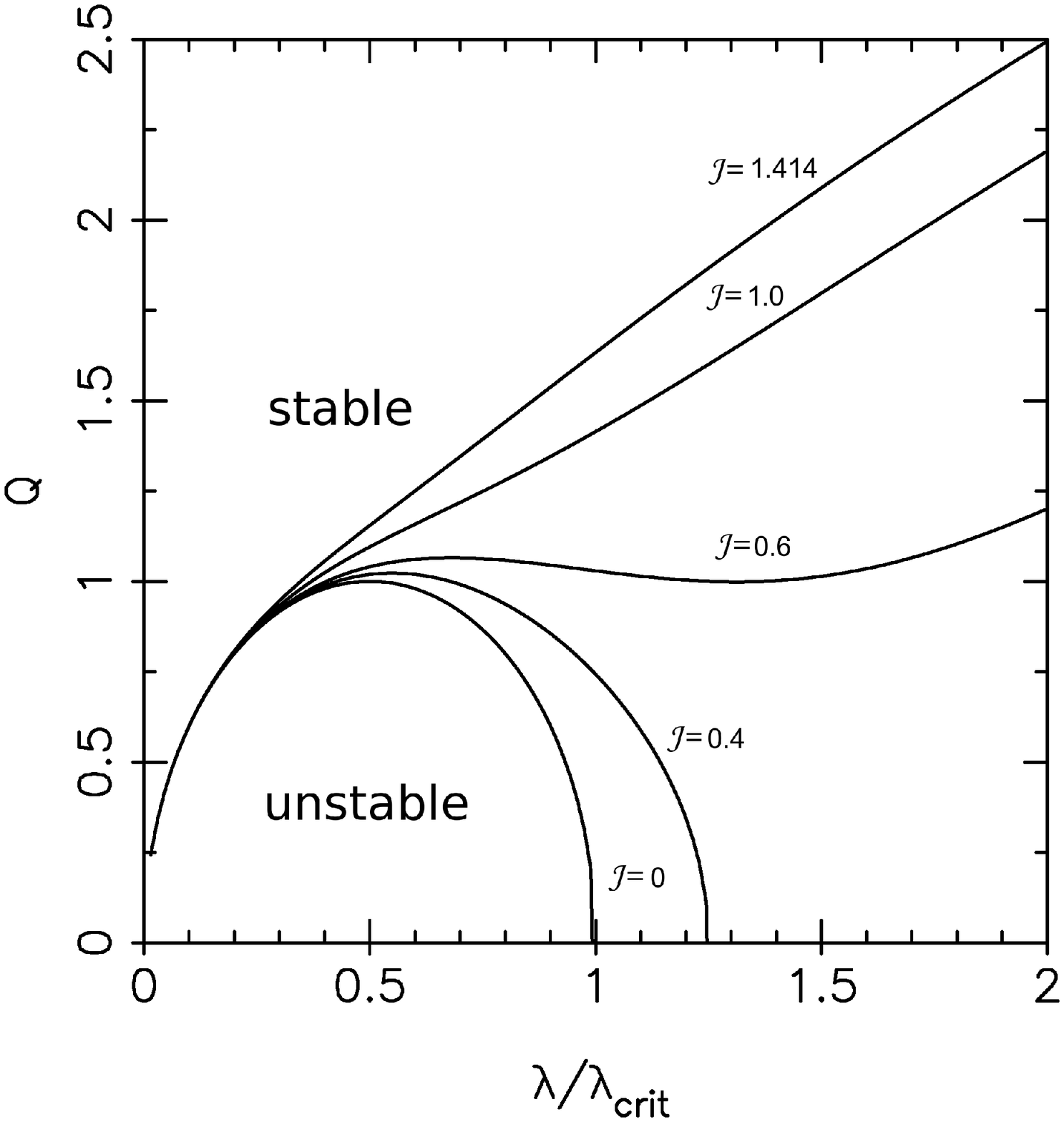}
\caption{
	Left: Neutral stability curves for tigiht-winding spiral instabilities based on the LS dispersion relation (red; Equation \ref{eq:LinShu}) 
	and LSK dispersion relation (black; Equation \ref{eq:LinShuKalnajs}). 
	The region below the curve is stable against tight-winding spiral instabilities.
	Right: Neutral stability curves for open spiral instabilities based on the BLL dispersion relation (Equation \ref{eq:BertinLauLin})
	with $\mathcal{J} = 0, 0.4, 0.6, 1.0$, and $1.414$.
	}
\label{fig:NeutralStableCurve}
\end{center}
\end{figure*}

Before explaining numerical results of the global mode analysis, 
let us consider the stability of fluid and stellar discs.
Consider first the case of axisymmetric perturbations \citep{Toomre1964,GoldreichLynden-Bell1965a}. 
Qualitatively, we consider the case where an axisymmetric disc receives a perturbation 
illustrated by the small arrows shown in Figure \ref{fig:discInstability}(a). 
The fluid and stars in the disc move radially, and try to make a ring structure, 
but the pressure (or velocity dispersion), centrifugal, and Coriolis forces 
suppress growth of this ring perturbation. 
Left panel of Figure \ref{fig:NeutralStableCurve} shows the neutral stability curves for tight-winding spirals 
($(\omega-m\Omega)^2=0$) showing $Q$-values as a function of $\lambda/\lambda_{\rm crit}$. 
As $Q$ for the disc is lowered, the disc moves from the stable to unstable regime. 
The wavelength which becomes unstable first is $p \lambda_{\rm crit}$, 
where $p= 0.5$ in the case of a fluid disc and $p=0.55$ in a stellar disc. 

In the case of spiral perturbations with a finite pitch angle (i.e., open spiral perturbations), 
the effects of the gravitational perturbation 
parallel to the spiral arm, and shear originating from differential rotation need to be taken into account. 
Again, we consider the case where an axisymmetric disc receives a perturbation 
given by the small arrows shown in Figure \ref{fig:discInstability}(b). 
In contrast to the case of Figure \ref{fig:discInstability}(a), 
the fluid and stars move in the azimuthal direction, and centrifugal and Coriolis forces do not appear.  
Thus, growth is not suppressed. 
The effect of the excessive centrifugal and Coriolis forces stabilises the perturbation 
with a long wavelength (Section 2.1.1 and Figure \ref{fig:DispersionRelation}), 
it is expected that the stabilization effect will become weak in a long wavelength regime
and that open spiral arms will become unstable. 

This qualitative expectation is checked quantitatively below.
\citet{LauBertin1978} derived the asymptotic dispersion relation 
of open spiral density waves in the fluid disc (Bertin-Lau-Lin dispersion relation; BLL dispersion relation): 
\begin{eqnarray}
 (\omega - m\Omega)^2 = 
 	\kappa^2 + k^2c_s^2\left[ 1 + \mathcal{J}^2\left(\frac{k_{\rm crit}}{k}\right)^2 \right] 
	\nonumber \\
	- 2\pi G \Sigma_0 |k| \left[1 +  \mathcal{J}^2\left(\frac{k_{\rm crit}}{k}\right)^2 \right],
 \label{eq:BertinLauLin}
\end{eqnarray}
where $k = \sqrt{k_R^2 + k_\phi^2}$, $k_R$, and $k_\phi = m/R$ are the wave-number, 
radial wave-number, and azimuthal wave-number, respectively \footnote{
Although the LS dispersion relation (Equation \ref{eq:LinShu}) is derived by the tight-winding approximation, 
i.e., $|R k_R| \gg 1$, the BLL dispersion relation (Equation \ref{eq:BertinLauLin}) is derived by an asymptotic
analysis based on the following ordering: $\epsilon_0^2 \ll 1$ and $(k/k_{\rm crit})^2 = \mathcal{O}(1)$.
Thus even very long waves with $|k_R| \ll 1$ can be described by means of a WKB treatment of
the gravitational potential, provided the quantity $m^2$ is taken to be formally large.
See \citet{Bertin2000} for more details.}.
We define two quantities 
\begin{eqnarray}
 && \mathcal{J} \equiv m \epsilon_{\rm 0} \left( \frac{4\Omega}{\kappa} \right)
 	\left| \frac{d\ln\Omega}{d\ln R} \right|,\\
 && \epsilon_{\rm 0} \equiv \frac{\pi G \Sigma_0}{R\kappa^2}.
\end{eqnarray}
where $\mathcal{J}$ indicates a stability parameter which depends on 
the disc mass relative to the total mass and a shear rate of the disc, and 
$\epsilon_{\rm 0}$ is a parameter which relates to the degree of the self-gravity.
In the LS dispersion relation,  as described in Section 2.1.1, the disc is stabilized by the rotational ($\kappa^2$)
and pressure effects ($k^2c_{\rm s}^2$) against self-gravity ($2 \pi G \Sigma_0 k$). 
In addition to these effects, the BLL dispersion relation includes within $\mathcal{J}$ the rate of shear, 
$\frac{d\ln\Omega}{d\ln R}$, as well as the self-gravity term, $\epsilon_{\rm 0}$. 

Based on the BLL dispersion relation, 
a neutral stability curve for spiral instabilities is given by  
\begin{eqnarray}
 Q^2 \geq 4 \left[ \frac{\lambda}{\lambda_{\rm crit}} 
 	- \frac{(\lambda/\lambda_{\rm crit})^2}{1+\mathcal{J}^2(\lambda/\lambda_{\rm crit})^2} \right],
 \label{eq:NeutralStableBLL}	
\end{eqnarray}
where $\lambda = 2\pi/k$ is a wavelength (right panel of Figure \ref{fig:NeutralStableCurve}).
If we set $\mathcal{J}=0$, the neutral stability curve is equivalent to 
ones for the LS dispersion relation.
From this neutral stability curve, a value of $Q$ larger than unity 
is required for stability against spiral disturbances with a larger value of $\mathcal{J}$. 
This means that open spiral arms are difficult to stabilise and will, more often than not, result in growth.

\begin{figure*}[htbp]
\begin{center}
\includegraphics[width=0.45\textwidth]{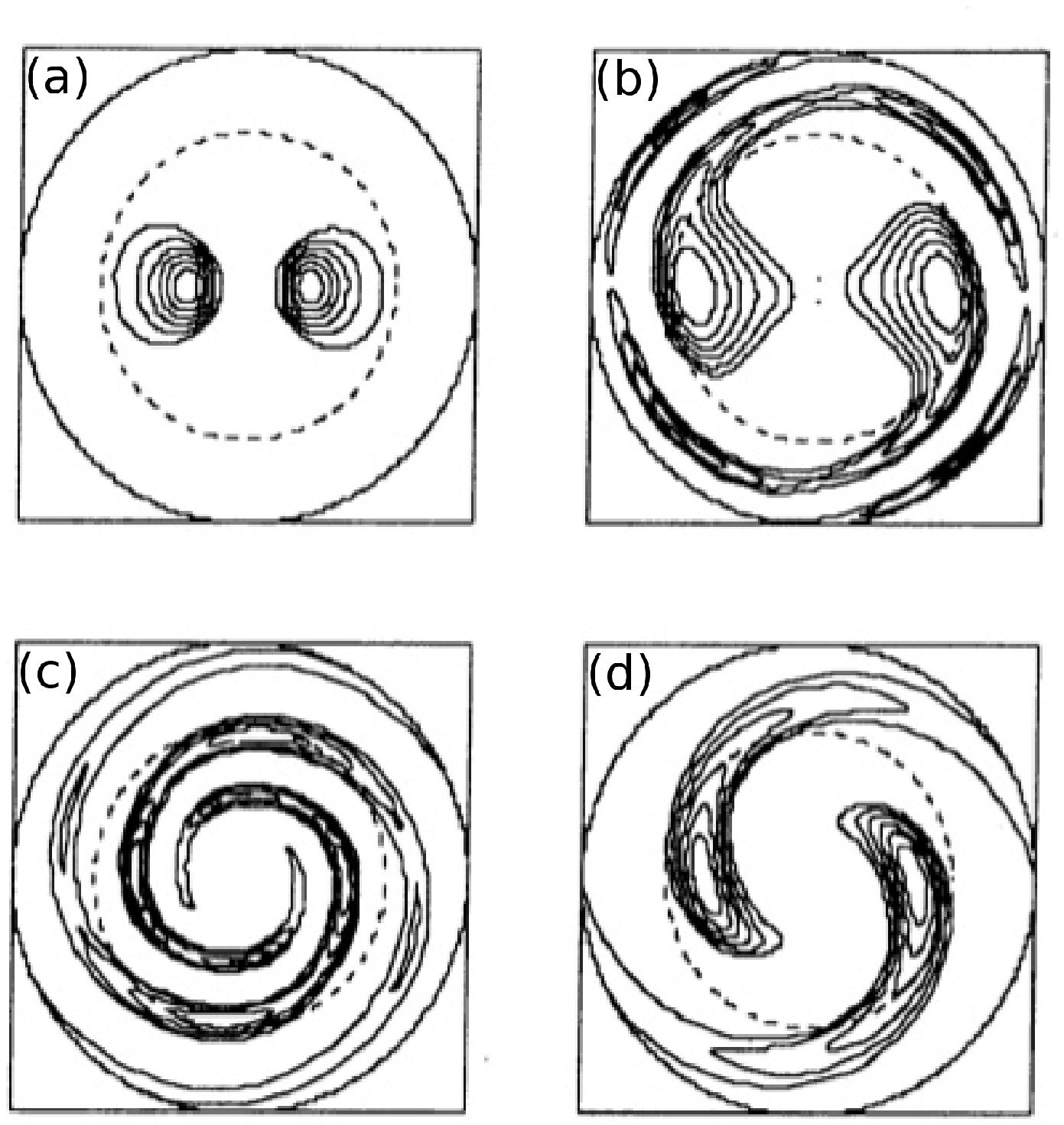}
\includegraphics[width=0.45\textwidth]{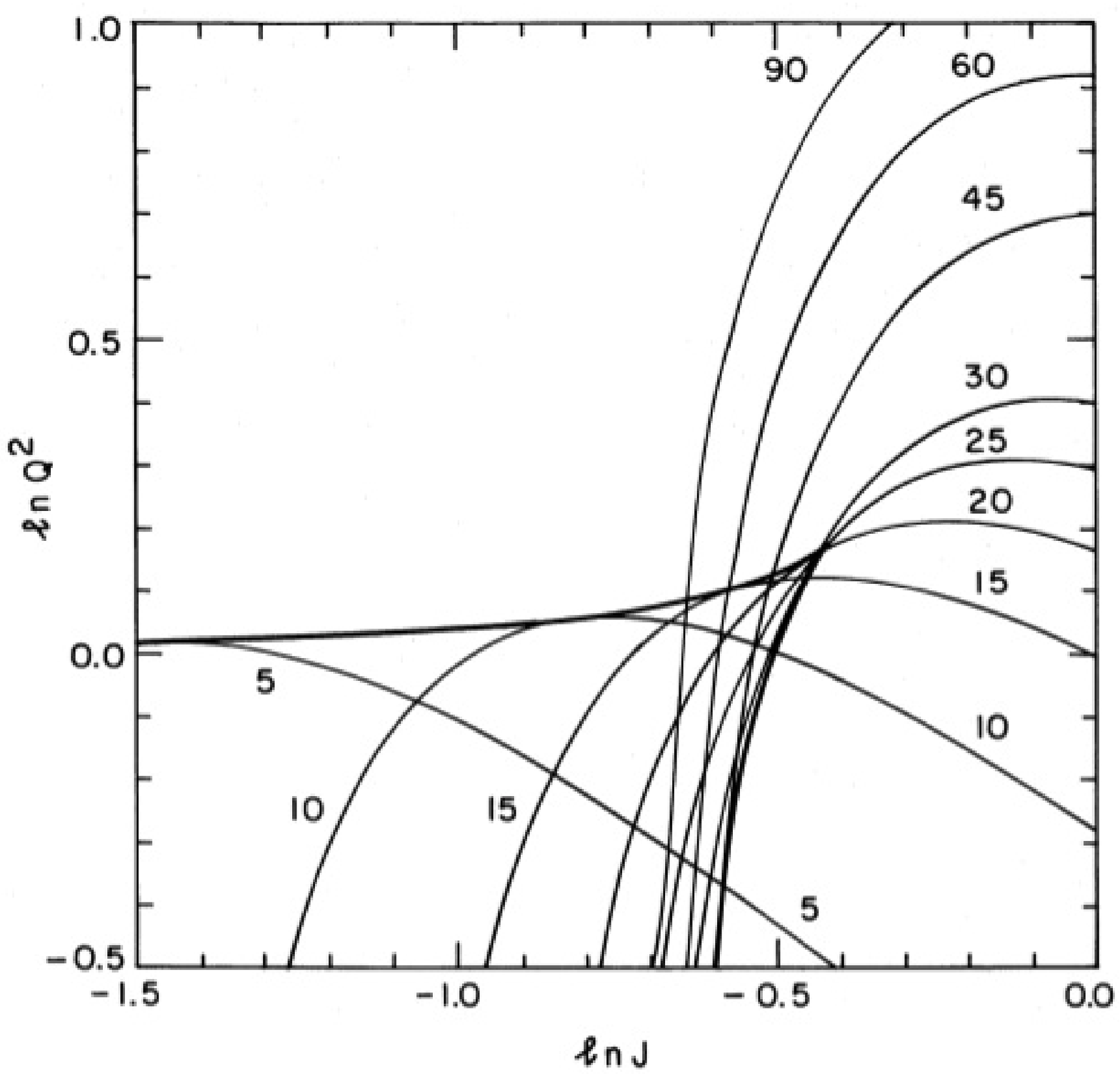}
\caption{
	Left: Density contours of global unstable modes for a rotating fluid disc where 
	(a) $\mathcal{J} = 0.604$ and $Q = 1.500$, (b), $\mathcal{J} = 0.538$ and $Q = 1.096$, 
	(c), $\mathcal{J} = 0.492$ and $Q = 1.002$, and (d) $\mathcal{J} = 0.858$ and $Q = 1.004$.
	Right: Curves of constant pitch angle $\alpha = \cot^{-1} \frac{k_\phi}{k_R}$ in the $(\mathcal{J},Q)$-plane.
	These curves are derived from the BLL dispersion relation (Equation \ref{eq:BertinLauLin}) 
	for the neutral stability condition (Equation \ref{eq:NeutralStableBLL})  with $\Gamma = 0$ (flat rotation curve).
	From \citet{Bertin+1989b}.
	}
\label{fig:globalmode}
\end{center}
\end{figure*}

Figure \ref{fig:globalmode} shows the numerically integrated density contours of the global unstable modes 
with different $\mathcal{J}$ and $Q$ values \citep{Bertin+1989b}. 
The pitch angle in the mode becomes smaller as the value of $\mathcal{J}$ 
decreases (panels (a), (b), and (c), respectively).
For the case where $\mathcal{J} $ and $Q$ are large (panel (a)), the bar mode becomes unstable. 
In the case of large $\mathcal{J}$ but small  $Q$ (panel (d)), 
only the spiral mode is unstable. 
This behaviour is in agreement with that expected from the BLL dispersion relation 
(right panel in Figure \ref{fig:globalmode}). 
The right panel in Figure \ref{fig:globalmode} shows 
curves of constant pitch angle $\alpha$ in the $(\mathcal{J},Q)$-plane. 
The pitch angle here (for the more general definition see Eqn~27) is given by 
\begin{equation}
 \alpha = \cot^{-1} \frac{k_\phi}{k_R},
\end{equation}
where $k_R$ and $k_\phi$ are the radial wave-number and azimuthal wave-number, respectively.
Thus, the unstable spiral mode is determined by two parameters, $\mathcal{J}$ and $Q$, 
given by the rotation curve of the galaxy disc. 
$\mathcal{J}$ controls the shape and growth rate of 
the unstable mode. The spiral mode appears for smaller $\mathcal{J}$, 
and the bar mode for larger $\mathcal{J}$ \citep{LauBertin1978}.

The number of spiral arms and their pattern speed cannot be determined 
in the framework of the tight-winding density-wave theory of Lin-Shu-Kalnajs (Section 2.1.1).
On the other hand, for global mode theory, 
if the unstable mode with the highest growth rate (trailing spiral modes) is assumed to be 
the spiral arms actually observed, then the spiral arms can be uniquely predicted from the equilibrium state of a galactic disc.
Therefore, it turns out that global unstable mode theory is a self-contained theory. 

However, there are some limitations in the global mode theory. 
Firstly, it is assumed that 
the spiral mode rotates as a rigid body, without changing its shape
in the global unstable mode analysis. 
\citet{Iye+1983} studied the global unstable modes of the fluid disc 
without the rigid-body rotation of the spiral modes, and reported 
the appearance of a global leading mode as well as global trailing modes,
which is a different result from the rigid-body case \citep{Aoki+1979}.
Recently, $N$-body simulations also show that spiral arms wind up over time 
by the shear of the galaxy disc \citep[][see Sections 2.2.1 and 4.2 for details]{Wada+2011,Baba+2013,Grand+2013}. 
Secondly, because global spiral modes grow up exponentially with time, 
the global mode theory requires self-regulated mechanisms such as 
damping effects in the stellar disc (e.g., Landau damping) 
and/or a gas component \citep[][see also Section 3.1]{LinBertin1985,BertinRomeo1988,Bertin+1989a}.
Finally, it is unclear that the global modes really accomplish a neutrally stable state.
The global mode theory hypotheses that the spiral arms are global neutral stability modes, 
which are accomplished by regulation mechanisms for the growth of density waves.
However, \citet{Lynden-BellKalnajs1972} showed that 
spiral waves transport angular momentum by the gravitational torque 
which changes the distributions of angular momentum and mass 
(i.e., induces migration of stars and gas).

\subsubsection{Simulations of long-lived spiral patterns}

Various studies have tried to reproduce non-barred quasi-stationary density waves 
using numerical simulations, but so far, no convincing isolated, long-lived grand design spirals have been produced. 
The $m=2$ case is of particular interest because low $m$ modes are most likely to be stable (see Section~4.5), 
hence simulations have tended to focus on trying to model galaxies with a 2 armed spiral structure.
As discussed in the previous sections, attaining a stable $m=2$ perturbation involves modelling 
a disc which is sufficiently self gravitating to be unstable to the $m=2$ mode, but stable to bar formation, 
and whereby density waves are able to be maintained by the presence of a $Q$ barrier before the ILR,  
where waves can be reflected back towards corotation (Section 2.1.2). 
Two armed spirals associated with bars, or interactions, are of course common outputs from simulations, 
and we discuss these in Sections 2.3 and 2.4. 

Early simulations of stellar discs all tended to form a bar (or oval distortion), 
and develop a strong $m=2$ spiral mode \citep{Miller1970, Hohl1971}. 
The formation of a bar was also predicted analytically in the case of 
a uniformly rotating disc \citep{Kalnajs1972, Kalnajs1974}. 
However with the adoption of an extended, massive (comparable to or more massive than the disc) dark matter halo, 
the bar mode was both predicted, and found to be suppressed \citep{OstrikerPeebles1973, Hohl1976}. 
Since then, simulations of isolated, non-barred galaxies have only produced 
multi-armed galaxies with transient spiral arms, as we discuss in Section 2.2. 
These types of galaxies can be produced readily with an $N$-body code.
By contrast, trying to model an $m=2$ spiral requires a long list of criteria to satisfy, 
and even then, $m=2$ spirals still appear to be transient, evolving to $m=3$ spirals and back again. 

\citet{Thomasson+1990}, and also \citet{ElmegreenThomasson1993}, 
performed calculations of a galaxy, where in addition to the conditions above,
they also enforced that the stellar velocity had to be maintained at a low value, 
and included a gas component. As will be mentioned in Section 3.2, and stated in Section 2.1.4, 
gas is likely required to allow energy from the spiral waves to dissipate.  
The galaxy tends to exhibit a pattern changing between 2 and 3 arms, and consequently has an asymmetric, 
rather than symmetric $m=2$ pattern at many time frames. 
Without a Q barrier, the spirals are shorter lived, whilst without cooling or gas, 
higher $m$ patterns become more prominent. 
\citet{DonnerThomasson1994} found similar results with a more consistent star formation scheme for the gas, 
and gas cooling and heating. \citet{Zhang1996} used the same setup as \citet{DonnerThomasson1994}, 
and found similar spiral patterns, but without including a  gas component. 
As well as changing to a 3 armed pattern, spirals which transition between 
an $m\geq2$ spiral and a barred spiral are likewise feasible to 
simulate by including gas accretion \citep{Combes2002}. 
But as yet no simulated galaxy retains a steady $m=2$ spiral.

\citet{Sellwood2011} tested some of the models claiming to find $m=2$ spirals. 
He performed $N$-body simulations designed to test directly whether a galaxy model 
corresponding to panel (c) of  Figure \ref{fig:globalmode} can 
in fact survive to support the slowly growing mode they predict should dominate.
He argued that this model evolves quickly due to multi-arm instabilities originating 
from swing-amplified noise (see Section 2.2.1) instead of producing quasi-stationary, 
two-armed spiral modes. This result suggests that dynamical evolution associated 
with shearing of spiral arms which is not considered in quasi-stationary density wave theories 
is important for generating the spiral arms in real galaxies. 
\citet{Sellwood2011} also tested some of the above models, 
which have proposed to exhibit long-lived spirals \citep{DonnerThomasson1994, Zhang1996}. 
He showed that the bisymmetric spiral arm represented 
as a mode is not a single long-lived pattern, 
but the superpositions of three or more waves that each grow and decay.

Whilst simulations have been unsuccessful in reproducing a stationary spiral pattern, $m=2$ or otherwise, recent work by \citet{D'Onghia+2013} and \citet{Sellwood2014} do report the existence of longer-lived `modes', which survive multiple rotations, and thus more resemble density wave theory. However these authors still state that their results are inconsistent with the idea that spirals are quasi-stationary density waves because the arms in their simulations still fluctuate in time. By contrast in global mode theory we would expect the arm shape to be unchanging for a number of rotations. For these long-lived spiral arms, the disc is required to be fairly gravitational dominated (\citealt{Sellwood2014} adopt $Q=1$) or include some perturbation(s) (\citealt{D'Onghia+2013}, see also \citealt{Salo2000}, Section~2.4.1).
\subsection{Dynamic spirals}

In this section we consider spiral arms which are transient, recurrent in nature. 
As we discussed in 2.2.1, the means of generating such arms is similar to that supposed in 
quasi-stationary density wave theory. However transient recurrent (or `dynamic') spiral arms are much easier to form. 
For example dynamic arms occur readily in numerical simulations, 
where we can in relate predictions from swing amplification theory to the properties of the spiral arms generated, 
and in turn observations (see Section 4). Moreover stationary arms are in essence a small subset of arms 
resulting from gravitational instabilities requiring very specialised conditions in the disc to maintain the arms, 
whereas dynamic spiral arms can be generated with essentially any disc configuration that is not strongly bar unstable.

Pioneering $N$-body simulations of the stellar discs by \citet{SellwoodCarlberg1984} 
have shown that spiral arms are transient and recurrent structures 
\citep{CarlbergFreedman1985,Bottema2003,Sellwood2010review,Sellwood2011,
Fujii+2011,Grand+2012a,Grand+2012b,Baba+2013,D'Onghia+2013,Roca-Fabrega+2013}. 
\citet{SellwoodCarlberg1984} argued that the spiral arms in $N$-body simulations generally fade out
over time because the spiral arms heat the disc kinematically and cause the $Q$ to rise.
Thus, the disc becomes stable against non-axisymmetric structure (Section 2.1.1). 
They suggested that continuous addition of a kinematically cold population 
of stars is necessary to maintain the spiral arms. This suggests that the gas can effectively cool the system and 
thus play an important role (Section 3.1). 
Recently, \citet{Fujii+2011} performed high resolution three-dimensional $N$-body 
simulations of pure stellar discs, and suggested that the rapid disappearance of the spiral arms 
may result from a low number of particles in previous simulations.
Instead, they revealed a self-regulating mechanism that maintains 
multi-arm spiral features for at least $10$ Gyr in a pure stellar disc (Figure \ref{fig:LongTermSpiralEvolution}).

Spiral arms in these $N$-body simulations are transient and recurrently reform. 
This is also the case for an $N$-body disc with a central bar \citep{Baba+2009,Grand+2012a}. 
The dominant spiral modes are time-dependent, 
reflecting a highly nonlinear evolution of spiral density enhancements, and 
radial changes (bottom panels in Figure \ref{fig:LongTermSpiralEvolution}). 
The arms are found to undergo a cycle -- breaking up into smaller segments with typical sizes of a few kpc, 
then reconnecting by differential rotation 
to reform large scale patterns \citep{Fujii+2011,Wada+2011}.
\citet{D'Onghia+2013} presented a similar argument that 
the evolution of the spiral arm is characterized by a balance between 
shear and self-gravity of the galactic disc: 
the shear tends to stretch and then break the spiral arms locally, 
whereas in regions where the self-gravity dominates, the spiral arm is over-dense and 
generates the segments making up the spiral arms. 
\citet{Baba+2013} pointed out that radial migration of stars around spiral arms 
are essential for damping of spiral arms, 
because excessive Coriolis forces originating from the growth of a spiral arm result in 
radial migration of the stars involved during the spiral arm evolution (their Figure 8).

In summary, these recent simulations of isolated disc galaxies conclude that the global spiral arms 
can appear to be long-lived visually, but they are assemblies of segments 
which break and then later reconnect with other segments of spiral arms.
In this sense, the spiral arms are in `dynamic equilibrium' between shear (or Coriolis force) 
and self-gravity rather than neutral stable waves assumed 
in the quasi-stationary density wave theory \citep{LinShu1964,BertinLin1996}.

\subsubsection{Swing amplified spirals}

We introduced swing amplification in Section 2.1.3 as a means of sustaining quasi-stationary density waves 
between the ILR and OLR. Here we describe the generation of dynamic spirals by swing amplification, 
but unlike the quasi-stationary density waves described in Section 2.1, 
there is no need for the waves to be reflected and maintained.

\begin{figure}[htbp]
\begin{center}
\includegraphics[width=0.42\textwidth]{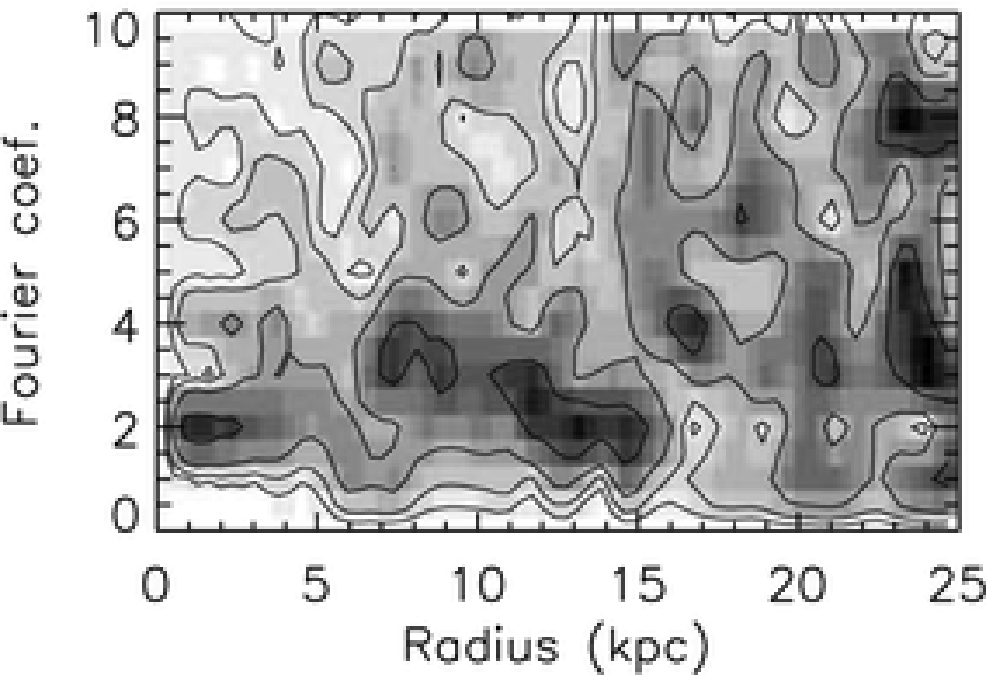}
\includegraphics[width=0.40\textwidth]{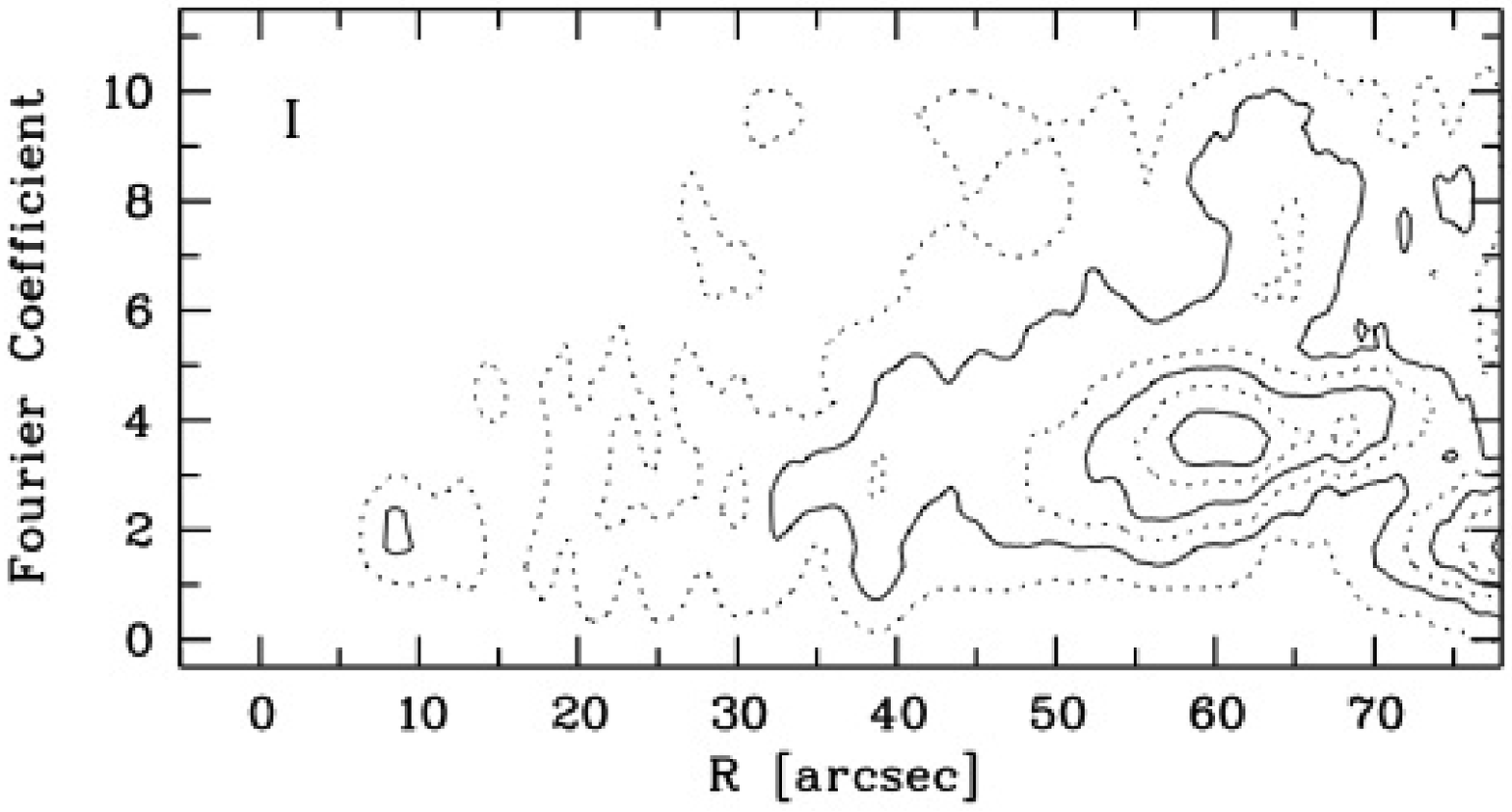}
\includegraphics[width=0.35\textwidth]{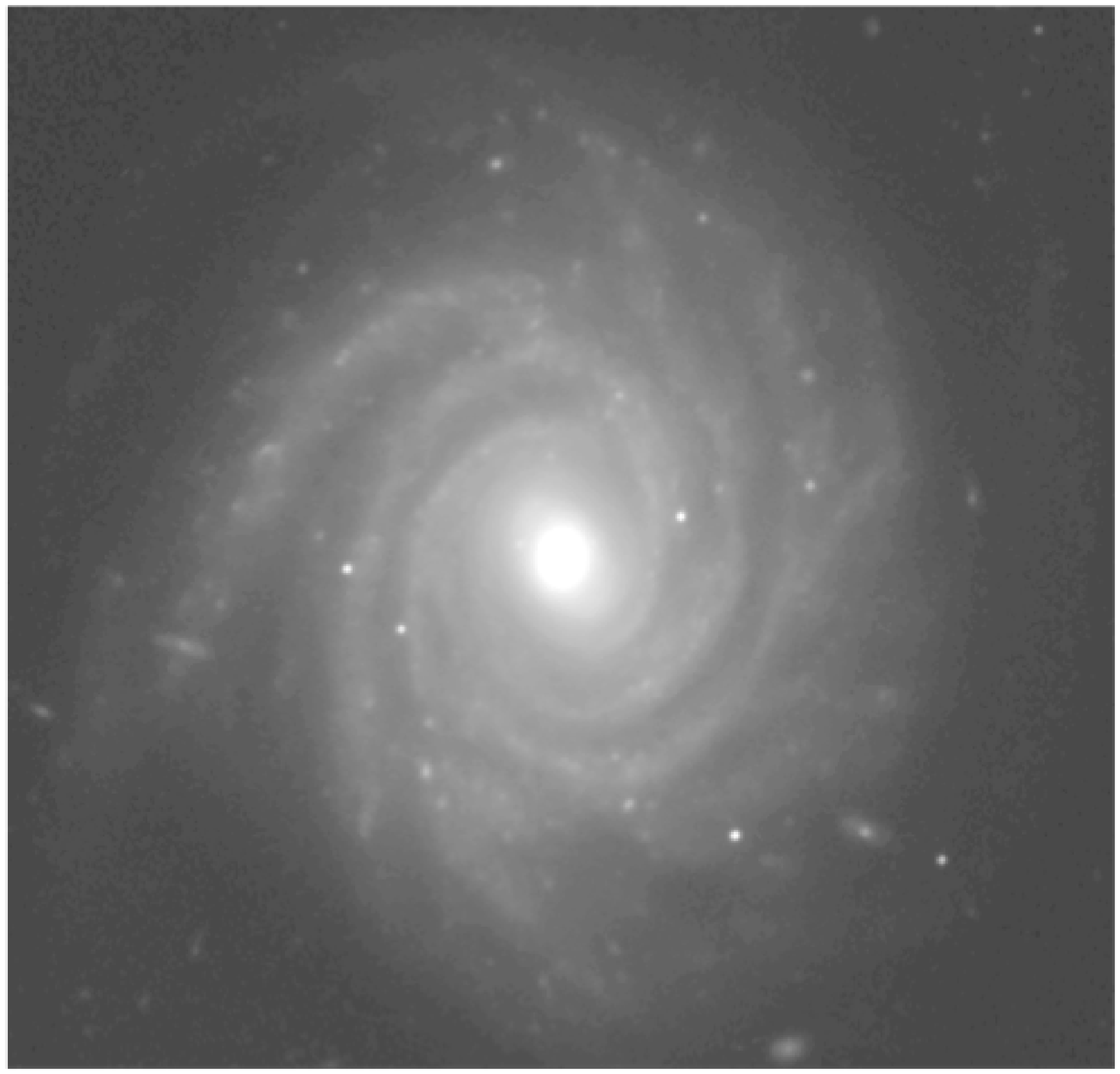}
\caption{
(top) Radial distribution of the number of spiral arms obtained by $N$-body simulations \citep{Bottema2003}.
(middle) Same as the top panel, but for observations of NGC 1288 \citep{FuchsMollenhoff1999}.
(bottom) $I$-band face-on view of NGC 1288 \citep{FuchsMollenhoff1999}.
}
\label{fig:SpiralRaidialDistribution}
\end{center}
\end{figure}

Full $N$-body simulations of stellar discs can test the predictions from swing amplification theory
\citep{SellwoodCarlberg1984,CarlbergFreedman1985,Bottema2003,Fujii+2011,D'Onghia+2013,Baba+2013}.
We can estimate the dominating number of spiral arms, $m$, as 
\begin{equation}
 m = \frac{\kappa^2 R}{2\pi G \Sigma_0 X} \approx \frac{\kappa^2 R}{4\pi G \Sigma_0},
\end{equation}
where $X \approx 2$ (Equation \ref{eq:X}) is assumed, and find this is  roughly consistent with 
the results of $N$-body simulations.
As described above, spiral arms typically develop most effectively when $1 < X < 2$, 
so taking $X \approx 2$ is appropriate (Figure \ref{fig:SWAfactor}). 
More generally, for a galaxy with a flat rotation curve ($\Gamma=0.5$) and a total mass within the disc of $M_{\rm tot}$
(mainly dark matter and stars), since $\kappa^2 = 2\Omega^2 \sim GM_{\rm tot}/R^2$, 
we also obtain the number of spiral arms as $m \sim \Sigma_{\rm tot}/\Sigma_0 = 1/f_{\rm disc}$.
\citet{CarlbergFreedman1985} performed $N$-body simulations of galactic discs
with various disc mass fractions and found that the number of spiral arms is strongly correlated with 
the disc mass fraction $f_{\rm disc}$. 
A similar result is also obtained by $N$-body simulations with much higher resolution 
\citep{Bottema2003,Fujii+2011,D'Onghia+2013}.
Also since $\kappa = \sqrt{2} \Omega \propto 1/R$ for the galaxy with a flat rotation curve, 
$m \propto 1/(R\Sigma_0)$ and the number of spiral arms tends to increase 
with radius in outer regions of exponential-discs. 
This agrees qualitatively with observations. As a specific example, 
we show the radial distributions of the number of spiral arms obtained by $N$-body simulations \citep{Bottema2003}
and observations of NGC1288 \citep{FuchsMollenhoff1999} in Figure \ref{fig:SpiralRaidialDistribution}.

\begin{figure}[htbp]
\begin{center}
\includegraphics[width=0.45\textwidth]{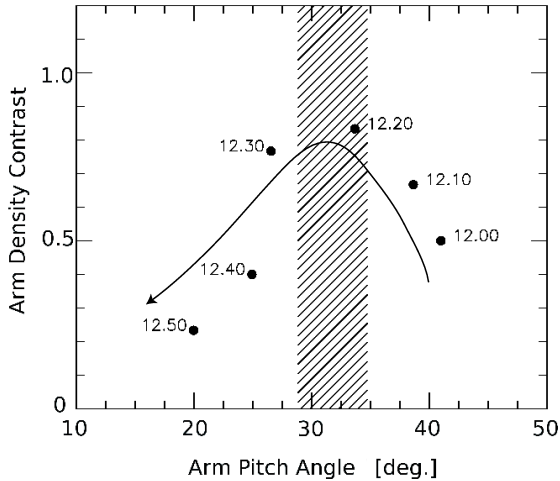}
\caption{
	Evolution of spiral arm on $\alpha-\bar{\delta}$ plane for $T_{rot} = 12.0-12.5$. 
	The hatched region corresponds to the predicted maximum pitch angle around 
	the analyzed region ($Q \approx 1.4$ and $\Gamma \approx 0.8$) 
	due to swing amplification (refer to Equation (98) in 
	\citet{Fuchs2001a}). 
 	From \citet{Baba+2013}.
	}
\label{fig:Baba+2013fig7}
\end{center}
\end{figure}

\begin{figure*}[htbp]
\begin{center}
\includegraphics[width=0.9\textwidth]{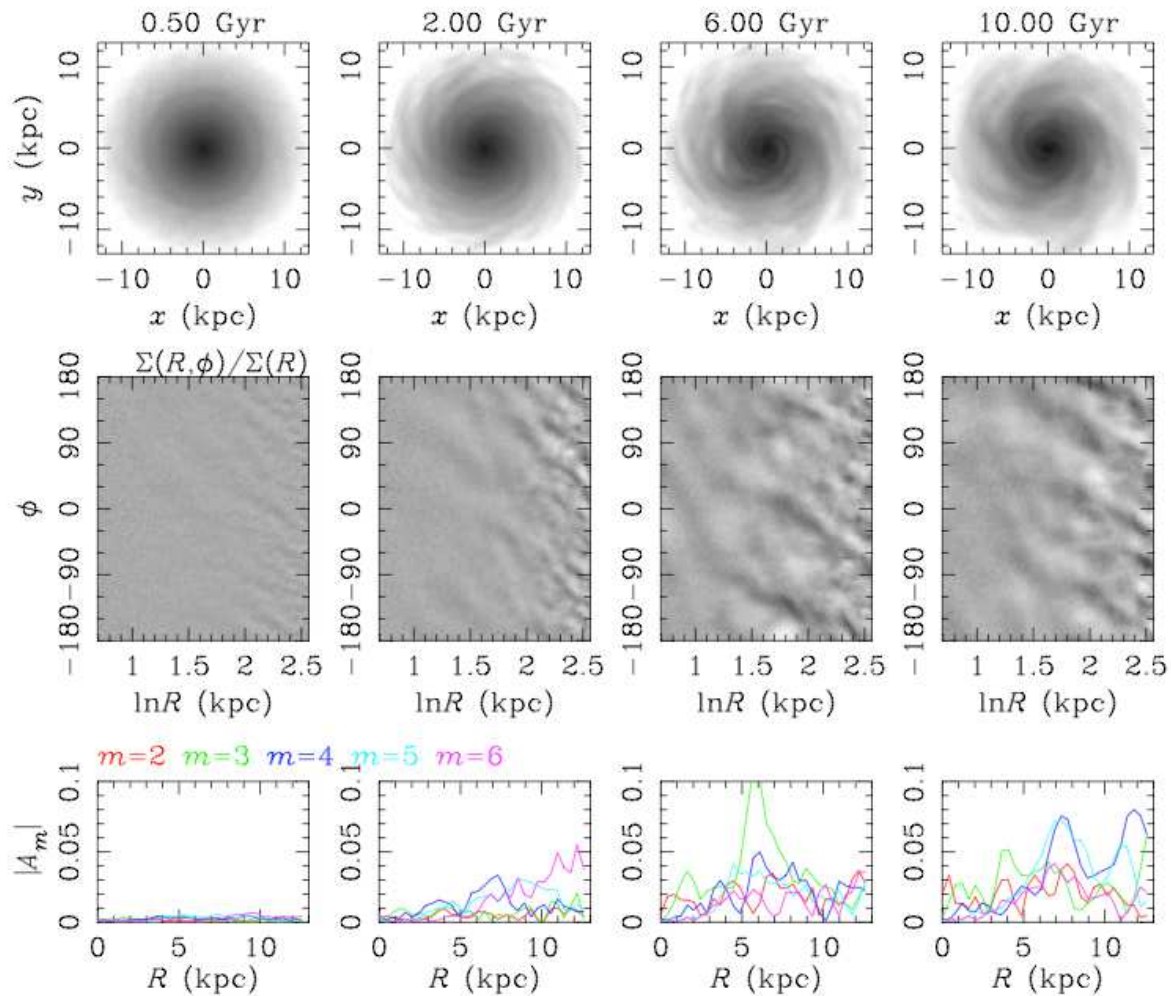}
\caption{
	Evolution of spiral arms with $N$ = 30M. 
	Top panels show the surface density, middle panels show the surface density normalized at each radius, 
	and bottom panels show the Fourier amplitudes.
	From \citet{Fujii+2011}.
	}
\label{fig:LongTermSpiralEvolution}
\end{center}
\end{figure*}

The pitch angle of the spiral arm is in reasonable 
agreement with the predictions of swing amplification theory.
Figure \ref{fig:Baba+2013fig7} shows the evolution of a stellar spiral arm along 
the pitch-angle - density-contrast ($\alpha-\bar{\delta}$) plane.
Due to differential rotation, one arm becomes more tightly wound as time goes by, 
and eventually disappears. In the meantime, new spiral arms with larger pitch angles start to grow. 
As the pitch angle of the spiral arm decreases from $\alpha \approx 40^\circ$ ($T_{\rm rot} = 12.0$) 
to $\alpha \approx 32^\circ$ ($T_{\rm rot} = 12.20$), the density contrast increases to a maximum, 
and the density contrast subsequently decreases with a decrease in the pitch angle. 
Thus, the spiral arm has a maximum amplitude when $\alpha \sim 32^\circ$.
This value is consistent with the expectation from swing amplification 
theory (hatched region in Figure \ref{fig:Baba+2013fig7}).
This behavior is similar to that reported in \citet{SellwoodCarlberg1984}. 
We compare the pitch angles of spiral arms produced by swing amplification theory with observations in Section~4.2.

In addition to the non-stationarity of stellar spiral arms, 
recent  $N$-body simulations have shown that  the pattern speed of the spiral arms 
decreases with radius in a similar manner to the angular rotation velocity of 
the disc \citep[See also Section 4.1;][]{Wada+2011,Grand+2012a,Grand+2012b,Baba+2013,Roca-Fabrega+2013}. 
Thus, the spiral arms are considered to be rotating with the rest of the disc at every radius, and are material arms. 
In the above models, the evolution of the spiral arms is governed by the winding of the arms,
which leads to breaks and bifurcations of the spiral arms.
\citet{SellwoodLin1989} and \citet{SellwoodKahn1991} instead argued that  
the dynamics originate from multiple wave modes of different pattern speeds 
constructively and destructively interfering with one another 
\citep{Sellwood2011,Quillen+2011,Roskar+2012,Sellwood2012}.

Finally, \citet{SellwoodCarlberg1984} also investigated the amplitudes of swing amplified spiral arms. 
They found that the growth of perturbations is in reasonable quantitative agreement with 
the prediction of swing amplification theory,  although the growth factor was slightly larger 
than a naive expectation from the level of particle shot noise (i.e., swing-amplified noise).
This discrepancy between numerical simulations and theoretical expectations 
is also seen in local stellar discs \citep{ToomreKalnajs1991}.
It may relate to non-linear effects of swing-amplified spiral instabilities. 
\citet{ToomreKalnajs1991} attributed the discrepancy to 
additional correlations between the particles that developed over a long period, 
i.e., the polarized disc response to random density fluctuations.
By contrast, \citet{Sellwood1989} showed that the amplitudes of spiral arms in 
global simulations of stellar discs seem to be independent of the particle number, 
rather than declining as $N^{-1/2}$ as would be predicted from the swing-amplified noise \citep{ToomreKalnajs1991}.
It should be noted that star clusters and GMCs in real galaxies can seed much larger  fluctuations
than shot noise from equal-mass particles \citep{D'Onghia+2013}.
\citet{Sellwood2011} also argued that spiral arms originating from swing-amplified 
shot noise are too low compared to observed spiral amplitudes. 
Instead, \citet{Sellwood2000} suggested that spiral arms
are vigorous large-scale modes originating from groove mode instabilities 
\citep{SellwoodLin1989,SellwoodKahn1991} (see Section~2.2.3).

\subsubsection{Corotation scattering and radial migration of stars}

Since the dynamic spiral arms do not have a single pattern speed 
but roughly follow the galactic rotation, or multi-wave patterns with different pattern speeds exist, 
these arms scatter stars everywhere in the disc via the corotation resonance
\citep{SellwoodBinney2002,Grand+2012a,Roskar+2012,Baba+2013,Grand+2014}.
Figure \ref{fig:OrbitEvolutionSpiral} shows the evolution of stars along the $\phi-R$ plane 
and the azimuth angle ($\phi$)-the angular momentum ($L_{z}$) plane.
The stars evolve in this plot due to changes in their angular momenta.
When the stars are captured by the density enhancement ($T_{\rm rot} \simeq 11.8-12.0$), 
they radially migrate along the spiral arms.
The stars approaching from behind the spiral arm (i.e., inner radius) tend to 
attain increased angular momenta via acceleration along the spiral arm, 
whereby they move to the disc's outer radius.
In contrast, the stars approaching ahead of the spiral arm (i.e., outer radius) tend to
lose their angular momenta via deceleration along the spiral arm, 
and they move to the disc's inner radius.
Along the $\phi-L_{z}$ plane, the stars oscillate both horizontally as well as vertically. 
Moreover, the guiding centers of the oscillations do not remain constant 
at the same value of $L_{z}$. 
This is essentially different from the epicycle motion in which $L_{z}$ is conserved.

The panels in the right column of Figure \ref{fig:OrbitEvolutionSpiral} show the so-called Lindblad diagram, 
where the angular momentum $L_{z}$ of each star is plotted against its total energy $E$.
The stars oscillate along the curve of circular motion by undergoing change in terms of 
both angular momentum and energy \citep{SellwoodBinney2002,Grand+2012b,Roskar+2012,Baba+2013}. 
This is because stars around the corotation point change their angular momenta without 
increasing their random energy \citep{Lynden-BellKalnajs1972}.

\citet{Grand+2012b} also noticed the slight heating of negative migrators  
and the slight cooling for positive migrators (their Figure 12).
\citet{Roskar+2012}, \citet{Minchev+2012}, 
and \citet{Baba+2013} also reported a similar effect of the radial migration 
of stars around the spiral arms upon disc heating.
Thus a non-negligible fraction of the particles that migrate outward have their orbits 
cooled by the spiral arm. 
This `dynamical cooling' can be important for recurrent spiral instabilities.

\subsubsection{Recurrent mechanisms for dynamic stellar spiral arms}

The mechanism by which spiral arms recur is unclear. 
 \citet{SellwoodKahn1991}, and \citet{SellwoodLin1989} proposed a feedback cycle 
 whereby narrow features in the angular momentum density of stars drive large-scale dynamic spiral arms. 
 The arms in turn lead to resonant scattering of stars, which serves as a seed for the next spiral arm formation. 
This large-scale spiral instability, which originates from the deficiency of stars over 
a narrow range of angular momenta (also corresponding to a change in the surface density for stars on a circular orbit), 
is called the `groove' instability. 
This feedback cycle was observed in $N$-body simulations of 
a low-mass disc with a near Keplerian rotation curve \citep{SellwoodLin1989}.
The phase space density is depopulated near the OLR of one wave, 
inducing a new large-scale spiral instability with a CR near the OLR of the first wave.
\citet{Sellwood2000} also reported that the distribution of the solar neighborhood stars 
on in angular momentum phase space
has similar fine structures \citep{Sellwood1994,Sellwood2010}, 
suggesting that this recurrent mechanism cycle may occur in real spiral galaxies. 
Scattering of stars by spiral arms at the ILR, in such a way to form a new spiral arm, 
is also observed in more massive discs with near flat rotation curves \citep{Sellwood2012}.
However, \citet{Sellwood2012} concluded that some other mechanism may be 
required for recurrent spiral instabilities, because he was not able to find 
evidence to support the groove-type cycle such as observed in less massive discs 
with a near Keplerian rotation \citep{SellwoodLin1989}.

\begin{figure*}[htbp]
\begin{center}
\includegraphics[width=0.9\textwidth]{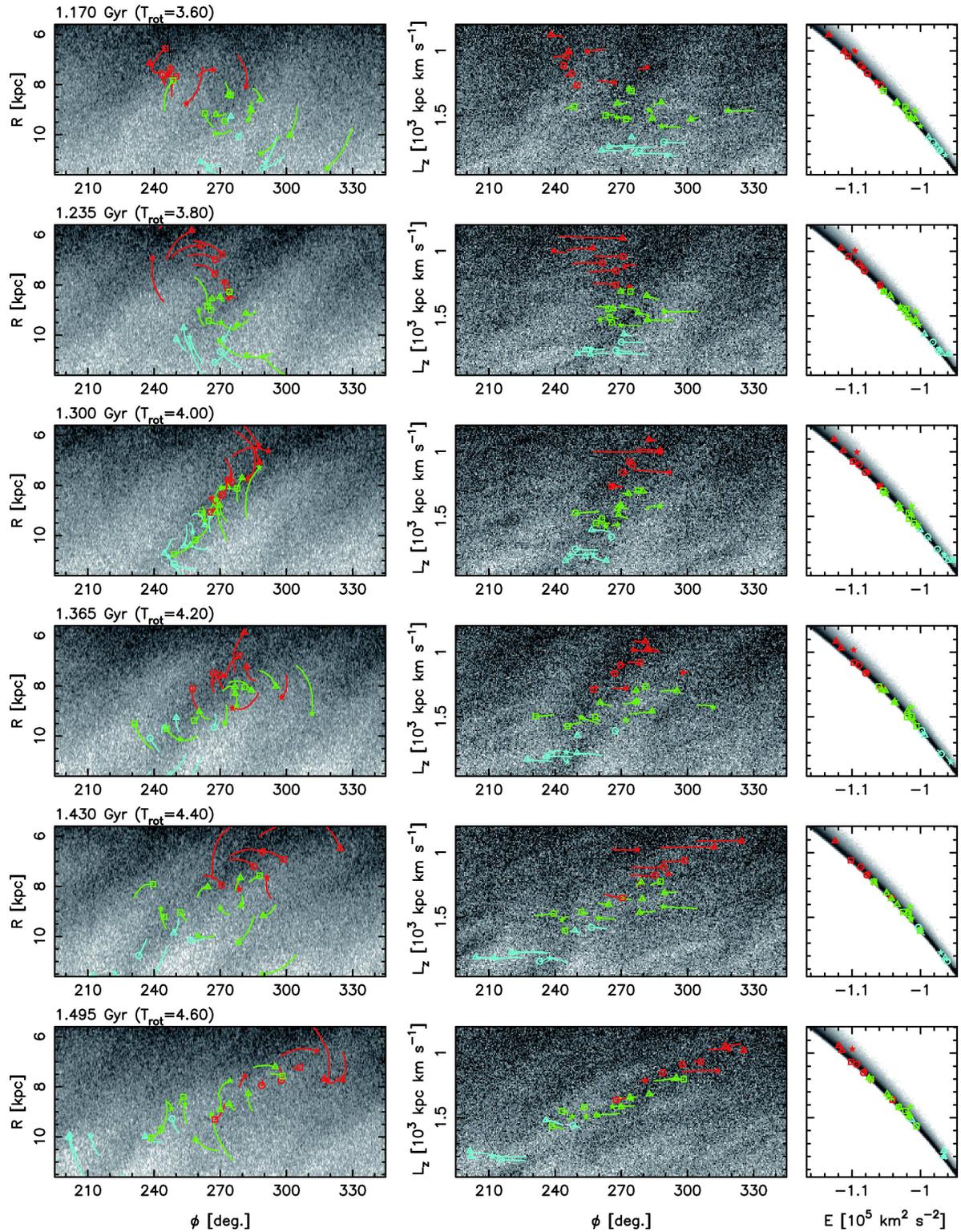}
\caption{
	Orbital evolution of stars in the spiral arm. 
	The stars associate around the spiral arm within a distance of $\pm 0.5$ kpc at $T_{\rm rot} = 4.0$. 
	Left columns: orbits on $\phi-R$ plane. Middle columns: orbits on $\phi-L_z$ plane. 
	Right columns: orbits on $E-L_z$ plane. 
	The colors denote the angular momentum at the time instants when the stars are associated with the spiral arm.
	From \citet{Baba+2013}.
	}
\label{fig:OrbitEvolutionSpiral}
\end{center}
\end{figure*}

\citet{Baba+2013} showed that oscillating stars successively undergo aggregation 
and disaggregation in energy-Lz space, 
thereby leading to the formation of structures referred to 
as `swarms of stars' along the $\phi-L_{z}$ and $R-\phi$ planes (the right column of Figure \ref{fig:OrbitEvolutionSpiral}). 
The non-steady nature of the spiral arms originates in the dynamical interaction between 
these swarming stars with a nonlinear epicycle motion, and the high-density regions, 
i.e., the spiral arms moving with galactic rotation.
This is entirely different from what is expected in stationary density waves, 
where these changes are limited to the CR and Lindblad resonances \citep{Lynden-BellKalnajs1972}. 
Thus, the gravitational interaction between the stars in the spiral arm and the spiral density 
enhancement changes the angular momentum and random energy of the stars, 
and this process in turn changes the structure of the spirals. 
During this process, the random energy of individual stars in the system does not 
increase monotonically. In other words, local interactions between the non-steady 
arms and stars increase or decrease the total energy of individual stars locally; 
however, the energy remains around its value for circular motion with 
the occurrence of a small dispersion. This is because the interaction causes the migration 
of the guiding centers of the stars without increasing their eccentricity or random energy. 
This `dynamical cooling' mechanism \citep{Grand+2012b,Roskar+2012,Minchev+2012,Baba+2013}
is essential to preventing heating of the stellar disc 
and erasure of the spiral arms, and the mechanism produces `swarms' of stars 
moving between non-steady spirals. 
The non-linear epicycle motion of the stars and their non-linear coupling with 
the density perturbation is the fundamental physics of the recurrently formed, 
non-steady spiral arms in a stellar disc.

\subsection{Bar driven spirals}

In many barred grand design spirals, the spiral arms start at the two ends of the bar. 
Two-armed spirals around strong bars are rather common, 
representing $\approx 70\%$ of typical field spirals, 
unlike unbarred field spirals where only $\approx 30\%$ are two-armed 
\citep{ElmegreenElmegreen1982}. 
Although this correlation suggests that the bar and spiral pattern have 
the same pattern speed and thus are related, 
the direct connection between bars and spirals is still unclear. 
There are three common interpretations \citep{Mo+2010}:
(1) the bar and spiral arms have a common pattern speed, 
(2) the bar and spiral arms have different pattern speeds and are independent patterns from each other, 
and 
(3) the bar and spiral arms have different pattern speeds but are coupled via some non-linear interactions.

\subsubsection{Spirals corotating with bars}

\begin{figure*}[htbp]
\begin{center}
\includegraphics[width=0.4\textwidth]{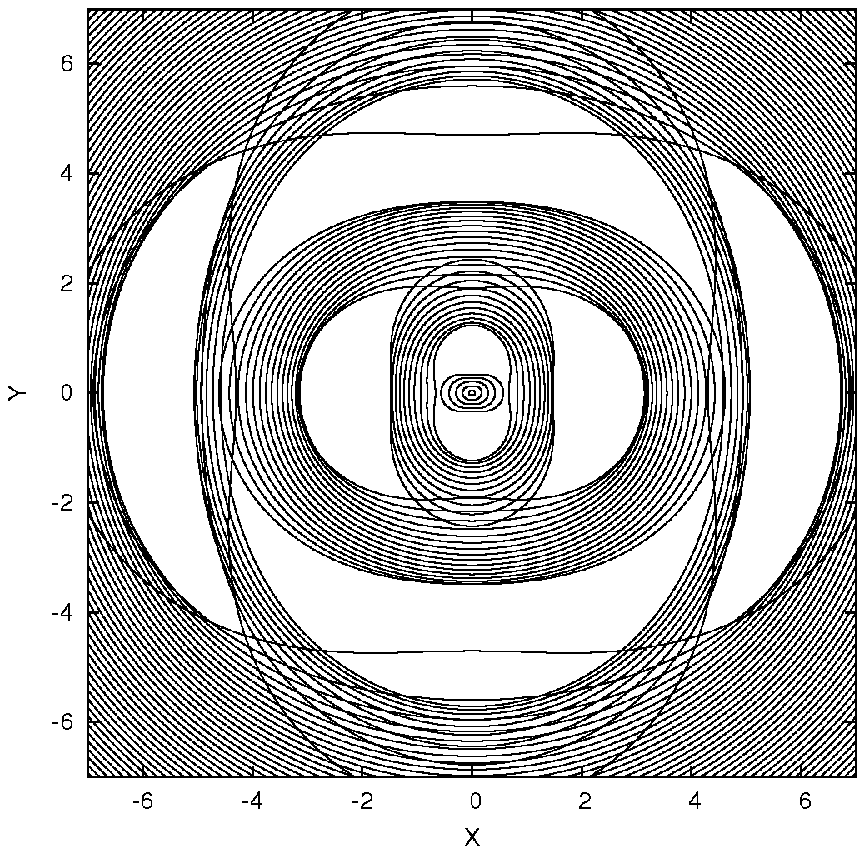}
\includegraphics[width=0.4\textwidth]{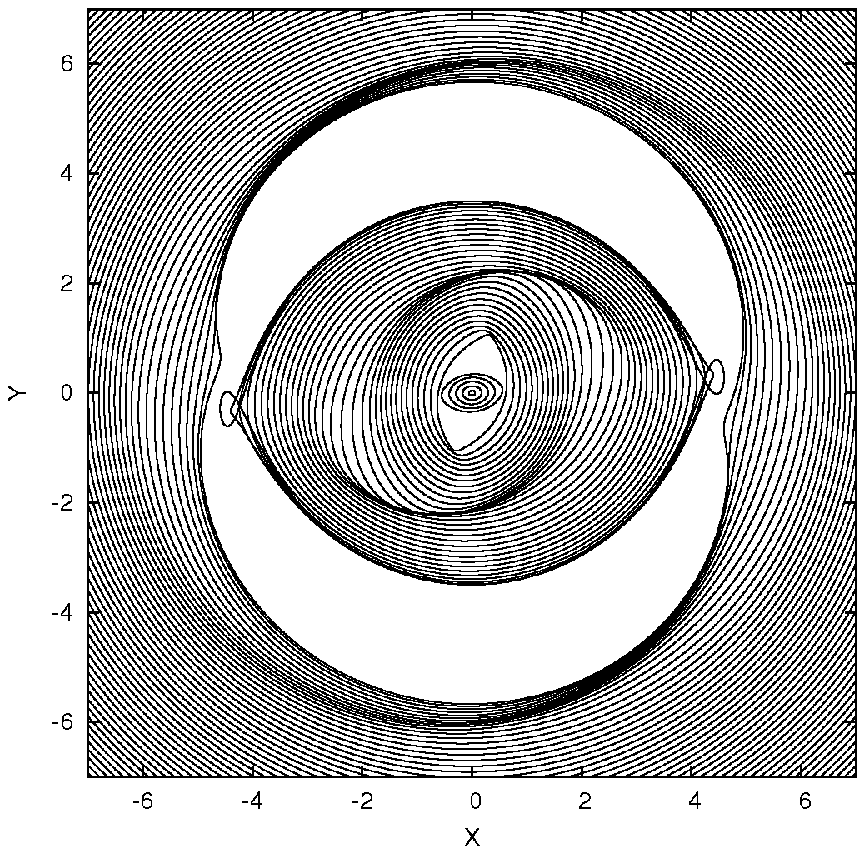}
\caption{
	Stellar closed orbits (left) and gaseous closed orbits (right) in a weak barred potential. 
	The radii of the inner ILR, outer ILR, CR, and OLR are at $0.8$, $2.4$, $4.6$, and $6.0$, respectively. 
	The gaseous closed orbits are calculated based on the damped orbit model by 
	\citet{Wada1994} who added the damping term (emulating the collisional nature gas)
	to equations of stellar orbits in a weak bar from Section 3.3 of \citet{BinneyTremaine2008}.
	Note that \citet{Wada1994} only showed a solution for radial direction.
	See the appendix of \citet{Sakamoto+1999} for a full set of the solutions.
	A similar introduction of a damping term was also made by \citet{SandersHuntley1976} and \citet{LindbladLindblad1994}.
	The stellar response to forcing by a steady bar cannot form spiral arms. 
	In contrast, the phase delay of epicycle motion in terms of the bar perturbation naturally 
	takes place as does in a damped oscillator affected by a periodic external force.
	This phase delay determines direction of spirals (i.e. trailing or leading) around the Lindblad resonance \citep{Wada1994}.}
\label{fig:OrbitBarPotentiall}
\end{center}
\end{figure*}

The first interpretation, which the bar and spiral arms have a common pattern speed,
is intuitive from observations that most spiral arms connect to the ends of the bar. 
\citet{SandersHuntley1976} studied the response of gas to a steady bar perturbation 
using hydrodynamical simulations.
They found that the gas eventually settled into a steady state with a prominent trailing spiral structure. 
Gaseous spiral arms driven by a bar have since been seen in many further simulations 
(e.g. \citealt{Schwarz1981,CombesGerin1985,Athanassoula1992b,
Wada1994,EnglmaierGerhard1999,Bissantz+2003,Rodriguez-FernandezCombes2008}).
The gas arms are a direct response of the bar forcing.
Figure \ref{fig:OrbitBarPotentiall} show the stellar closed orbits (left) 
and gaseous closed orbits (right) in a weak bar potential. 
Stellar orbits are always parallel or perpendicular to the bar, 
whilst the gas orbits change their orientation with radius, due to the effects of dissipation. 
The elliptical gaseous orbits are inclined to the bar potential in a trailing sense outside corotation.
Thus, dissipation associated with the gas viscosity plays a critical role in driving gaseous spirals.
Note that simulations in which a gas disc embedded in a `steady' bar potential is replaced 
by a collisionless disc of test star particles also gives rise to a prominent trailing spiral structure 
but the stars never settle into a steady spiral structure. 
But stellar spiral arms can be excited by a `growing' bar \citep{Hohl1971}. 

Manifold theory or manifold flux-tube theory is proposed 
as a way of determining the orbits of stars in spiral arms driven by a bar
\citep{Romero-Gomez+2006,Romero-Gomez+2007,Athanassoula+2009a, 
Athanassoula+2009b,Athanassoula+2010,Athanassoula2012,Voglis+2006a,Voglis+2006b,Tsoutsis+2008,Tsoutsis+2009}. 
According to this theory, the backbone of barred spirals are bunches of untapped stars 
(so-called Lyapunov orbits) escaped from the unstable Lagrangian points\footnote{
The direction in which the (chaotic) orbit can escape from the unstable Lagrangian points 
is not all direction but is set by the invariant manifolds. 
Manifolds can be thought of as tubes that guide the motion of particles whose energy is equal to theirs
\citep{Romero-Gomez+2006,Athanassoula+2009a}.
}, which are located on the direction of the bar major axis, outside the bar but near its ends. 
This means that, contrary to the quasi-stationary density waves, 
the stars do not cross the barred spiral arms but they move along them 
or they are spatially well confined by the manifolds\footnote{
Although, there is another view of the manifold theory: 
the locus of all points with initial conditions at the unstable manifolds that reach 
a local apocentric \citep{Voglis+2006a,Voglis+2006b,Tsoutsis+2008,Tsoutsis+2009} 
or pericentric \citep{Harsoula+2011} passage, but the details are beyond the scope of this review.
} \citep[Figures 1 and 4 of][]{Athanassoula2012}.
\citet{Athanassoula+2010} and \citet{Athanassoula2012} compared the properties of spiral arms predicted by 
manifold theory with $N$-body simulations. They found good agreement between the manifold theory and 
the simulations \citep[see Figure 12 of][]{Athanassoula+2010}, 
and in the simulations stars moved along the spiral arms as predicted \citep[see Figure 4 of][]{Athanassoula2012}.
The manifold theory predicts that the relative strength of the non-axisymmetric forcing in the region
around and beyond CR influences the winding of spiral arms, in the sense that in strongly barred galaxies 
the spirals will be more open than in less strongly barred ones \citep{Athanassoula+2010}.
This trend was corroborated in observed barred spiral galaxies \citep{Martinez-Garcia2012}.

One observational indication of the bar-driven spiral scenario 
is that grand-design spirals are more frequent 
in barred galaxies than in unbarred galaxies \citep{ElmegreenElmegreen1982}. 
However there are still many multi-armed (e.g. NGC 1232, NGC 3344, NGC 3953, NGC 6946, IC 342, Figure \ref{fig:BarredSpiralObs}), 
and flocculent (e.g. NGC 1313, NGC 5068) spirals that exhibit bars\footnote{though the latter seem more difficult to find.}.
Early type barred galaxies tend to have stronger bars and grand-design or multiple spiral arms, 
while late type barred galaxies have weaker bars and flocculent spiral arms (Fig.13 of \citealt{Elmegreen+2011}). 
Several studies have examined correlations between bar strengths and spiral arm strengths, 
with some finding clear evidence of bar driven spirals \citep{Block+2004, Salo2010}, 
and others finding little or no evidence \citep{SeigarJames1998, Durbala+2009, Kendall+2011}.
Thus observations suggest that 
the bar-driven spiral scenario is not necessarily valid for all barred spiral galaxies.

\subsubsection{Decoupling between spirals and bars}

The second possibility is that bars and spiral arms can be independent patterns.
In this case, spirals in barred galaxies are associated with a spiral density wave, 
but probably with a pattern speed different from that of the bar 
\citep{SellwoodSparke1988,RautiainenSalo1999}. 
Indeed, \citet{SellwoodSparke1988} have 
demonstrated $N$-body simulations of a stellar disc, and 
shown that multiple pattern speeds are quite common in disc galaxies, 
with the spiral structure typically having a much lower pattern speed than the bar. 
In other words, bars and spiral arms can be independent features.
This implies a more or less random distribution of the phase difference between the bar 
and the start of the spiral arms, 
which seems to be in conflict with observations. 
However, as pointed out by \citet{SellwoodSparke1988}, 
contour plots of the non-axisymmetric density in their simulations show that 
the spiral arms appear to the eye to be joined to the ends of the bar for most of the beat frequency. 
This suggests that the observed correlation between bars and spirals might simply be an illusion.

\subsubsection{Non-linear coupling between spirals and bars}

A third interpretation for the origin of spiral arms in barred galaxies is 
a non-linear coupling between bars and spiral density waves, 
where the bar and spiral arm have different pattern speeds
\citep{Tagger+1987,Sygnet+1988,MassetTagger1997,Minchev+2012}. 
This mechanism assumes some small overlap between the corotation (CR) of 
the bar and the inner Lindblad resonance (ILR) of the spiral density wave. 
Using the tight-winding and epicyclic approximations for density waves, 
\citet{Tagger+1987} and \citet{Sygnet+1988} showed that this overlap enables 
the transfer of energy and angular momentum between the bar, spiral density wave and beat (m=0 and m=4) waves. 
The bar is stabilized at a finite amplitude by transferring energy and 
angular momentum to the spiral density wave, and the non-linear coupling drives beat waves.
This theoretical argument on the non-linear coupling was also studied by $N$-body simulations 
of stellar discs \citep{MassetTagger1997,RautiainenSalo1999}, where
the predicted strong beat waves were observed. 
This scenario is similarly supported by more recent $N$-body/SPH simulations of galactic discs \citep{Minchev+2012}.
Notably though, the derived beating waves exhibit chaotic behaviour rather than a stationary spiral pattern.

\begin{figure*}[htbp]
\begin{center}
\includegraphics[width=0.326\textwidth]{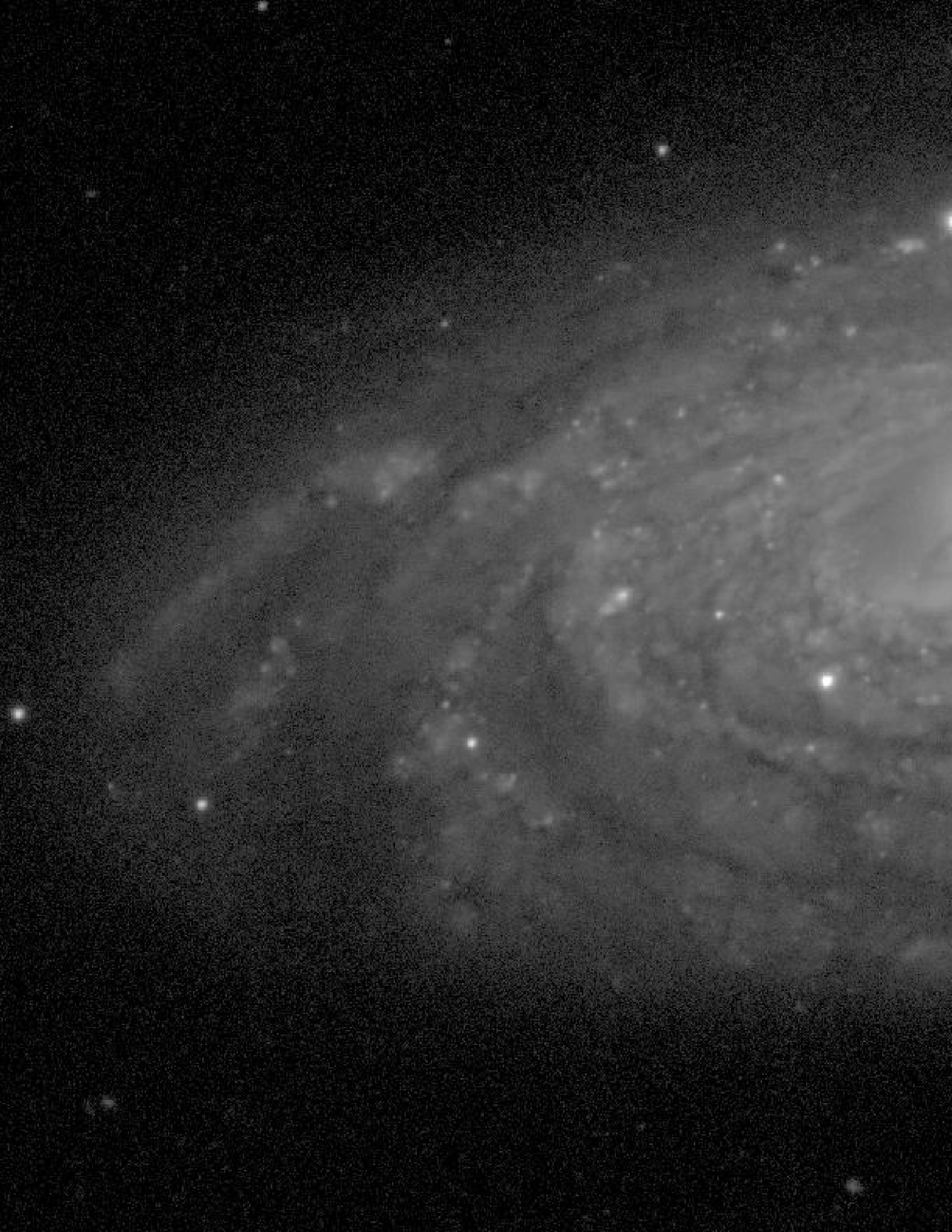}
\includegraphics[width=0.300\textwidth]{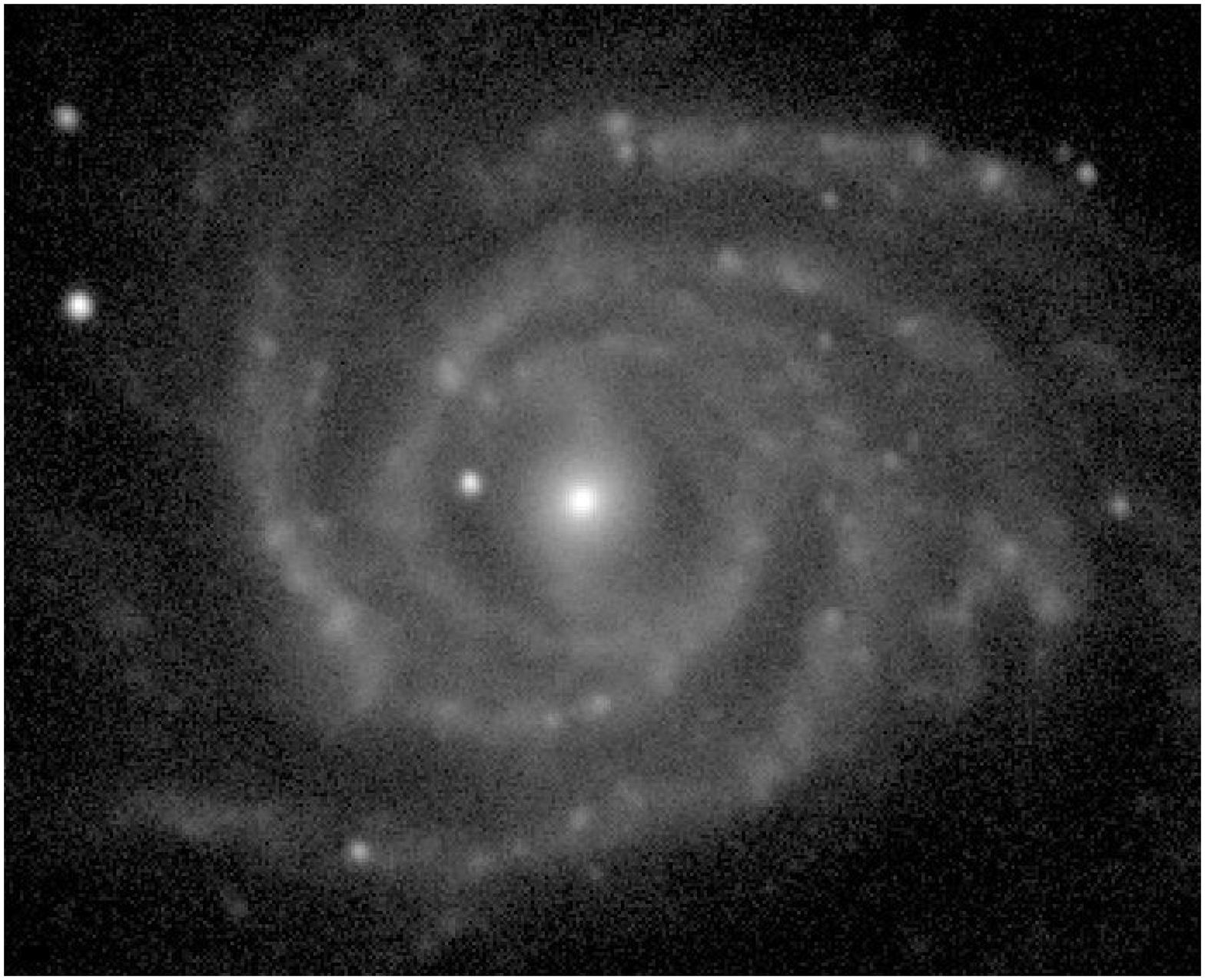}
\includegraphics[width=0.309\textwidth]{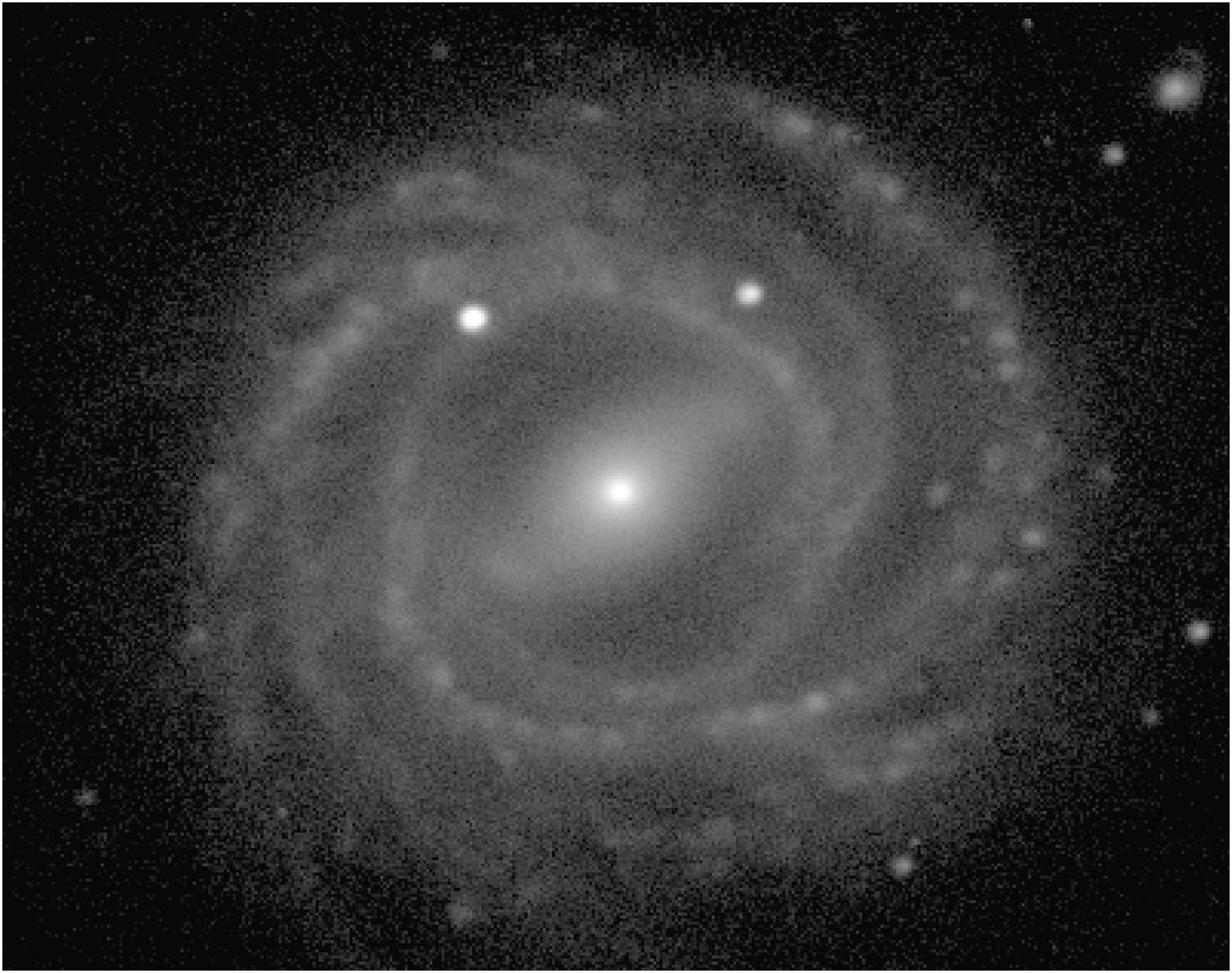}
\caption{
	$B$-band images of NGC 3953 (left), NGC 3124 (middle) and NGC 3450 (right).
	From {\it The de Vaucouleurs Atlas of Galaxies} \citep{Buta+2007}.
	}
\label{fig:BarredSpiralObs}
\end{center}
\end{figure*}

Sometimes the spiral arms do not start from the ends of the bar 
but exhibit a clear phase difference (e.g., NGC 1365). 
Similarly, some observed barred galaxies, such as NGC 3124 \citep{Buta+2007,Efremov2011} 
and NGC 3450 \citep{Buta+2007}, show the curved, leading ends of the stellar bar (Figure \ref{fig:BarredSpiralObs}).
Similar morphology can be seen in $N$-body simulations of barred galaxies 
due to the oscillations between trailing and leading ends of the bar 
(e.g. \citealt{Fux1997,RautiainenSalo2000,Martinez-ValpuestaGerhard2011}).
\citet{Martinez-ValpuestaGerhard2011} suggested that the oscillations could be related to 
the oscillations seen in the bar growth in $N$-body simulations (e.g. \citealt{Dubinski+2009}) 
through angular momentum transfer to disc stars (e.g. \citealt{Sellwood1981}) 
and to non-linear coupling modes between 
the bar and spiral density wave as mentioned above.

\subsubsection{Non-stationary spiral arms in barred galaxies}

\citet{Grand+2012a} performed $N$-body/hydrodynamic 
simulations of a Milky Way-sized barred galaxy and analyzed the spiral pattern speed. 
They found that the spiral arms are transient features and their pattern speeds decrease 
with radius in a similar manner to the angular velocity, but the pattern speed is slightly higher than the angular velocity of the disc. 
These results suggest that spiral arms in barred galaxies could be 
neither rigid-body rotating patterns predicted by the quasi-stationary density wave theory 
nor independent features, but transient features boosted by the bar.
The non-stationarity of spiral arms in barred galaxies is also reported by other $N$-body/hydrodynamics 
simulations \citep{Fux1997,Baba+2009}. \citet{Baba+2009} argued that 
non-stationary, winding spiral arms in a simulated barred spiral galaxy 
originate via swing amplification (Section 2.1.3).
In contrast, \citet{Roca-Fabrega+2013} reported that simulated spiral arms in strongly barred galaxies 
have a pattern speed almost constant in radius. 
More interestingly, they reported that the spiral pattern speed is close to 
disc rotation only when the bar is weak, as obtained by \citet{Grand+2012a}, 
but becomes almost constant when the bar has fully formed. 
These results suggest that the relation between bars and spiral arms can change 
during the evolutionary stages of bars,
although there is no observational evidence to support, or contradict this picture.

\subsection{Tidal interactions}

\subsubsection{Historical overview}

Tidal encounters are frequent across all astronomy, with interacting galaxies providing some of the clearest examples. Early attempts to categorise interacting, and other more unusual galaxies, showed many examples of galaxies with tidal tails, bridges and clear spiral structure \citep{Vorontsov-Velyaminov1959, Arp1966} prompting the morphology of galaxies to be associated with tidal effects (e.g. \citealt{vandenBergh1959, Lindblad1960, Hodge1966, Toomre1969}). The idea that tidal interactions may be responsible for spiral arms was in fact first demonstrated 20 years earlier, by \citet{Holmberg1941}. In a now famous experiment, \citet{Holmberg1941} modelled the interaction of two galaxies by representing the galaxies by a series of lightbulbs. The lightbulbs have initial velocities associated with them due to the initial velocities of each galaxy assumed for the interaction, and their rotation curves.  A photocell is used to measure the total amount of light at any particular point in the galaxies. Since light obeys a inverse square law the same as gravity, the total light received by the photocell is equivalent to the total gravitational force at that point in the galaxy. This force, or rather acceleration, is then used to calculate how far to move the given lightbulb. This step is then repeated for all the lightbulbs used, and the whole process repeated for many steps. The results of this experiment showed clearly the development of tidal spiral arms. 

From the 1960s, actual numerical calculations of interacting galaxies were able to be performed \citep{Pfleiderer1961, Pfleiderer1963, Tashpulatov1970, Toomre1972, Eneev1973}, although they were still limited to test particle simulations using a restricted 3 body approach, which neglects stellar self gravity. 
These simulations focused mainly on the origin of tails and bridges in galaxies, rather than spiral arms themselves. Nevertheless, \citet{Toomre1972} still represents one of the most comprehensive studies of galaxy interactions, spanning over all possible alignments of the two galaxies in space, unequivocally showing that bridges and tails were indeed the result of tidal interactions. These simulations also reproduced a number of known systems remarkably well, including M51 and the Mice. 

\subsubsection{Tidally induced arms: stationary, kinematic or material arms?}
\begin{figure*}
\begin{center}
\includegraphics[width=0.48\textwidth]{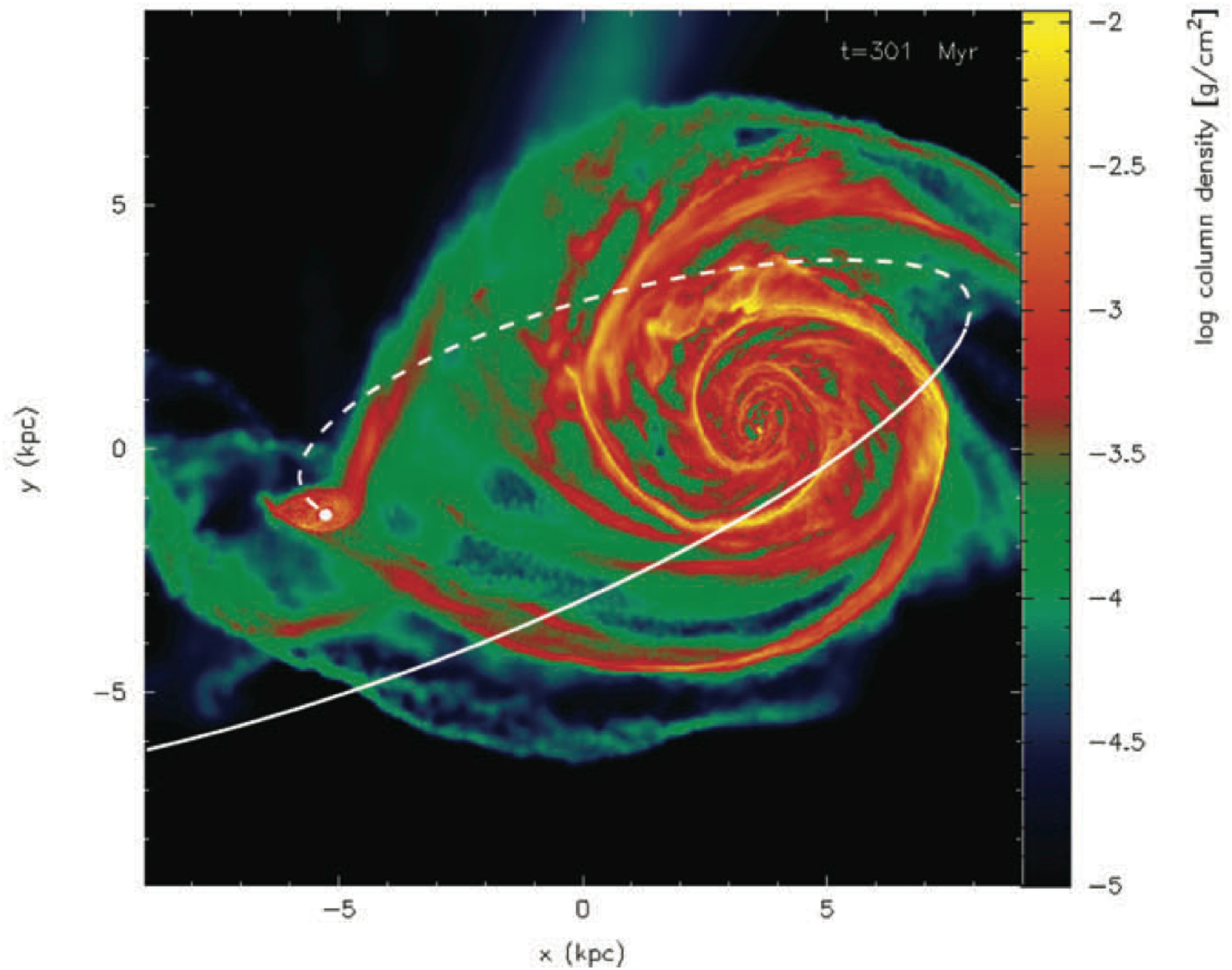}
\includegraphics[width=0.45\textwidth]{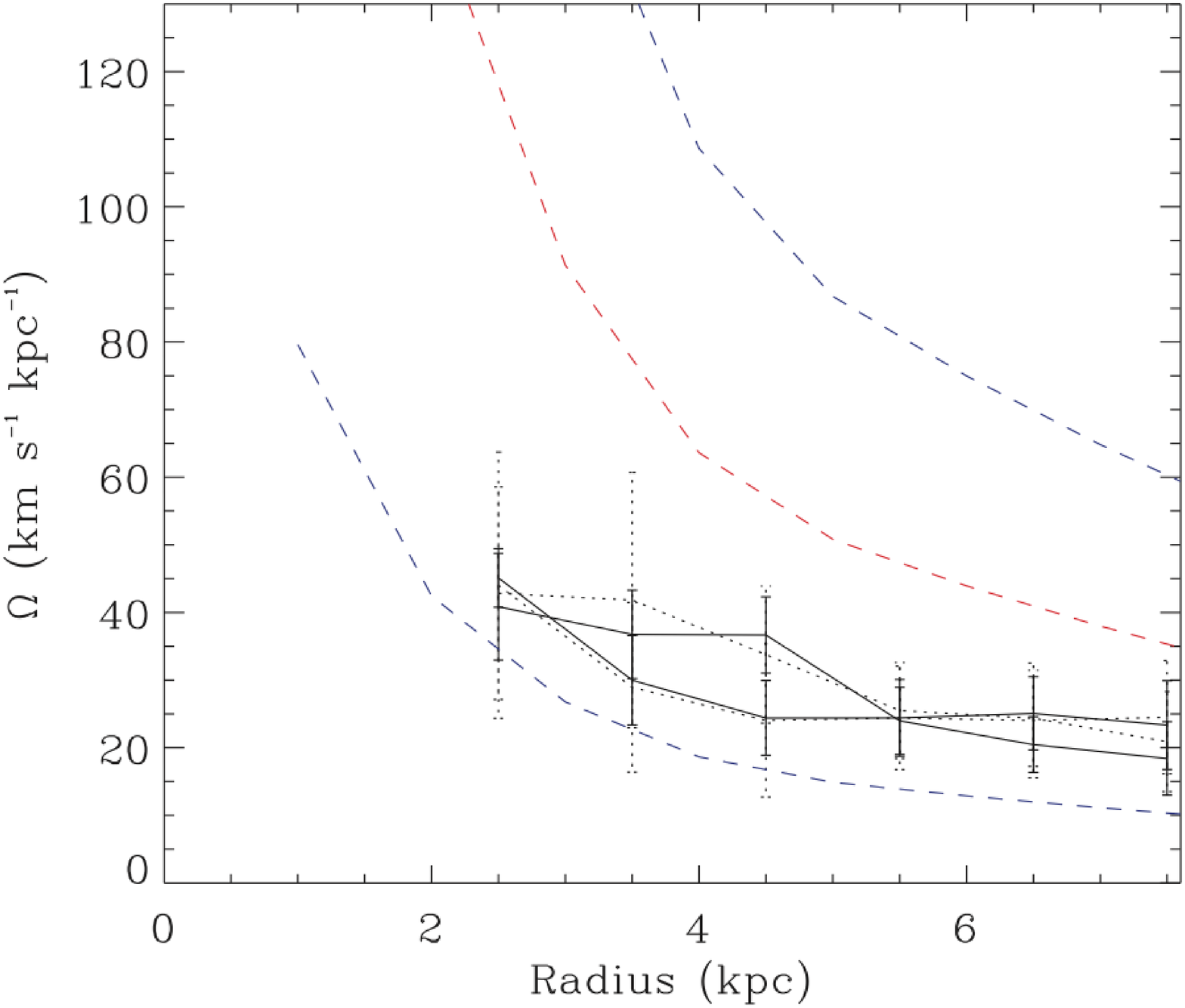}
\caption{
	Simulation of M51 (left panel) showing the present day appearance of the galaxy, 
	the orbit (dashed line) and the position of the perturber (white dot). 
	The pattern speeds of the two spiral arms are shown on the right hand panel, with error bars (dotted lines). 
	The angular velocity of the stars is also shown (red dashed line) and $\Omega \pm \kappa/2$ (blue dashed lines).
	From \citet{Dobbs+2010}.
	}
\label{fig:Tidal}
\end{center}
\end{figure*}

Since the 1970s, full N body simulations, were able to model interactions with much higher resolution, and demonstrate that tidal interactions could account not only for tails and bridges at large galactic radii, but also spiral arms penetrating to the centre of a galaxy \citep{Hernquist1990b, Sundelius1987, DonnerThomasson1994, Salo2000, Dobbs+2010} (see Figure~\ref{fig:Tidal}, left panel). Whilst these studies clearly demonstrate $m=2$ spiral arms, a more critical question is whether the spiral arms are representative of the quasi-stationary spiral arm picture, are kinematic density waves, or material arms.
For material arms  there is no difference between the spiral arms and the underlying differential rotation of the stellar disc -- the pattern speed of the arms is that of the disc, i.e. $\Omega_{\rm p}(R)=\Omega(R)$. Whilst material arms may describe the outer arms, or tidal tails  of galaxies (e.g. \citealt{Toomre1969, Meidt2013}) they are not found to characterise the arms over the main part of the stellar disc. \citet{Sundelius1987} demonstrated using numerical simulations that tidally induced spiral arms are density waves rather than material arms, whilst observations have shown that the velocity fields of interacting galaxies do not correspond to material arms (e.g. \citealt{Rots1975}).  

Secondly the arms may be kinematic density waves. Kinematic density waves are not actually waves, in the sense that they don't propagate through the disc, and have zero group velocity. But gas and stars do flow through the arms, although unlike quasi-stationary density waves, the spiral pattern is not fixed.  For purely kinematic density waves, self gravity of the stars can be ignored (kinematic density waves can be induced even when $Q=\infty$), the stars behaving simply as test particles. The influence of a perturber can be treated by the impulse approximation, and induces elliptical stellar orbits. Such orbits are not generally closed, but we can choose an angular speed such that the orbit is closed. For an $m=2$ perturbation, we can choose a rotating frame such that the orbit is closed after half the epicylic frequency (or after each time an arm is encountered) i.e. 
\begin{equation}
\Omega_{\rm p}(R)= \Omega(R)-\frac{1}{2} \thinspace  \thinspace \kappa(R), 
\label{eqn:tidal}
\end{equation}
to a first order approximation (see \citealt{BinneyTremaine2008}).
If the orbits are aligned along the same major axis, then the perturbation produces a bar. If however, the orbits are offset as a function of radius, then they naturally produce a spiral pattern \citep{Kalnajs1973}. Increasing the offset makes the spirals more tightly wound. In the case of a moving (prograde) perturber, the orbits are not aligned, and a trailing spiral pattern develops with a pattern speed given by equation 24 (trailing since Comment 39: $\Omega(R)-\thinspace  \thinspace \kappa(R)/2<\Omega$). The aphelia of the ellipses corresponding to the densest parts (arms) in the disc. Calculations of a perturber passing a galaxy in the non self gravitating case by \citet{Oh+2008} demonstrate that the induced arms do indeed exhibit this pattern speed. Thus the pattern speed decreases with radius, but less so than the material arms case. The locations of the spiral arms can also be determined analytically, by applying the impulse approximation to the stellar orbits, and computing the Jacobian matrix from the derivatives the resultant orbit equations \citep{Struck-Marcell1990, Donner1991, Gerber1994, Appleton1996}. The surface density of the response to a tidal perturbation is then 
\begin{equation}
\Sigma_0=\Sigma_0 \frac{R_0}{R} |J|^{-1}
\end{equation}
\citep{Gerber1994} where $R_0$ is the original (unperturbed) radius of the (circular) orbit and $J$ is the Jacobian. The points where $J=0$ are caustics, and correspond to the locations of the induced arms.

Alternatively the spiral arms may be quasi-stationary density waves. In the self gravitating case, the effect of self gravity is to make the spiral pattern more rigid, increasing $\Omega_{\rm p}$. Swing amplification may also act to enhance the density of the arms still further. However it is not established whether self gravity is sufficient to make the spiral pattern fully self gravitating, and develop into a quasi-stationary density wave. Both \citet{Oh+2008}, and \citet{Dobbs+2010} find that although the pattern speed is higher in their models than given by Equation~\ref{eqn:tidal}, and decreases less with radius, the arms are not completely rigid and still wind up with time (see Figure~\ref{fig:Tidal}, right panel). 
\citet{Sundelius1987} also find the development of spiral density waves in the absence of swing amplification, whilst \citet{Oh+2008}, and \citet{Dobbs+2010} find swing amplification only has a minor effect. Salo et al. 2000 also find that the pattern speed is radially decreasing and again slightly higher than given by Equation~\ref{eqn:tidal}, and again generally suppose that swing amplification has only a minor role in generating the arms. They do however find a more constant pattern in the centre kpc or so of their simulation of M51, and suppose that here Lin-Shu-Kalnajs (LSK) waves operate, the lack of an ILR in their simulation meaning waves can penetrate to the centre without being absorbed (in a number of their models, a bar forms in the centre, similar to the actual M51). 

Overall the nature of spiral arms in tidally interacting galaxies likely represent the behaviour of the underlying disc. Galaxies with dynamic spiral arms likely do not exhibit fixed spiral patterns when tidally interacting as they are not gravitationally dominated. Galaxies with more massive discs, and likely bars, may well exhibit fixed patterns, at least in the central gravitationally dominated regions.

\subsubsection{Prograde and retrograde encounters, and the orbit of the perturber}

In addition to the nature of the spiral arms induced, we can also consider how the orbit of the perturber affects the tidal perturbation. The simulations of \citet{Toomre1972}, and later \citet{Howard1993} demonstrated that retrograde encounters have a relatively small effect on a galactic disc, whereas prograde encounters are very effective at producing spiral arms, naturally of a trailing nature. Other analysis showed that tidal interactions could produce a leading spiral pattern, with one predominant leading arm \citep{Kalnajs1971, Athanassoula1978, Thomasson1989}. For the prograde case, as discussed earlier with respect to kinematic density waves, the angular speed of the perturber at closest approach during its orbit will likely be nearest to the Inner Lindblad resonance ($\Omega(R)-\kappa(R)/2$), and hence particularly effective at inducing an $m=2$ perturbation. For the retrograde case, the angular speed exhibits the opposite sign, has little correspondence to any resonance \citep{Toomre1969}, but will likely be closest to an m=1 perturbation ($\Omega_{\rm p}=\Omega(R)- \kappa(R)$) corresponding to one leading (generally $\kappa(R)>\Omega(R)$) arm \citep{Byrd1989}. Leading spiral arms are found to be rare in simulations, requiring a strong perturbation, and a large halo mass \citep{Thomasson1989}.  The latter is important to prevent swing amplification, which would convert any leading perturbation into a trailing one. Observationally, NGC 4622, a ring galaxy, is the only galaxy found to exhibit leading spiral arms \citep{Buta1992, Buta2003}. 

Various simulations have also investigated the impact on the galaxy from perturbers of different masses \citep{Byrd1992, Oh+2008, Struck+2011}. The simulations find that a perturber typically needs to be at least 0.01 times of the mass of the main galaxy to have an effect, ideally closer to 0.1 times the mass to produce a clear grand design pattern, although naturally there is a degeneracy with the pericenter of the orbit \citep{Oh+2008}. \citet{Toomre1972} showed that a perturbing galaxy has greatest impact when orbiting in the plane of the main galaxy, but changing the angle of inclination of the perturbing galaxy has little qualitative effect.

Another factor determining the dynamics of tidally induced spirals is the number of orbits of the perturber, and thus whether it is bound. For M51, the best matched orbit of M51 and NGC5195 currently involves two orbits of the NGC 5195 around M51, after which the two galaxies merge \citep{Salo2000b, Theis2003}. Consequently the dynamics are quite chaotic, whilst the orbit induces different sets of spiral arms resulting in noticeable kinks along the spiral arms as observed today \citep{Salo2000, Dobbs+2010}. The spiral arms consequently show clear departures from logarithmic spirals. \citet{Oh+2008}, and \citet{Struck+2011} present simulations where a perturber is on an unbound orbit, and passes the galaxy only once. In this case, the dynamics are less chaotic, and the arms smoother.

\subsubsection{Longevity of tidally-induced spirals}
If $m=2$ spirals are difficult to produce except by tidal interactions or bars, as we conclude from Section~2.1.5, their lifetime is an important characteristic. Grand design $m=2$ spirals are common, hence tidal interactions must induce relatively long-lived spiral arms if they are the main source of such galaxies.

Assuming their pattern speed is not fixed, tidal arms are expected to have a pattern that winds up slower than local transient arms discussed in the previous section, but to be shorter lived than quasi-stationary spiral arms.
For tidally induced spiral arms, there are two main questions regarding how long they last. The first is how much they wind up over time, the second is how long the arms take to decay or damp. To think about the first issue, we can consider the pitch angle of the arms, defined as the angle between the tangent of the spiral arm and a circle, i.e.
\begin{equation}
\tan \alpha =  \frac{1}{R}\frac{dR}{d\phi},
\end{equation} 
where the derivation is evaluated along the spiral arm.
If we consider the change in $\phi$ at a later time as $\phi(R,t)=\phi_0+\Omega_{\rm p}(R)t$ then the pitch angle can be written as:
\begin{equation}
\cot \alpha = \left| R\, t\, \frac{d\Omega_{\rm p}}{dR} \right|
\end{equation} 
\citep{BinneyTremaine2008}. For material arms, $\Omega_{\rm p}=\Omega$ and for a flat rotation curve of $v_c=200$ km s$^{-1}$, the pitch angle will be $\sim 1^{\circ}$ after about 1 Gyr. This is considerably lower than observed pitch angles. For material arms, the pattern winds up on a timescale of order $t\sim |dR/(R  d\Omega)|=1/|d\Omega/d\ln R|$, in the above example $\lesssim 100$ Myr.

For kinematic tidal arms, in the absence of self gravity, $\Omega_{\rm p}(R)=\Omega(R)-\kappa(R)/2$ (Section 2.4.1), 
and so $|d\Omega_{\rm p}/d\ln R| \ll |d\Omega/d\ln R|$. Hence the spiral pattern is expected to last somewhat longer. In the presence of self gravity, $\Omega_{\rm p}$ versus $R$ can become even shallower. For example, if we take Figure  \ref{fig:Tidal} (right panel), in 1 Gyr, the pattern winds up at a rate $\sim 4$ times slower than the above example for material arms, and is expected to have a pitch angle of 5 or 6$^{\circ}$ after 1 Gyr. These values are, as would be predicted, at the lower end of observed values \citep{SeigarJames1998, Seigar+2006}.

Simulations of tidally induced spiral arms confirm this behaviour, with the pattern winding up and simultaneously decaying on a timescale of $\sim 1$ Gyr \citep{Oh+2008, Struck+2011}. Similar to the case of dynamic spiral arms (Section~2.2), 
\citet{Struck+2011}  also found that the arms persisted for longer with higher resolution simulations. 
\citet{Struck+2011} also supposed that galaxy encounters in groups and clusters are likely frequent, 
and with spiral arms persisting for $\sim1$ Gyr, tidally induced spiral galaxies common.

\subsection{Stochastic star formation}
The idea of the generation of spiral arms by stochastic self-propagating star formation (SSPSF)
was developed principally by \citet{Mueller1976}, and \citet{Gerola1978}. Each generation of star formation is presumed to trigger new star formation in neighbouring regions, by the production of shocks from supernovae winds. Then, due to differential rotation, the newly formed stars are sheared into material spiral arms. Like the picture of  \citet{GoldreichLynden-Bell1965a}, the spiral arms are new stars, but in their case the spiral arms were associated with gravitational instabilities in the gas, rather than supernovae. This mechanism is not supposed to account for grand design galaxies, but tends to produce flocculent spiral arms \citep{Gerola1978, Jungwiert1994, Sleath1995}. Spiral arms are not long-lived in this model, rather they are continually created and destroyed.

The conclusion of these studies is that SSPSF is a secondary effect, rather than a primary means of generating spiral arms. In general, stellar discs in observed galaxies are not smooth, spiral structure is seen in the old stellar population, which is much amplified by the response of the gas. In fact, \citet{Mueller1976} concluded that stochastic star formation would not produce global spiral structure, but rather in conjunction with other mechanisms, such as density waves, would add an irregular structure to the galaxy. For example supernovae and triggered star formation are likely to help produce much wider spiral arms in tracers such as HI, CO and H$\alpha$ than expected from the gas response to a spiral shock. In the absence of feedback, the width of spiral arms is too narrow compared to observations \citep{Douglas2010}.

Local regions of likely SSPSF have been observed mostly in the LMC \citep{Westerlund1966, Feitzinger1981, Dopita1985, Kamaya1998}. In particular SSPSF seems most successful in irregular galaxies \citep{Hunter1985, McCray1987, Nomura2001}, which are not dominated by rotation, or other mechanisms for producing spiral structure. Direct triggering of molecular cloud formation by supernovae has been suggested observationally, for higher latitude regions in the Milky Way \citep{Dawson2011}, and in the LMC \citep{Dawson2013}. However there is little evidence that SSPSF is a global driver of spiral arms. After the 1980s interest in SSPSF appears to have waned. It is now feasible to perform hydrodynamical models of galaxies, including stellar feedback. These seem to indicate that instabilities and collisions dominate star formation in spiral galaxies, unless the level of feedback is unrealistically high \citep{Dobbs2011}. 

\subsection{Exotic mechanisms}
An alternative means of generating spiral arms, aside form gravitational instabilities in the stellar disc or tidal interactions with visible perturbers, is from asymmetries in the dark matter distribution. This may take the form of gravitational instabilities induced by asymmetries in the dark matter profiles of galaxies \citep{Khoperskov2013}, or tidal perturbations from dark matter sub halos \citep{Tutukov2006, Dubinski2008, Kazantzidis2008, Chang2011}. For the latter case, although the masses of the sub halos are assumed to be small ($\sim 10^6$ M$_{\odot}$), they are extremely close to, if not passing through the galactic disc during their orbits. Cosmological simulations predict a multitude of sub halos, albeit with a greater frequency than observed. The effect of dark matter sub halos is thus at present entirely speculative. Simulations predict that the impact of such halos should be detectable \citep{Dubinski2008, Chang2011}, but whether such effects are distinguishable from other perturbations (e.g. non-dark matter perturbers, previous low mass mergers, bar instability, gravitational instabilities) is an open question.

\section{Behaviour of gas in spiral galaxies}

So far we have only considered the response of the stars in galactic discs, but the response of the gas is important for considering spiral structure. The velocity dispersion in the gas is less than that of the stars, so the gas response to any perturbations in the stellar disc is highly amplified. Thus even small overdensities in the stars can result in clear spiral arms in the gas (for example see the IR maps in \citet{Elmegreen2011} compared to optical images). And furthermore as the gas forms young stars, in the optical we are dominated by the spiral pattern in the gas not the stars. Therefore processes in the gas, and star formation, will have shaped the spiral patterns that we observe. 

\subsection{Stability of a star and gas disc}
As described in Section 2.1, gas or stars in a disc are expected to undergo local axisymmetric gravitational instabilities according to the criteria in Equations \ref{eqn:gasQ} and \ref{eqn:starQ}. For a disc of stars and gas, local, transient instabilities in the stars are expected to be coupled by a similar response in the gas. Similar to the dispersion relations and stability criteria derived for gas and stars separately, we can also derive similar expressions for a disc of gas and stars. 
We note that, like in Section 2.1, the derivations here assume the tight winding approximation.

\citet{JogSolomon1984} first tackled the problem of a galactic disc of stars and gas, by treating the disc as a two-fluid system. They derived the following dispersion relation
\begin{equation}
\begin{split}
(\omega^2-\kappa^2-k^2 c_s^2+2 \pi G k \Sigma_{s0}) \\
\times (\omega^2-\kappa^2-k^2 c_g^2+2 \pi G k \Sigma_{g0}) \\
-(2\pi Gk \Sigma_{s0})(2\pi Gk \Sigma_{g0})=0
\end{split}
\end{equation}
where $c_s$ and $c_g$ are the velocity dispersion of the stars and gas respectively, and $\Sigma_{s0}$ and $\Sigma_{g0}$ are the surface densities of the stars and gas respectively. Thus the stars and gas are treated as co-existing fluids with different surface densities and velocity dispersions. They also determined a local stability criterion.

\citet{BertinRomeo1988} then determined a global stability criteria for a two-fluid disc. They defined a marginal stability curve given by
\begin{equation}
 \begin{split}
Q_H^2=\frac{2 \lambda}{\beta} [(\alpha+\beta)-\lambda(1+\beta)+\\
\sqrt{\lambda^2(1-\beta)^2-2 \lambda (1-\beta)(\alpha-\beta)+(\alpha+\beta)^2}]
\end{split}
\end{equation}
where $\lambda=\frac{k_g}{|k|}$, $\alpha=\rho_c/\rho_h$ and $\beta=\sigma_c^2/\sigma_h^2$, where $\rho$ is density, $\sigma$ is the radial dispersion, and subscripts $c$ and $h$ represent cold and hot components. Potentially, $Q_H$ can exhibit two peaks, one arising from instability in the gas (at smaller wavelengths), and one from the stars \citep[see Figure~3 of][]{BertinRomeo1988}. Stability globally at all wavelengths then requires that $Q^2$, where $Q$ corresponds to the standard criterion (Equation~\ref{eqn:gasQ}) for the hot component, is greater than the maximum of $Q_H^2$. 

\citet{Rafikov2001} derived the dispersion relation for a disc with a fluid, and a collisionless component (see also \citealt{Romeo1992}):
\begin{equation}
2\pi G k \frac{\Sigma_{g0}}{\kappa^2+k^2c_g^2-\omega^2}+2\pi G k \frac{\Sigma_{s0}\mathcal{F}}{\kappa^2-\omega^2}=1
\end{equation}
 where $\mathcal{F}$ is defined as for Equation~9. 
 Then, for the disc to be stable (requiring $\omega^2(k)>0$ for all $k$), he derived the criterion 
 \begin{equation}
 \frac{1}{Q_{sg}}=\frac{2}{Q_s}\frac{1}{q}[1-e^{-q^2}I_0(q^2))] + \frac{2}{Q_g}R\frac{q}{1+q^2R^2}>1
 \label{eqn:qsg}
 \end{equation}
 where 
 \begin{equation}
 \begin{split}
 Q_s=\frac{\kappa \sigma_s}{\pi G \Sigma_{s0}}, \qquad   Q_g=\frac{\kappa c_g}{\pi G \Sigma_{g0}}, \\
 q=k\sigma_s/\kappa, \medspace \medspace \rm{and} \medspace \medspace R=c_g/\sigma_s.
 \nonumber
 \end{split}
 \end{equation}
Note that $Q_g$ is not the same as shown earlier for Equation~\ref{eqn:gasQ}, and this is again a criterion for local instabilities.  \citet{Wang1994} also present a simplified approximate stability criterion, $(Q_s^{-1}+Q_g^{-1})^{-1}$, with $Q_s$ and $Q_g$ defined as above.

These equations still do not represent a multiphase medium, and are for a thin disc. Some authors have tried to incorporate a more realistic ISM \citep{Romeo2010, Romeo2011, Romeo2013}, but we do not consider these further here.
One of the main results arising from these criteria though is that small changes in the gas can change the stability of the disc significantly, compared to relatively large changes in the stellar component \citep{JogSolomon1984, Rafikov2001}.

Following these derivations, \citet{Li2005} investigated the stability criterion of \citet{Rafikov2001} using numerical simulations of an isothermal disc. 
They found gravitational collapse when $Q_{sg}<1.6$, and vigorous star formation when $Q_{sg}<1$. 
With a multiphase medium, gravitational collapse will always occur in a disc of stars and gas with realistic surface densities. 
Many such simulations have shown the development of dynamic spiral arms simultaneously in the gas and stars, 
and the formation of molecular clouds and star formation within them (e.g. \citealt{RobertsonKravtsov2008, Hopkins2011, Wada+2011}). 

\subsection{Damping of spiral arms}
Large-scale shocks (namely spiral shocks or galactic shocks) are 
predicted in the gas as the result of spiral density waves (see Section 3.5), 
or even generic turbulence in the spiral arms, so they are naturally expected to lead to energy dissipation.

By computing the energy change across the shock, and momentum conservation, 
\citet{Kalnajs1972} showed that the rate of change of energy density of tightly winding quasi-stationary spiral density waves is
\begin{equation}
\dot{E_w}=\frac{\Omega_{\rm p}}{\Omega-\Omega_{\rm p}} \dot{E_s}
\end{equation}
(see also \citealt{BinneyTremaine2008}), where $\dot{E_s}$ is the energy dissipation in the shock and $\dot{E_w}$ is the energy change in the density wave.
The energy dissipation, $\dot{E_s}$ is negative, 
hence $E_w$ increases when $\Omega_{\rm p}<\Omega$ (i.e., $R<R_{\rm CR}$) and 
decreases when $\Omega_{\rm p}>\Omega$ (i.e.,  $R>R_{\rm CR}$). 
By noting that $E_w$ is negative in the first case, and positive in the second case \citep{BinneyTremaine2008}, it is evident that the induced shock always damps the density wave \citep{Kalnajs1972}. The damping timescale, $-E_w/\dot{E_w}$, is estimated to be between $\sim10^8-10^9$ yrs depending on the calculation of the energy terms, and the nature of the shock \citep{Kalnajs1972, RobertsShu1972, Toomre1977}. Another basic consequence of damping is that in the case of quasi-stationary spiral density waves, 
the streamlines predicted to describe the gas trajectories (\citealt{Roberts1969}, see Section 3.5) will not be closed \citep{Kalnajs1972}.

Following this result, a model of self regulated spiral structure was put forward by \citet{BertinRomeo1988}, also following discussion in \citet{RobertsShu1972}. Spiral perturbations in the disc are predicted to grow with time (see Section 2.1.4). 
Hence \citet{BertinRomeo1988} proposed that gas damps the spiral perturbations, which are then regenerated on timescales comparable to the damping timescale. They point out that in the absence of gas, the stars would instead continue to heat until the disc becomes stable to spiral perturbations \citep{LinBertin1985}.   
     
For dynamic spirals, damping was also thought to be important.
As mentioned in Section 2.2, early simulations (e.g. \citealt{SellwoodCarlberg1984}) found that stellar discs heated up with time, as supposed in the previous paragraph, and consequently the spiral arms disappear after several galactic rotations. 
Furthermore, galaxies which have little gas did not appear to have any spiral structure, 
suggesting that gas damping is always a requisite for spiral structure \citep{BinneyTremaine2008}.
However, \citet{Fujii+2011} and \citet{D'Onghia+2013} showed that spiral arms were able to survive much longer (see Section 2.2). \citet{Fujii+2011} demonstrated that stellar heating was too high in previous lower resolution calculations, partly due to two-body effects\footnote{Note that \citet{Sellwood2012} disagrees two body effects are important, rather he supposes the main difference with higher resolution is that there is lower amplitude noise, which results in weaker spiral arms and less heating.}. Thus they showed that it was possible for galaxies to exhibit stellar spiral arms in the absence of gas. Indeed such galaxies, with spiral arms but no recent star formation or large gas reservoir, are now observed \citep{Masters2010}.  

\subsection{Physical processes in the ISM}
The gas in galaxies is subject to many processes other than spiral shocks, including cloud-cloud collisions, hydrodynamic instabilities (see also Section 3.5.2) and stellar feedback, as well as gas self gravity, thermodynamics and magnetic fields. Even in a purely smooth stellar disc, these processes still lead to a considerable degree of substructure in the gas, if not long spiral arms (e.g. \citealt{ShettyOstriker2006, Tasker2009, Dobbs2011}). In the presence of spiral arms, these processes will still clearly occur, often preferentially in, or modified by the spiral arms.

The quasi-periodic spacing of gas structures along spiral arms observed in some galaxies has long been supposed associated with a gravitational origin of Giant Molecular Clouds (GMCs) or Associations (GMAs)
 \citep{Shu+1972, Woodward1976, Elmegreen1979, Cowie1981, Elmegreen1983, Balbus1985, Kim2002, ShettyOstriker2006}. 
 The dispersion relation for a gas disc, Equation \ref{eq:LinShu} is often used 
 to derive expressions for the expected mass and spacing of GMCs along a spiral arm.  
If we consider the gas which collapses on the shortest timescale, this occurs when $d\omega/dk=0$, 
at a wavenumber $k= \pi G \Sigma_{g0}/c_s^2$. 
The corresponding wavelength is then
\begin{equation}
\lambda_{\rm max}=\frac{2c_s^2}{G \Sigma_{g0}}.
\label{eqn:cloudspacing}
\end{equation}
This is the predicted separation of the clouds. The mass of the clouds is then 
\begin{equation}
M=\Sigma_{g0} \left(\frac{\lambda_{\rm max}}{2}\right)^2=\frac{c_s^4}{G^2\Sigma_{g0}}.
\label{eqn:cloudmass}
\end{equation}
The spiral arms provide a denser environment, which can make the gas susceptible to instabilities at wavelengths where it would not be unstable in the absence of spiral arms. Also, as $\Sigma_{g0}$ increases, and $c_s$ likely decreases in the spiral arms, the properties of the GMCs change (though technically Equation~6, should be applied over large scales rather than localised to a spiral arm).

Whilst the dispersion relation adopts a number of caveats, e.g. a thin disc, these masses and spacings have been shown to approximately agree with simple numerical simulations of a gravitationally unstable isothermal medium \citep{Kim2002, ShettyOstriker2006, Dobbs2008}. These calculations ignored the multiphase nature of the ISM (though see \citealt{Elmegreen1989}), which with the inclusion of thermal instability and turbulence, may lead the thermal term to actually promote rather than prevent instability \citep{Elmegreen2011}. In a medium of clouds and diffuse gas, self gravity can also act to increase cloud collisions \citep{Kwan1987} which would not necessarily lead to the same masses and separations as Equations~ \ref{eqn:cloudspacing} and \ref{eqn:cloudmass}. Finally these estimates of the mass and separation generally require  that the maximum cloud mass is reached before feedback disrupts the cloud, or the cloud moves out of the spiral arms (see also \citealt{Elmegreen1994,Elmegreen1995}). 

Cloud collisions occur regardless of spiral arms due to the random dispersion of the clouds, but are much more frequent in the spiral arms. As will be discussed in Section 3.5.2, dissipative collisions of either smaller molecular clouds or cold HI can lead to the formation of more massive GMCs. Structure is always present in the ISM, so gas entering the spiral arms will exhibit some structure (though the gas need not be molecular). Even for a homogenous warm medium, rapid cooling in the spiral shock quickly leads to thermal instabilities and the formation of structure \citep{Dobbs+2008, Bonnell2013}. Like gravitational instabilities, cloud collisions in the spiral arm induce a spacing between GMCs. In this case, the spacing predominantly depends on the strength of the shock the gas encounters, which in turn depends on the spiral forcing or amplitude and the sound speed. The separation of GMCs is proportional to the epicyclic radius, which represents the radii of the disc over which material can be brought together to a single point, or into a single cloud \citep{Dobbs2008}. Stronger shocks produce more massive, widely spaced clouds. In this sense the behaviour is opposite to gravitational instabilities.

Parker instabilities have also been proposed to form GMCs in spiral arms \citep{Mouschovias1974, Mouschovias2009, Elmegreen1982} and shown to produce density enhancements of factors of several, which may be sufficient to induce a phase change in the ISM. Density enhancements solely due to Parker instabilities are finite, and thus likely to be overwhelmed by gravitational instabilities \citep{Elmegreen1982, Kim2001, Santillan2000, Kim2002}. However there is some evidence of loops caused by Parker instabilities in the Galactic Centre, where magnetic fields are strong \citep{Fukui2006}.

All these processes lead to considerable substructure in the gas on size scales up to the most massive GMCs. Either fragmentation (via gravitational instabilities) or agglomeration of clouds leads to a mass spectrum from masses of $<100$ M$_{\odot}$ up to 
giant molecular associations of $10^{7-8}$ M$_{\odot}$. 
In the case that $\Omega_{\rm p}<\Omega$ (i.e., $R<R_{\rm CR}$), 
complexes formed by all these methods leave the arms and are sheared out into trailing spurs (see next section) by differential rotation.  

As well as processes which lead to the accumulation of gas into clouds, stellar feedback also has a substantial effect on the gas. 
Although spiral shocks may account for the very narrow dust lanes in galaxies, the width of the shocked region, both from \citet{Shu+1972} and simulations (e.g. \citealt{WadaKoda2004,DobbsBonnell2006, ShettyOstriker2006}) is very narrow compared to the width of CO arms in nearby galaxies. 
Comparing with the Canadian Galactic Plane Survey (CGPS)), \citet{Douglas2010} found that HI velocity longitude maps from simulations without feedback produced too narrow spiral arms compared to the Milky Way. 
Stellar feedback also produces bubbles and holes in the ISM. 
\citet{Dobbs2011}, and also \citet{Shetty2008}, showed that with large amounts of feedback, 
it is possible to largely erase the pattern of the original imposed stellar spiral potential. Thus the substructure associated with that of the stellar feedback becomes comparable to the imposed spiral pattern (similar to the stochastic star formation scenario, Section~2.5). 

\subsection{Substructure along spiral arms}

\begin{figure*}
\begin{center}
\includegraphics[width=.8\textwidth]{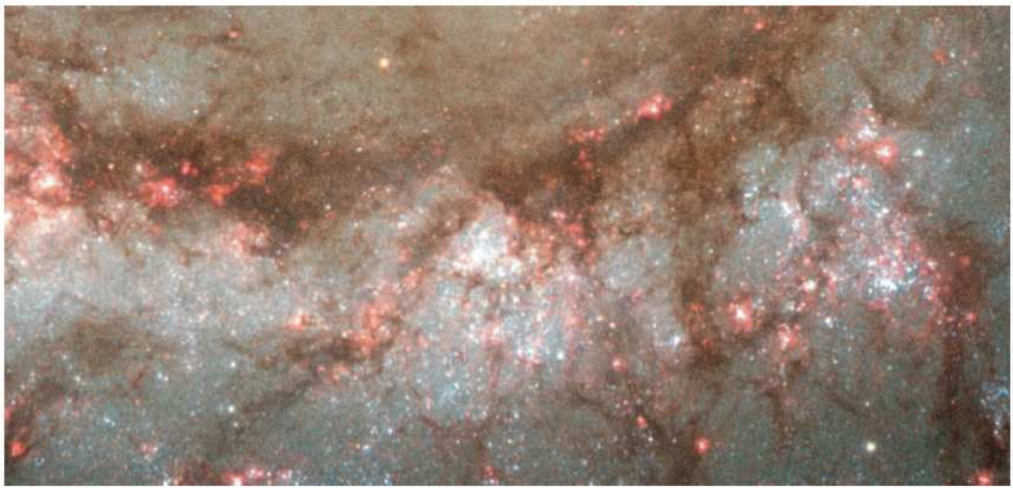}
\caption{A section along the southern spiral arm of M51, from the Hubble Heritage image. Gas flow is predominantly left to right in the figure. The spiral arm spans the figure, with 2 massive complexes along the dust lanes of the spiral arms, containing HII regions, suggesting that star formation occurs very quickly once clouds form. Below the spiral arm, are narrow lanes of gas and dust, also connected with HII regions. We term these features spurs in this paper. Some spurs extend to the next spiral arm. Bridges, which would be more associated with a bifurcation in the arms, are not particularly evident in M51. The figure is taken from \citet{Elmegreen2007} and is originally form a Hubble Heritage image, and is reproduced with permission from AAS \copyright. 
	}
\label{fig:m51observations}
\end{center}
\end{figure*}

Substructure reflects both giant molecular clouds, as well as branches, spurs and feathers which extend at clear angles away from the (typically) trailing side of the arm (see Figure~\ref{fig:m51observations}). Branches, spurs and feathers are observed in many spiral galaxies, and occur in numerical simulations. As we will see, the formation of these features is different for the different spiral arm models.

There are no formal definitions of branches, spurs and feathers. Spurs and feathers in particular have multiple meanings in the literature. Branches generally describe long structures which may go from one arm to another, and/or where one arm bifurcates into two. Consequently it may not be clear in an observed galaxy whether a feature is actually a branch or a spiral arm (including the Local Arm, \citealt{Carraro2013}). Spurs and feathers tend to be shorter features, and often describe quasi-periodic rather than isolated features. 
In their observational study, \citet{LaVigne2006} used feathers to refer to dust lanes which extend between spiral arms, and spurs to describe strings of star formation in the inter arm regions. However these `feathers' typically harbour the regions of star formation or young stars, so theoretically there is no clear need to distinguish between these two types of feature. \citet{Chakrabarti2003} use an alternative notation, whereby spurs are leading features and feathers trailing. Although they found both in numerical simulations, it is not clear observationally whether such leading features are seen in actual spiral galaxies. Finally \citet{DobbsBonnell2006} referred to spurs as any relatively short (i.e. less than one inter arm passage), narrow trailing features seen in the gas, the definition we adopt here. 

\subsection{Quasi-stationary density waves}
\begin{figure}
\begin{center}
\includegraphics[width=.45\textwidth]{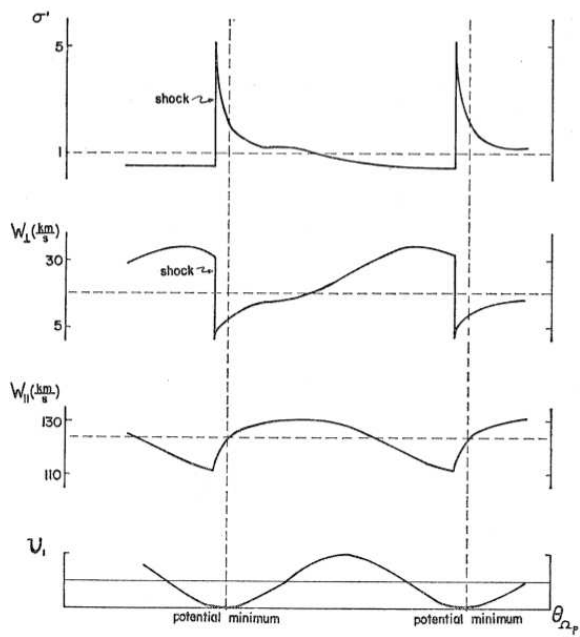}
\caption{Illustration of a typical shock solution for the gas response to a steady spiral density wave, from \citet{Roberts1969}. 
Gas flows from left to right.
The figure shows density (top), velocity perpendicular to the spiral arms (second), velocity parallel to the spiral arms (third), 
and the potential (last), versus the azimuthal angle around the galaxy. Figure reproduced with permission from AAS \copyright. }
\label{fig:roberts}
\end{center}
\end{figure}

The response of gas to spiral arms has been considered most in the context of quasi-stationary spiral density waves, where, in the case of a simple sinusoidal stellar potential, an analytic solution for the response of the gas can be obtained. Motivated by the suggestion that narrow dust features seen in external galaxies might be due to shocks, \citet{Fujimoto1968} first examined the response of gas to a spiral potential. He confirmed that the gas would indeed be likely to undergo a shock. \citet{Roberts1969} extended this analysis and, with a small correction to the work of \citet{Fujimoto1968}, obtained four equations which can be solved to obtain the velocities, spatial coordinates and density of a parcel of gas as it moves round the disc (i.e. along a streamline). These equations demonstrate that the properties of the shock, and indeed whether there is a shock, depend on the amplitude of the spiral potential ($F$), the sound speed and/or the turbulent velocity of the gas, $\sigma_{g}$, the pitch angle and location in the disc. For example, for warm gas and moderate forcing, a narrow shock is expected ahead of the minimum of the potential (Figure~\ref{fig:roberts}). If the gas is cold however, a very narrow shock is expected after the minimum of the potential. Magnetic fields are not found to greatly affect the solution, the shock is merely weaker in the magnetic case \citep{RobertsYuan1970}.

Another interpretation of spiral shocks was put forward by Kalnajs, and shown in \citet{Toomre1977}. Here the spiral forcing is considered analogous to a series of pendulums. The pendulums are assumed to oscillate like harmonic oscillators and bunch up periodically at certain intervals. The bunching up of the pendulums is analogous to parcels or clouds of gas crowding together at the spiral arms. 
\citet{Toomre1977} supposed that gas clouds pile up at the locations of the spiral arms, a little like a traffic jam. A simple calculation of test particles in a spiral potential obeys this behaviour. In the case of spiral density waves with gas pressure,  \citet{Shu+1973} found that for $c_s=8$ km s$^{-1}$, the forcing required to produce a shock is around a few \%.

\citet{Shu+1972} also investigated the dynamics of spiral shocks in the context of cloud collisions. They assumed a steady state solution and solved the jump conditions at the shock, in order to study the resultant shock structure for a two phase medium, consisting of cold clouds of a given filling fraction surrounded by warm intercloud medium. The different phases exhibit different density enhancements (of around 10 and 40), as expected, and they were able estimate a width of the shocked region of 50 pc, essentially the length scale after which the medium re-asserts an equilibrium state. Processes such as cloud collisions, and supposed gravitational fragmentation, led the authors to suppose a dynamic, rather than quasi-stationary shock scenario. Furthermore clouds will have dispersions relative to each other, and enter the spiral arms at different locations, and velocities.   

Since the 1960s and 1970s, there have been many studies that have underlined the complex response of gas to spiral density waves, and departures from the \citet{Roberts1969} picture. The gas structure along the arms arises through i) resonances, ii) instability of the spiral shock, and, iii) additional physical processes such as self gravity, cloud collisions etc. which were discussed in Section 3.3. Resonances are intrinsic to the underlying stellar potential, although they can be enhanced by self gravity in the gas. Processes included in ii) and iii) depend on the properties of the gas. But essentially all induce the formation of secondary, or substructure within the gaseous spiral arms. 

\subsubsection{Resonances}
Resonances are one means to generate gaseous substructure along spiral arms, specifically for the case of quasi-stationary density waves.
Resonances occur when the epicyclic frequency, $\kappa$ of the stellar orbits are some integer multiple of the angular frequency in the rotating frame of the spiral potential, or vice versa, thus
\begin{equation}
m(\Omega-\Omega_{\rm p})= \pm\frac{\kappa}{n}
\end{equation}
where $\Omega_{\rm p}$ is the pattern speed of the spiral, and $m$ is the number of spiral arms in the stellar disc. 
In this case stars perform $n$ radial oscillations every encounter with the $m$-armed spiral pattern.
Determining the presence and location of resonances in the disc implicitly assumes that the pattern speed, $\Omega_{\rm p}$, does not vary with radius or time.
In the vicinity of resonances, the behaviour of stellar and or gas orbits are abruptly altered and become nonlinear (e.g. \citealt{Contopoulos1986, Contopoulos1988}). The primary resonances at the ILR and OLR ($n=1$) are, as discussed in Section 2.1.2, associated with the boundary of where the spiral density waves exist. Thus we are predominantly interested in resonances within these radii. 
\citet{Shu+1973} supposed that the gas would be expected to experience perturbations due to resonances, and showed analytically that gas undergoes a secondary compression to a spiral potential at the ultraharmonic resonance ($n=2$)
\footnote{For $m=2$ spirals, the ultraharmonic resonance is called the     $4:1$ resonance.}. 
\citet{Woodward1975} demonstrated the nonlinear response of gas at the location of the ultraharmonic resonance with 1D calculations, and there have since been many further 2D and 3D (Smoothed Particle Hydrodynamics (SPH) and grid code) calculations \citep{Bertin1993, Patsis1994, Patsis1997, Chakrabarti2003}. In particular \citet{Patsis1994} showed the bifurcation of the spiral arms at the 4:1 resonance (as also predicted by \citealt{Art1992}), provided there is a spiral forcing of $F~\gtrsim5$~\%. 

\citet{Chakrabarti2003} showed the development of more varied features, including branches (bifurcations) and shorter leading and trailing features (spurs / feathers), again occurring primarily near the 4:1 resonance, with the morphology of the feature dependent primarily on the level of forcing of the spiral potential. \citet{Chakrabarti2003} suppose that flocculence in spiral galaxies could be due largely to such resonant features, an idea recently followed up by \citet{Lee2012}, where they investigate the possibility that higher order resonances lead to the formation of multiple spurs along the arms. There is a notable difference between the work of \citet{Lee2012}, and GMC formation by gravitational instabilities in the gas or cloud-cloud collisions (which are subsequently sheared into spurs). For the former, the location of the spurs does not change over time, the GMCs always forming and dispersing in the same place in the spiral arms (seemingly less likely in a dynamic environment). For other GMC formation mechanisms there is no expectation that clouds form in the same place.

\subsubsection{Stability and structure of the shock} 
Even in the non-magnetic, non self gravitating regime, several authors have questioned the stability of spiral shocks. From analytical work, \citet{Mishurov1975} first proposed that the flow through a spiral shock could be unstable. 
In contrast \citet{Nelson1977} solved the fluid equations numerically in 1D, and predicted that the flow should be stable (see also \citealt{Dwarkadus1996}) although their solutions indicate some asymmetric features. \citet{WadaKoda2004} pointed out that the latter studies adopted a tightly wound pattern, and a flat rotation curve. They performed 2D numerical simulations with different pitch angles and rotation curves, and found the spiral arms to be Kelvin-Helmholz unstable when a more open spiral pattern was used. The instability is most readily seen as spurs along the spiral arms. \citet{KimOstriker2006} found that in 3D numerical models, Kelvin-Helmholtz instabilities were suppressed, although \citet{Kim+2014} suggest an alternative `wiggle instability' mechanism.

\citet{DobbsBonnell2006} (see also \citealt{Dobbs2008}) supposed a different mechanism for producing structure, and spurs in particular, along the shock in the purely hydrodynamical, non self-gravitating case. They supposed that any substructure in the gas gets amplified as it passes through a shock. Thus, like the cloud collisions in the \citet{Toomre1977} model, clouds, or structure in the gas, get forced together by orbit crowding in the spiral shock.  A similar idea was shown in \cite{RobertsStewart1987}. Although they do not perform hydrodynamic calculations, clouds in their models undergo dissipative collisions. Like \citet{Toomre1977}, clouds can be forced together and move apart after the shock, but unlike Toomre the presence of dissipation means some clouds are effectively compressed together and retain structure after the shock. \citet{DobbsBonnell2006} showed that this process was only valid in the presence of cold gas, when the ISM is subject to thermal instabilities \citep{Dobbs+2008}  unless there is very large spiral forcing. For a warm medium, the spiral shock is relatively weaker and the pressure smoothes out any structure in the gas. The same process could have also plausibly operated (rather than Kelvin Helmholtz instabilities) in the calculations by \citet{WadaKoda2004} and \citet{KimOstriker2006}. 

\begin{figure}[h]
\begin{center}
\includegraphics[width=.45\textwidth]{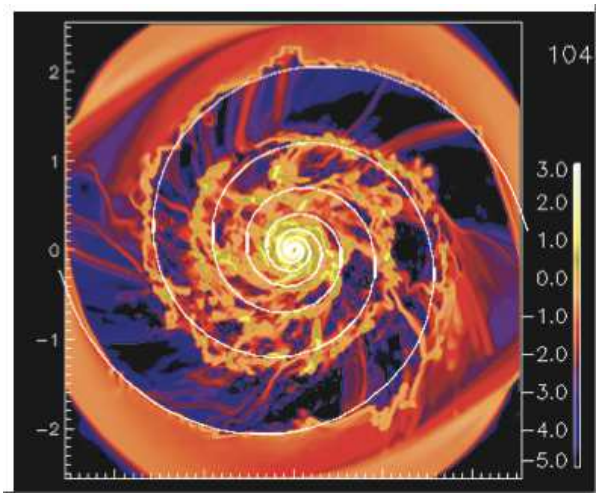}
\caption{The response of gas to an $m=2$ fixed spiral potential is shown, from \citet{Wada2008}. The minima of the spiral potential are indicated by the white lines. The simulation include a multiphase medium, and stellar feedback, so the response of the gas is highly complex. No clear continuous shock is found, and the density peak of the gas does not have a continuous offset from the minimum, although typically the density peak is after (on the trailing side of) the potential minimum.}
\label{fig:wadashock}
\end{center}
\end{figure}

In the presence of self gravity, \citet{Lubow1986} showed using 2D calculations that the gas experiences a reduced shock from the stellar potential. In an extreme case, where all the gas is situated in self gravitating clouds, the behaviour of the clouds would resemble the zero pressure case, similar to billiard balls entering the potential. \citet{Wada2008} performed full 3D hydrodynamical simulations with self gravity and a multi-phase medium, finding that the intermittency of dense gas entering the spiral potential leads to a non steady state, where the gas spiral arms are neither continuous, nor exhibit a constant offset from the arms (see Figure~\ref{fig:wadashock}, and e.g. also \citealt{DobbsPringle2013}). Rather the arms switch back and forth with time. Consequently \citet{Wada2008} does not call the response of the gas a shock in this context. The behaviour of the gas is quite different from the original \citet{Roberts1969} picture largely because the gas is far removed from a homogenous flow. Also, simulations with a multiphase medium typically do not exhibit a shock or peak in density before the spiral potential. Typically the gas density peaks after, or coincident with the minimum of the potential, because the cold gas (within a multi-phase medium) shocks later. \

Figures~\ref{fig:m51observations} and \ref{fig:wadashock} illustrate a number of the points made in Section 3.5, for a section of spiral arm in M51 and a numerical simulation respectively. The various processes in the ISM, including instabilities, turbulence and feedback lead to a `shock', or dust lanes that are very much more structured and broad than the simple analytic case. Nevertheless the response of the gas is still much sharper than the underlying potential or old stellar population. With gravity and cooling, trailing spurs are very easy to make from arm GMCs. 
As discussed in the next section, we would expected these features regardless of whether the arms are tidally induced and slowly winding up or truly stationary, the only difference for the dynamic arms being the absence of trailing spurs.

\subsection{Tidally induced spiral arms}
As discussed in section 2.4, the main difference for tidally induced spirals compared to the QSSS case is likely to be a radially dependent \citep{Oh+2008, Dobbs+2010}, rather than constant pattern speed\footnote{although there are some exceptions to this view, \citet{Salo2000} find a constant pattern speed in the centre of their models of M51, whilst \citet{Meidt+2008} adopt a model of multiple patterns for M51.}. Consequently resonance related substructure is not likely to feature in tidally induced spirals.

At large radii, the spiral arms may well be material arms, with little gas flow through the arms \citep{Meidt2013}. However at most radii, as mentioned in section 2.4, the arms are likely to be density waves,  with a pattern speed lower than that of the rotation speed (unlike the local, transient arms). Similar to the quasi-stationary density wave picture, the gas will flow through the arms and experience a spiral shock. GMCs are expected to form in the arms by the means described in Sections 3.3 and 3.5, being sheared out into spurs as they leave the arms.  
The structure of galaxies such as M51 largely reflects this behaviour.

The simulations of \citet{Salo2000} and \citet{Dobbs+2010} did not achieve the resolution required to study GMC formation, or spur formation, in their models of interacting galaxies . The simulations do predict that unlike the quasi-steady state spiral galaxy, the gas and stars are not found to be offset from each other. In their model of M51, \citet{Dobbs+2010} also showed that the double, and relatively close passage of the perturbing galaxy introduced somewhat chaotic dynamics, inducing large radial inflow and outflow motions (see also \citealt{Shetty+2007}). The chaotic dynamics also mean that the spiral arms may move with respect to the gas on relatively short ($\sim$ 10 Myr) timescales.  

\subsection{Gas flow in dynamic spiral arms}
The main difference in the behaviour of gas in galaxies with dynamic spiral arms, 
compared to quasi-stationary or tidally induced arms, is again due to the pattern speed. 
Dynamic stellar spiral arms do not exhibit significantly different rotation from the rest of the galaxy. 
Rather the spiral arms exhibit corotation everywhere \citep[e.g.,][see Section~2.2.2]{Baba+2013}. 
Thus the gas, stars and spiral arms will have the same angular velocities and there is no gas flow through the arms. 
Similar to the case of tidally induced spiral arms, the dynamics of the spiral arms means substructure due to resonances cannot occur.

In dynamic arms, the gas is still subject to the gravitational potential of the arms. As shown in \citet{DobbsBonnell2008} and \citet{Wada+2011}, gas effectively falls in to the minimum of the potential, from both sides of the spiral arm. In massive gas rich discs, the spiral arms may themselves be a manifestation of gravitational instabilities in the gas, so gas infall is coincident with arm formation. 
For dynamic arms, a systematic offset is not expected between the density peak of the gas, and the stellar minimum \citep{DobbsBonnell2008, Wada+2011}. Because the gas does  not flow through the spiral arm, the gaseous arm remains until the stellar arm disperses. Even then, since the velocity dispersion in the gas will be less than that of the arms, it will still be likely that a gaseous arm remains, even when the stellar arm has dispersed \citep{DobbsBonnell2008}

Gas can still clearly undergo shocks as it falls into the minimum of the potential, particularly if it has cooled. Shocks in nearby spiral arms were associated with the dynamic spiral arm picture, as well as the QSSS scenario in early observations (e.g. \citealt{Quirk1973}). 
The presence of quasi-regular spurs however seems less likely in galaxies with dynamic spiral arms. 
As stated in Section 3.5.2, spurs are usually the result of GMCs in spiral arms being sheared out as they move into the inter arm regions. However if the gas does not pass out of the spiral arms, this mechanism is not feasible, and indeed spurs are not typically seen in simulations \citep{DobbsBonnell2008, Wada+2011}. Larger features, such as branches, are possible though. As mentioned above, when the stellar spiral arm dissolves, the gaseous arm may remain intact for longer. The gaseous arm thus represents a feature without a stellar counterpart, can appear as a branch between spiral arms \citep{DobbsBonnell2008}. 

Again, GMC formation will occur as described in Section 3.3. So far, there is no noticeable difference between the properties of GMCs in simulations of galaxies with global spiral spiral arm versus local transient arms \citep{Hopkins2011, Dobbs2011}, and likely not different to tidal arms. However there are differences in comparison to the clouds in a completely smooth stellar disc, where there are no stellar spiral arms, and processes such as cloud-cloud collisions are more limited. In the case without spiral arms, the maximum cloud mass is smaller (e.g. a few $10^5$ M$_{\odot}$ compared to a few $10^6$ M$_{\odot}$), whilst clouds tend to preferentially exhibit prograde rotation rather than randomly orientated rotation.

\section{Observational evidence for different mechanisms of generating spiral structure}
Although not yet conclusive, there are a growing number of observational tests for whether galaxies display quasi-stationary density waves, tidally induced or bar driven spirals, or instability induced dynamic spiral arms. Most of the observational tests relate to the pattern speed of the spiral arms, and whether the distribution of gas and stars match the predictions for a fixed pattern speed or not. Only the quasi-stationary density wave picture adopts a constant pattern speed, whilst the other mechanisms induce arms with radially decreasing pattern speeds.

\subsection{Pattern speeds}
The pattern speed is difficult to measure directly, but there have been many attempts to estimate pattern speeds in galaxies. Most measurements have assumed that the pattern speed is constant, thereby immediately adopting the assumption that the spiral pattern is a quasi-stationary spiral density wave. The simplest means of determining the pattern speed is to estimate the location of corotation \citep[e.g. assumed coincident with the outer extent of the arms, or a dip in HI or CO at a certain radius][]{Shu+1971, Rots1975b, Elmegreen+1989, Sempere1995}. Supposed locations for the ILR (e.g. assumed coincident with the inner extent of the spiral arms, rings or inter arm features) and/ or the OLR can be similarly used to estimate the pattern speed \citep{LinShu1967, Gordon1978, Elmegreen+1989}. These methods suffer from uncertainties, both observationally and theoretically about where spiral arms begin and end (see e.g. \citealt{Contopoulos1986, Elmegreen1998, Englmaier2000}), and of course whether these really are resonance features. 

Another indirect test of the pattern speed is the location of the spiral shock, and star formation relative to the minimum of the stellar potential.  For a fixed spiral pattern, the spiral shock will lie one side of the minimum of the stellar potential within corotation, and the opposite side outside corotation. The width of this offset can be used to determine the pattern speed, if a constant sound speed is assumed. \citet{Gittins2004} demonstrated this method using numerical simulations, where they presumed the spiral shocks will correspond to dust lanes. Likewise, assuming a timescale for star formation to occur, the molecular clouds (CO) and H$\alpha$ will have a predicted offset. \citet{Egusa2004} used this prediction to derive a constant pattern speed for NGC~4254, following which they are able to derive pattern speeds for 5 out of a sample of 13 galaxies \citep{Egusa+2009}. Difficulties in obtaining pattern speeds for many spirals, and the large scatter with this method, likely arise because the spirals are transient, and the offsets local and non-systematic. \citet{Tamburro+2008} instead measured offsets between atomic hydrogen and recent star formation (24 $\mu$m maps) to simultaneously fit $\Omega_{\rm p}$ and the star formation timescale for M51, the latter found to be 1-4 Myr. Repeating their analysis however, \citet{Foyle+2011} found no evidence for a systematic offset, and thus a constant pattern speed. One difference may be that \citet{Foyle+2011} try to fit a pattern speed over the entire galaxy, whereas \citet{Tamburro+2008} studied localised regions. Such differences indicate the large uncertainties in observationally determining the behaviour of spiral arms, and that the assumption of a constant pattern speed may be invalid.

As well as using morphological features of spiral galaxies to determine corotation, the kinematics can also be used, either from the residual velocity fields or changes in the directions of streaming motions \citep{Canzian1993, Sempere1995, Elmegreen1998}. \citet{Canzian1993} showed that the spiral residual velocity field shows a single spiral feature inside corotation, and 3 spiral features outside corotation. 
\citet{Font+2011} also used the velocity field to determine the location of resonances from the locations where the residual velocities are zero, from high resolution H$_{\alpha}$ data.

So far all the methods described assume that the pattern speed is constant. 
The \citep{Tremaine1984} method uses the continuity equation for gas flow across the spiral arms, which relative to the rest frame, and integrating over each direction, gives an expression of the form
\begin{equation}
\Omega_{\rm p} \int \Sigma x dx = \int \Sigma v_y dx
\end {equation}
\citep{Merrifield+2006}. Here, $\Sigma$, $v_y$ and $x$ are all observables (of the relevant tracer) which means $\Omega_{\rm p}$ can be determined. For the continuity equation to be valid, this method tends to use HI or CO to avoid problems with extinction, although a small fraction of the gas will be turned into stars. The Tremaine-Weinberg method also assumes a steady state, i.e. that the spiral arm is not changing over the timescale that gas passes through the arm, and assumes a thin disc, but is in principle not limited by the shape of the arm.

This technique has been used mainly for barred galaxies, due mainly to the simpler geometry, but also for a number of spiral galaxies \citep{Sempere1995, Zimmer2004, Rand2004}. The method can be extended to allow for a radial dependent pattern speed \citep{Westpfahl1998, Merrifield+2006}. In this case, the pattern speed can be determined by solving a matrix equation over different positions within a small (e.g. 0.5 kpc width) region along a spiral arm. The pattern speed is then computed for other regions at different radii. Using this method, a number of studies have found radially decreasing pattern speeds \citep{Merrifield+2006, SpeightsWestpfahl2011, SpeightsWestpfahl2012} including for M81 \citep{Westpfahl1998}. 
\citet{Meidt+2008,Meidt+2009} also found radial dependent pattern speeds in M51, M101 and a number of other galaxies, 
but attributed these to different patterns with different pattern speeds at different radii, rather than a continuously decreasing pattern speed. Differences in pattern speed are likely at the transition from a bar to spiral arms (e.g. \citealt{Meidt+2008}, see also Section~2.3), but this would not explain discrete changes in pattern speed at larger radii, or in the absence of a bar. 

To date, no examples of galaxies with a constant pattern speed have been found with the radially varying Tremaine-Weinberg method. Most galaxies show a slowly decreasing pattern speed in the outer regions. The Tremaine-Weinberg method appears sufficient to establish that patterns speeds vary radially (i.e. the pattern speed varies much more than the error bars), but not whether the pattern speed varies continuously or consists of multiple segments each rotating at a constant pattern speed \citep{Meidt+2008,Meidt2008b}.

\begin{figure}[htbp]
\begin{center}
\includegraphics[width=0.45\textwidth]{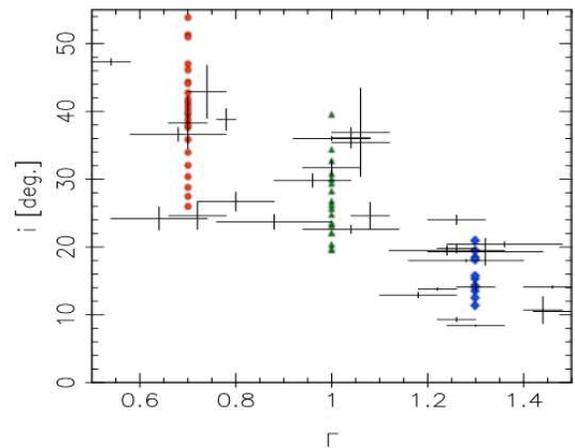}
\caption{
	The pitch angle is shown versus shear, from \citet{Grand+2013}. The coloured points represent simulated values taken from \citet{Grand+2013}, whilst the crosses are observed values, from Table~3 of \citet{Seigar+2006}.
	}
\label{fig:Grand+2013}
\end{center}
\end{figure}

\subsection{Pitch angles}
We can also consider whether the pitch angles of spiral arms of observed galaxies match predictions from models and theory.
\citet{Grand+2013} analyzed the pitch angles of transient stellar spiral arms in 
galaxy models with different shear rates ($\Gamma$), and showed 
that the higher shear rates produce more tightly wound spiral arms. It is also clear that the pitch angle of the spiral arms decreases with time.
Figure \ref{fig:Grand+2013} shows the pitch angles of simulated spiral arms plotted against
the shear rate of model galaxies. 
The observed correlation of real spiral galaxies are also overlaid on this figure.
This trend and scatter are both consistent with the observations \citep{Seigar+2005,Seigar+2006}.
Though spiral arms wind up by differential rotation, typical pitch angles depend on the
shear rate of disc galaxies suggesting that swing amplification is important for generating 
spiral arms because swing-amplified spiral arms reach maximum amplitudes at a specific 
pitch angle depending on the shear rate (Section 2.1.3 and Figure \ref{fig:Baba+2013fig7}).

The quasi-stationary density wave theory may also satisfy the pitch angle-shear rate correlation qualitatively 
\citep{LinShu1964,Roberts+1975}, since \citet{LinShu1964} demonstrated that the pitch angle 
of quasi-stationary density waves is lower for higher central mass concentration, i.e., a higher shear rate. 
Observations by \citet{Block+1994} support the scenario that spiral arm properties are intrinsic to 
a galaxy dependent on galaxy morphology and gas content (see also Section 4.5).

\subsection{Stellar cluster ages}
\begin{figure*}
\begin{center}
\includegraphics[width=.8\textwidth,angle=90]{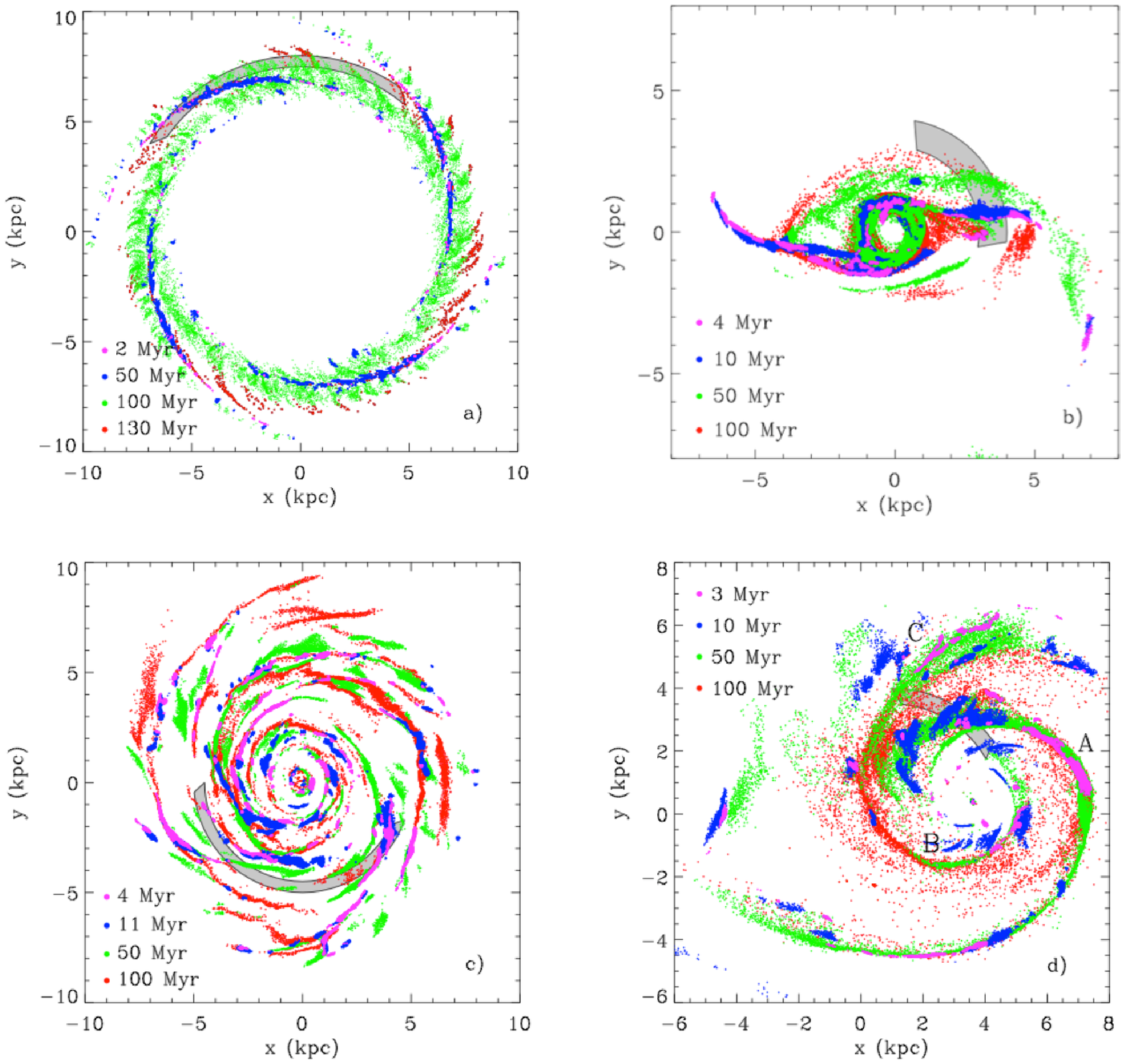}
\caption{The spatial distribution of clusters of different ages is shown for different galaxy models: fixed spiral potential (top left), barred galaxy (top right), dynamic spiral arms (lower left) and a model of M51 (lower right). For the fixed potential and bar. there is a transition of stellar ages moving away from the spiral arms / bar. For the flocculent galaxy, star clusters of similar age tend to be located in a spiral arm, and the ages do not show clear transitions, rather they are more random. From \citet{DobbsPringle2010}.}
\label{fig:clusterages}
\end{center}
\end{figure*}

The ages of stellar clusters can also be used as a test of the underlying dynamics in galaxies. This method was described in \citet{DobbsPringle2010}. The ages of clusters should clearly increase with distance away from the spiral arm (in the leading direction) for a constant pattern speed. Likewise a similar pattern is expected for a bar. However for the case of dynamic spiral arms due to local instabilities there is no flow of material through the spiral arms. Hence no age pattern is expected, rather stars of similar ages lie along a spiral arm (see Figure~\ref{fig:clusterages}). 
The numerical models of \citet{DobbsPringle2010} were relatively simple, and did not include for example stellar feedback. More complicated models have since been performed by \citet{Wada+2011} and \citet{Grand+2012a,Grand+2012b}. They confirmed the case that for dynamic arms, there is no clear age pattern, testing multi-armed galaxies both with and without bars. \citet{Dobbs2014} extended this idea further by looking specifically at stellar age spreads in GMCs, and suggest again that different age distributions may reflect how the spiral arms are generated.  

\citet{DobbsPringle2010} also tested a model of M51, where they found that although the spiral arms are kinematic density waves, and there is flow of material through them, the dynamics of the interaction were rather chaotic and produced a chaotic distribution of stellar ages. Observations of stellar ages in M51 have since confirmed a similar picture \citep{Foyle+2011, Kaleida2010, Chandar2011}. Studies of galaxies undergoing less violent interactions have not been tested. 

From the numerical models, the patterns for the dynamic spirals appear most robust, because the peaks in the number of stars corresponding to the arms are significantly higher than the noise (e.g. a factor of $\sim$ 10). For the case of the stationary density wave, the pattern is potentially more confusing, because the peaks in the number of young stars decrease and broaden away from the arm,  so they are less distinct. Furthermore after a relatively short time (10s Myrs), the young stars will catch up with the next spiral arm.

Several observational studies have examined whether age patterns exist in a number of nearby galaxies \citep{Sanchez2011, Foyle+2011, Ferreras+2012}. With the exception of one or two galaxies e.g. M74, the observations generally find little evidence of age patterns. 
As well as stellar clusters, a number of studies have also used colour gradients across the spiral arms as a measure of a transition in stellar ages \citep{Efremov1982, Regan1993, Beckman1990, Gonzalez1996, Martinez2009}. However again, with the exception of one or two cases, there is rarely a clear trend in the colour gradients.

\subsection{Resonances and interference patterns}
As well as looking at the pattern speeds, or age spreads, it is also possible to look for specific features that result from the quasi-stationary density wave picture. Patterns of star formation along the spiral arms have been seen in some galaxies, and attributed to resonances, for example a dip in star formation at corotation \citep{Cepa1990, Knapen1992}.
As already mentioned, features associated with resonances are expected at certain radii in a disc. In a few galaxies, symmetric spurs, or breaks in the spiral arms are relatively convincing, e.g. NGC 1566 \citep{ElmegreenElmegreen1990}. Features such as outer rings due to bars can also be seen (e.g. \citealt{Buta1991}). However in many cases subtle features associated with spiral arms may simply be due to the shearing of clouds, or bridges where spiral arms in the gas remain whilst corresponding features in the stars have dissipated (Section 3.7). 

In the global model theory of spiral arms, see Section 2.1.4,
the spiral pattern is thought to correspond to an interference pattern resulting from leading and trailing waves in the stellar disc.  \citet{Elmegreen1989} found some signs of leading waves for M51, M81 and M100, but equally the patterns they found could be simply due to the underlying complex structure of the disc.

\subsection{Observations of grand design and flocculent structure}
Any theory(ies) of spiral structure also need to explain the observed frequency of grand design and flocculent spirals. 
One of the arguments for quasi-stationary density waves has been the large number of galaxies with $m=2$ spiral structure. In density wave theory,  density waves with $m>3$ are less likely to be stable \citep{LinShu1967, Toomre1977,Thomasson+1990} compared with $m=2$, explaining the preference for $m=2$ spirals. An alternative explanation is that the $m=2$ spirals are all tidally induced or bar driven. This hypothesis was made by \citet{KormendyNorman1979}, who found the majority of grand design galaxies had bars or companions.  Some isolated galaxies exhibited arms which did not extend to the flat part of the rotation curve, which meant that they could potentially be longer lived spirals, as the winding problem is avoided. Other isolated grand design galaxies in their sample were thought to have undergone recent mergers. 

Observations also show that the frequency of grand design galaxies doubles in clusters or groups compared to otherwise \citep{ElmegreenElmegreen1983}. There are thus few isolated grand design galaxies, but even then, determining whether there are examples which could not be explained by tidal interactions is difficult, partly due to the difficulty of establishing truly isolated galaxies and those that have not undergone a recent merger (see e.g. \citealt{Verley2007}). It is also not established either theoretically, or using cosmological simulations, whether interactions are likely to be frequent enough to account for the observed number of grand design spirals. As discussed in Section 2.4.4, spiral galaxies can be expected to retain $m=2$ structure for $\sim1$ Gyr after an interaction.

One argument for the existence of long-lived spiral arms is the finding that some galaxies that appear flocculent or multi-armed in the optical exhibit an underlying grand design pattern in  the old stellar population, i.e. as seen in the K band \citep{BlockWainscoat1991,Block+1994, Block+1996, Thornley1996, ThornleyMundy1997a, GrosbolPatsis1998, Seigar+2003}. Whilst these structures could be tidally induced, \citet{Block+1994} suggested that in fact the stellar disc supports low $m$ modes whereas the gas (and young stars) does not since low $m$ modes are damped at the ILR. Thus in this scenario the two components of the disc are assumed to be decoupled. Given that theory, and simulations, show that gas shocks at the spiral arms, presumably there is still some relation of the gas to the stars, many of the examples in \citet{Block+1994} simply show an extra optical arm. 
An alternative scenario, in the dynamic arm picture is that these galaxies, which are sufficiently massive to exhibit only a small number of spiral arms, are transitioning between 2 and 3 armed patterns, and the features in the optical are remainders of spiral arms where the stellar pattern has dispersed, but the gas arm (which is clearly denser and colder) still persists.   
\citet{Chakrabarti2003} provided an alternative picture whereby resonances may be responsible for generating substructure from an underlying stationary $m=2$ pattern, particularly for highly flocculent galaxies, although other means of generating substructure (e.g. instabilities, stellar feedback) may be just as likely responsible  (see reference to \citealt{Elmegreen+2003a} below). 
The most recent observations found that most flocculent galaxies do not exhibit grand-design structure \citep{Elmegreen+2011} and those that do have very weak spiral arms \citep{Elmegreen1999}, but the co-existence of different patterns still needs to be explained.

Conversely all galaxies may be flocculent galaxies, which merely develop an overwhelming $m=2$ mode during tidal perturbations, or with a bar (Sections 2.3 and 2.4). \citet{Colombo2014} found evidence for an underlying flocculent spiral in the grand-design spiral M51, proposed for the old stars as well as the gas (CO). \citet{Elmegreen+2003a} suggested that both grand design and flocculent spirals (as seen in the old stars) exhibit a similar structure in the gas and young stars (independent of the underlying old stellar population) which is driven by turbulence in the disc. 

\subsection{The Milky Way}
The spiral structure of our Galaxy is reviewed thoroughly in Benjamin (2014), so we only briefly discuss the Milky Way here. 

The number of spiral arms in our Galaxy is still debated somewhat (see e.g. \citealt{Vallee2005}), but is most frequently considered to be either 2 or 4. There are 4 main spiral arms; the Perseus, Sagittarius, Scutum-Crux, and Norma spiral arms, and at least one bar. There is in addition the Outer Arm, which may be the outer part of one of the inner arms, and the Local or Orion arm, which is much shorter, and may be a bridge or spur rather than a real arm.

A large number ($m>2$) of arms would support the view that the Galaxy better resembles a flocculent, rather than grand design spiral, with multiple dynamic arms induced by local gravitational instabilities. An alternative interpretation is that the Galaxy has two main spiral arms (the Perseus and Scutum-Centaurus arms), with the other two arms lesser features, perhaps only present in gas and young stars \citep{Drimmel2000}. Such a scenario could arise in the quasi-stationary density wave picture if the secondary arms are resonance related features \citep{Martos2004}. In the \citet{Churchwell2009} map of the Galaxy, the secondary arms appear to start at the ends of the bar, and / or be connected with the inner 3 kpc arms, and the main arms also start at the bar.

As well as gravitational instabilities induced locally by perturbations in the stellar distribution, or GMCs, the Galaxy is surrounded by low mass companions, and contains one or two bars. Hence there is no shortage of mechanisms to generate spiral arms. \citet{Purcell2011} showed that a recent passage of the Sagittarius galaxy could have induced spiral arms, though their simulations did not show the detailed spiral structure. Either the bar(s) or interactions could lead to an $m=2$ pattern. One of the most striking pieces of evidence that there is a symmetric $m=2$ pattern, suggestive of Lin-Shu density wave theory is the recent discovery of an outer HI arm, which is found to match up exactly with the inner Sagittarius arm, assuming a continuous $m=2$ logarithmic spiral pattern \citep{Dame2011}. Some other models of the Galaxy tend to show kinked arms, rather than continuous spiral arms (e.g. \citealt{Taylor1993}). 

Numerical simulations have also been performed to examine the structure of the Milky Way, by comparing $l-v$  maps of simulations with those observed \citep{Wada1994,Fux1999,EnglmaierGerhard1999, Rodriguez-FernandezCombes2008, Baba+2010, DobbsBurkert2012,Pettitt+2014}. 
\citet{DobbsBurkert2012} showed that the nearest spiral arm, in their instance from adopting a symmetric $m=2$ spiral, likely corresponds to the `molecular ring'. However generally it is difficult to reproduce the outer Milky Way with logarithmic spirals \citep{EnglmaierGerhard1999,Pettitt+2014}. Fitting the $l-v$ map from a simulation of a bar and dynamic spiral arms appears more successful \citep{Baba+2010}. \citet{Baba+2009} also analysed the velocities of gas and young stars from $N$-body$+$hydrodynamical simulations, and concluded that the high peculiar (non-circular) velocities they obtain, in general agreement with those observed in the Galaxy, arise from dynamic rather than stationary spiral arms.

Generally, the spiral pattern remains uncertain for the Milky Way, particularly as little is known about the spiral structure on the other side of the Galactic Center. There is also no conclusive observational evidence yet on the nature of the dynamics of the spiral arms. 
The Very Long Baseline Interferometer (VLBI) astrometry \citep[e.g., VERA;][]{Honma2013}, as well as future space missions for infrared astrometry 
\textsc{GAIA} \citep{Perryman+2001} and \textsc{JASMINE} \citep{Gouda2012}, may well be able to provide a better indication of the nature of the spiral structure of our Galaxy.

\subsection{Spiral arm triggering of star formation}

A related question to the inducement of spiral arms in galaxies is whether the spiral arms induce star formation. This possibility has been considered in the quasi-stationary density wave picture, where the spiral arms induce a shock in the gas, thus raising the gas to the densities where it becomes molecular and self gravitating \citep{Fujimoto1968,Roberts1969}. In this scenario, the star formation is significantly higher with the presence of spiral shocks than it would be in a galaxy with no, or weak spiral arms. Some evidence in support of spiral arm triggering is observations by \citet{SeigarJames2002}, that show a correlation between arm strength and star formation. However other work suggests there is no difference in the star formation rate between grand design and flocculent galaxies \citep{Elmegreen1986, Stark1987, Foyle+2011, Eden2012}. Instead, the spiral arms are supposed merely to gather gas which would anyway form stars, into the spiral arms \citep{Vogel1988}, with the increase in gas densities and star formation in the arms offset by lower values in the inter arm regions. Numerical simulations support this picture \citep{Dobbs2011}, finding only a factor of $\sim2$ increase with spiral arms compared to without. The act of the spiral arms simply gathering up gas is also consistent with the picture of shocks being highly dynamic, and intermittent, as described in Section 3.5.2.  However there is a tendency to form more massive clouds, and more stable clouds in the spiral arms, which may well lead to higher star formation rates in stronger spiral arms \citep{Dobbs2011}.  

\section{Summary and Discussion}
The origin of spiral arms in galaxies is a longstanding problem in astrophysics. Although, by no means solved, here we summarise the progress  of different theories and observations. 

\subsection{Quasi-stationary density wave theory}
In the late 1960s and 1970s the problems of maintaining quasi-stationary spiral density waves were emerging, 
and the WASER mechanism/ swing amplification proposed to maintain standing waves in the disc. This approach has been developed further, for example investigating damping and gas dissipation to maintain a steady state, as described in Sections 2.1 and 4.2. As also described in Section 2.1, solutions and stability criteria for non-uniformly rotating discs have also been investigated. However as yet there has not been any demonstration that the WASER mechanism works, and that standing waves can develop. $N$-body simulations of galaxies were just developing in the 1970s and 80s, but the picture has remained largely unchanged. Instead, as discussed in Section 2.1.5, spirals in simulations appear to be dynamic features, more associated with the swing amplification mechanism for generating spiral features. Some simulations find longer lasting modes than predicted by swing amplification alone, but the overall spiral pattern is still transient, recurrent in nature \citep{D'Onghia+2013,Sellwood2014}. Others specifically designed to support a standing wave between the ILR and OLR, still find a pattern that changes from $m=2$ to $m=3$ and is ultimately transient recurrent \citep{Sellwood2011}. The simulations of \citet{Salo2000} also resemble 
the density waves proposed by density wave theory \citep{Kalnajs1965,LinShu1966}. In this case, self gravity of the disc is high, and the tidally induced features in their models may indeed be sufficiently self-gravitating to allow propagating waves. However it is still not clear that these waves are maintained, or indeed any clear necessity that the density waves need to be maintained.

We note that the success of these models in reproducing density wave theory depends to some extent on the interpretation of quasi-stationarity, and whether current simulations satisfy quasi-stationarity. However for those simulations with longer lived spirals, it has not been shown that that the spirals satisfy global mode theory (e.g. \citealt{Bertin+1989a,Bertin+1989b}) and do not exhibit a steady shape over their lifetime. Observationally we do not readily distinguish between very transient spiral arms, and spiral arms which are ultimately still transient, but survive multiple rotation periods.

\subsection{Dynamic spirals}
Spiral arm formation from swing amplified instabilities was demonstrated nearly 30 years ago in simulations by \citet{SellwoodCarlberg1984}, and still remains a clear mechanism for producing spiral arms. Typically the dynamic spiral arms produced resemble multi-armed or flocculent galaxies, but as discussed in the previous section, it also possible to produce low $m$ patterns.
In recent work, as described in Section 2.2, more details of this mechanism have emerged, for example non-linear evolution, radial migration of stars, the behaviour of the arms and how they corotate with the gas. The simulations have also demonstrated that the predictions of the number and properties of spiral arms are in agreement with the theory. The simulations have recently shed light on a long-standing conundrum with regards the longevity of spiral patterns generated in this way. High resolution calculations \citep{Fujii+2011, D'Onghia+2013} demonstrate that in fact the heating of the spiral arms due to dissipation is much less than previously thought. Thus it possible for such spiral patterns to last for much longer, up to $\sim10$ Gyr. Coupled to this, observations also demonstrate that spiral galaxies exist with little or no star formation \citep{Masters2010}, so there is no longer a need for a gas component, or cold accretion onto the galaxy.

\subsection{Tidal interactions}
Tidal interactions were certainly recognised as a means of producing spiral arms by the 1980s, but it was not clear whether the induced arms would correspond to kinematic density waves or stationary waves, and whether tidal interactions could produce spiral structure extending to the centres of galaxies. It is now clear from simulations that tidal interactions can readily reproduce grand design structure, although are unlikely to account for multi-armed or flocculent patterns. The dynamics of the arms is dependent on the self gravity of the disc. In the absence of self gravity, the arms are kinematic waves. With increasing self gravity, the arms become more rigid, less susceptible to winding, and with a higher pattern speed. In particular, the central parts of galaxies which are most dense are most susceptible to developing a more rigid pattern, and in some cases a bar. Simulations have shown that tidally induced spirals can last around a Gyr, thus certainly in galaxy groups interactions may well be frequent enough to explain the presence of $m=2$ spirals.

\subsection{Bar driven spirals}
As described in Section~2.3, there are now numerous means by which bars can induce spiral arms, and consequently different behaviour of the spiral arms in relation to the bar. As yet however, there is no clear indication which scenario, whether manifold theory, bar induced spirals,  different patterns for the bar and arms, or nonlinear coupling prevails. And, as discussed in Section 2.3, the behaviour of the spiral arms. Whether they have near constant pattern speeds, or are trailing in nature more similar to the swing amplified model of arm formation, is different between different simulations, and in any case is likely to evolve with time. The range of morphology in observed barred galaxies suggests that spirals in barred galaxies have multiple origins. 

\subsection{Other mechanisms}
As we have stated in Section 2.5, the stochastic star formation mechanism has fallen out of favour. Self propagating star formation likely leads to structure in the gas and new stars in galaxies, which produces a much more irregular and flocculent appearance than the underlying old stars. However, simulations that adopt  a smooth (structureless) stellar disc and follow the gas and new star formation with hydrodynamics do not find very realistic spiral patterns. At least some structure is required in the stars, for example from swing-amplified noise or perturbations. 

Dark matter halos are certainly a plausible means of generating spiral structure but at present we have no way of telling where they are or what effects they are having (if any) on the dynamics of stellar discs.

\subsection{Observations}  
As discussed in Section~4, current observational tests do not yet rule out any of the proposed mechanisms for determining spiral structure. However we note that now the resolution of observational data is such that tests on determining  the origin of spiral arms are becoming feasible, and results increasingly reported in the literature. We emphasised that in the past, observational results have often been limited by the assumption of a constant pattern speed. Applications of the radially dependent Tremaine-Weinberg method have shown that the arms in both grand-design and flocculent galaxies exhibit radially decreasing pattern speeds. Mapping the ages of stellar clusters appears to be a useful test of distinguishing galaxies where gas does not flow through the spiral arms, as is the case for local swing amplified instabilities. Distinguishing the nature of pattern speeds in clear grand design galaxies may be a good way of testing the rigidity of spiral arms. Examining clusters in galaxies that appear to be isolated grand design galaxies (or multi-arm galaxies with a prominent $m=2$ pattern), may be a good test of whether the arms originate from swing amplified instabilities, or are density waves.

Interestingly, the gas response seems  to first order independent of the nature of the spiral arms (whereas bars for example induce such large shear that star formation appears to be suppressed). Gas and young stars dominate the observed structure, but the spiral potential merely gathers the gas together in the arms than change the gas properties or star formation rate. Thus, other processes in the ISM, such as turbulence, gravity and cloud collisions may have a greater role on the gas dynamics and star formation than spiral arms.

\begin{acknowledgements}
We would like to thank the referee for a very helpful, and thorough report.
CLD acknowledges funding from the European Research Council for the 
FP7 ERC starting grant project LOCALSTAR.
JB would like to thank Keiichi Wada, Masafumi Noguchi, Shugo Michikoshi, Shunsuke Hozumi, 
and Kana Morokuma-Matsui for their valuable comments.
JB was supported by the HPCI Strategic Program Field 5 ``The Origin of Matter and the Universe.'' 
\end{acknowledgements}

\bibliographystyle{apj}
\bibliography{ms}

\begin{thebibliography}{358}
\expandafter\ifx\csname natexlab\endcsname\relax\def\natexlab#1{#1}\fi

\bibitem[{{Aoki} {et~al.}(1979){Aoki}, {Noguchi}, \& {Iye}}]{Aoki+1979}
{Aoki}, S., {Noguchi}, M., \& {Iye}, M. 1979, PASJ, 31, 737

\bibitem[{{Appleton} \& {Struck-Marcell}(1996)}]{Appleton1996}
{Appleton}, P.~N. \& {Struck-Marcell}, C. 1996, \fcp, 16, 111

\bibitem[{{Arp}(1966)}]{Arp1966}
{Arp}, H. 1966, \apjs, 14, 1

\bibitem[{{Artymowicz} \& {Lubow}(1992)}]{Art1992}
{Artymowicz}, P. \& {Lubow}, S.~H. 1992, \apj, 389, 129

\bibitem[{{Athanassoula}(1978)}]{Athanassoula1978}
{Athanassoula}, E. 1978, \aap, 69, 395

\bibitem[{{Athanassoula}(1984)}]{Athanassoula1984}
---. 1984, PhR, 114, 319

\bibitem[{{Athanassoula}(1992)}]{Athanassoula1992b}
---. 1992, MNRAS, 259, 345

\bibitem[{{Athanassoula}(2012)}]{Athanassoula2012}
---. 2012, MNRAS, 426, L46

\bibitem[{{Athanassoula} {et~al.}(2009{\natexlab{a}}){Athanassoula},
  {Romero-G{\'o}mez}, {Bosma}, \& {Masdemont}}]{Athanassoula+2009b}
{Athanassoula}, E., {Romero-G{\'o}mez}, M., {Bosma}, A., \& {Masdemont}, J.~J.
  2009{\natexlab{a}}, MNRAS, 400, 1706

\bibitem[{{Athanassoula} {et~al.}(2010){Athanassoula}, {Romero-G{\'o}mez},
  {Bosma}, \& {Masdemont}}]{Athanassoula+2010}
---. 2010, MNRAS, 407, 1433

\bibitem[{{Athanassoula} {et~al.}(2009{\natexlab{b}}){Athanassoula},
  {Romero-G{\'o}mez}, \& {Masdemont}}]{Athanassoula+2009a}
{Athanassoula}, E., {Romero-G{\'o}mez}, M., \& {Masdemont}, J.~J.
  2009{\natexlab{b}}, MNRAS, 394, 67

\bibitem[{{Athanassoula} \& {Sellwood}(1986)}]{AthanassoulaSellwood1986}
{Athanassoula}, E. \& {Sellwood}, J.~A. 1986, MNRAS, 221, 213

\bibitem[{{Baba} {et~al.}(2009){Baba}, {Asaki}, {Makino}, {Miyoshi}, {Saitoh},
  \& {Wada}}]{Baba+2009}
{Baba}, J., {Asaki}, Y., {Makino}, J., {Miyoshi}, M., {Saitoh}, T.~R., \&
  {Wada}, K. 2009, ApJ, 706, 471

\bibitem[{{Baba} {et~al.}(2010){Baba}, {Saitoh}, \& {Wada}}]{Baba+2010}
{Baba}, J., {Saitoh}, T.~R., \& {Wada}, K. 2010, PASJ, 62, 1413

\bibitem[{{Baba} {et~al.}(2013){Baba}, {Saitoh}, \& {Wada}}]{Baba+2013}
---. 2013, ApJ, 763, 46

\bibitem[{{Balbus} \& {Cowie}(1985)}]{Balbus1985}
{Balbus}, S.~A. \& {Cowie}, L.~L. 1985, \apj, 297, 61

\bibitem[{{Bardeen}(1975)}]{Bardeen1975}
{Bardeen}, J.~M. 1975, in IAU Symposium, Vol.~69, Dynamics of the Solar
  Systems, ed. {A.~Hayli}, 297--+

\bibitem[{{Beckman} \& {Cepa}(1990)}]{Beckman1990}
{Beckman}, J.~E. \& {Cepa}, J. 1990, \aap, 229, 37

\bibitem[{{Bertin}(1983)}]{Bertin1983}
{Bertin}, G. 1983, A\&A, 127, 145

\bibitem[{{Bertin}(1993)}]{Bertin1993}
---. 1993, \pasp, 105, 640

\bibitem[{{Bertin}(2000)}]{Bertin2000}
---. 2000, {Dynamics of Galaxies}, ed. {Bertin, G.}

\bibitem[{{Bertin} {et~al.}(1977){Bertin}, {Lau}, {Lin}, {Mark}, \&
  {Sugiyama}}]{Bertin+1977}
{Bertin}, G., {Lau}, Y.~Y., {Lin}, C.~C., {Mark}, J.~W.-K., \& {Sugiyama}, L.
  1977, Proceedings of the National Academy of Science, 74, 4726

\bibitem[{{Bertin} \& {Lin}(1996)}]{BertinLin1996}
{Bertin}, G. \& {Lin}, C.~C. 1996, {Spiral structure in galaxies a density wave
  theory}, ed. G.~{Bertin} \& C.~C. {Lin}

\bibitem[{{Bertin} {et~al.}(1984){Bertin}, {Lin}, \& {Lowe}}]{Bertin+1984}
{Bertin}, G., {Lin}, C.~C., \& {Lowe}, S.~A. 1984, in ESA Special Publication,
  Vol. 207, Plasma Astrophysics, ed. T.~D. {Guyenne} \& J.~J. {Hunt}, 115--120

\bibitem[{{Bertin} {et~al.}(1989{\natexlab{a}}){Bertin}, {Lin}, {Lowe}, \&
  {Thurstans}}]{Bertin+1989a}
{Bertin}, G., {Lin}, C.~C., {Lowe}, S.~A., \& {Thurstans}, R.~P.
  1989{\natexlab{a}}, ApJ, 338, 78

\bibitem[{{Bertin} {et~al.}(1989{\natexlab{b}}){Bertin}, {Lin}, {Lowe}, \&
  {Thurstans}}]{Bertin+1989b}
---. 1989{\natexlab{b}}, ApJ, 338, 104

\bibitem[{{Bertin} \& {Romeo}(1988)}]{BertinRomeo1988}
{Bertin}, G. \& {Romeo}, A.~B. 1988, A\&A, 195, 105

\bibitem[{{Binney} \& {Tremaine}(2008)}]{BinneyTremaine2008}
{Binney}, J. \& {Tremaine}, S. 2008, {Galactic Dynamics: Second Edition}, ed.
  {Binney, J.~\& Tremaine, S.} (Princeton University Press)

\bibitem[{{Bissantz} {et~al.}(2003){Bissantz}, {Englmaier}, \&
  {Gerhard}}]{Bissantz+2003}
{Bissantz}, N., {Englmaier}, P., \& {Gerhard}, O. 2003, \mnras, 340, 949

\bibitem[{{Block} {et~al.}(1994){Block}, {Bertin}, {Stockton}, {Grosbol},
  {Moorwood}, \& {Peletier}}]{Block+1994}
{Block}, D.~L., {Bertin}, G., {Stockton}, A., {Grosbol}, P., {Moorwood},
  A.~F.~M., \& {Peletier}, R.~F. 1994, A\&A, 288, 365

\bibitem[{{Block} {et~al.}(2004){Block}, {Buta}, {Knapen}, {Elmegreen},
  {Elmegreen}, \& {Puerari}}]{Block+2004}
{Block}, D.~L., {Buta}, R., {Knapen}, J.~H., {Elmegreen}, D.~M., {Elmegreen},
  B.~G., \& {Puerari}, I. 2004, AJ, 128, 183

\bibitem[{{Block} {et~al.}(1996){Block}, {Elmegreen}, \&
  {Wainscoat}}]{Block+1996}
{Block}, D.~L., {Elmegreen}, B.~G., \& {Wainscoat}, R.~J. 1996, \nat, 381, 674

\bibitem[{{Block} \& {Wainscoat}(1991)}]{BlockWainscoat1991}
{Block}, D.~L. \& {Wainscoat}, R.~J. 1991, \nat, 353, 48

\bibitem[{{Bonnell} {et~al.}(2013){Bonnell}, {Dobbs}, \& {Smith}}]{Bonnell2013}
{Bonnell}, I.~A., {Dobbs}, C.~L., \& {Smith}, R.~J. 2013, \mnras, 430, 1790

\bibitem[{{Bottema}(2003)}]{Bottema2003}
{Bottema}, R. 2003, MNRAS, 344, 358

\bibitem[{{Bournaud} \& {Combes}(2002)}]{Combes2002}
{Bournaud}, F. \& {Combes}, F. 2002, \aap, 392, 83

\bibitem[{{Burbidge} \& {Burbidge}(1964)}]{Burbidge1964}
{Burbidge}, E.~M. \& {Burbidge}, G.~R. 1964, \apj, 140, 1445

\bibitem[{{Buta} \& {Crocker}(1991)}]{Buta1991}
{Buta}, R. \& {Crocker}, D.~A. 1991, \aj, 102, 1715

\bibitem[{{Buta} {et~al.}(1992){Buta}, {Crocker}, \& {Byrd}}]{Buta1992}
{Buta}, R., {Crocker}, D.~A., \& {Byrd}, G.~G. 1992, \aj, 103, 1526

\bibitem[{{Buta} {et~al.}(2003){Buta}, {Byrd}, \& {Freeman}}]{Buta2003}
{Buta}, R.~J., {Byrd}, G.~G., \& {Freeman}, T. 2003, \aj, 125, 634

\bibitem[{{Buta} {et~al.}(2007){Buta}, {Corwin}, \& {Odewahn}}]{Buta+2007}
{Buta}, R.~J., {Corwin}, H.~G., \& {Odewahn}, S.~C. 2007, {The de Vaucouleurs
  Altlas of Galaxies} (Cambridge University Press)

\bibitem[{{Byrd} \& {Howard}(1992)}]{Byrd1992}
{Byrd}, G.~G. \& {Howard}, S. 1992, \aj, 103, 1089

\bibitem[{{Byrd} {et~al.}(1989){Byrd}, {Thomasson}, {Donner}, {Sundelius},
  {Huang}, \& {Valtonen}}]{Byrd1989}
{Byrd}, G.~G., {Thomasson}, M., {Donner}, K.~J., {Sundelius}, B., {Huang},
  T.~Y., \& {Valtonen}, M.~J. 1989, Celestial Mechanics, 45, 31

\bibitem[{{Canzian}(1993)}]{Canzian1993}
{Canzian}, B. 1993, \apj, 414, 487

\bibitem[{{Carlberg} \& {Freedman}(1985)}]{CarlbergFreedman1985}
{Carlberg}, R.~G. \& {Freedman}, W.~L. 1985, ApJ, 298, 486

\bibitem[{{Carraro}(2013)}]{Carraro2013}
{Carraro}, G. 2013, ArXiv 1307.0569

\bibitem[{{Cepa} \& {Beckman}(1990)}]{Cepa1990}
{Cepa}, J. \& {Beckman}, J.~E. 1990, \apj, 349, 497

\bibitem[{{Chakrabarti} {et~al.}(2003){Chakrabarti}, {Laughlin}, \&
  {Shu}}]{Chakrabarti2003}
{Chakrabarti}, S., {Laughlin}, G., \& {Shu}, F.~H. 2003, \apj, 596, 220

\bibitem[{{Chandar} {et~al.}(2011){Chandar}, {Whitmore}, {Calzetti}, {Di Nino},
  {Kennicutt}, {Regan}, \& {Schinnerer}}]{Chandar2011}
{Chandar}, R., {Whitmore}, B.~C., {Calzetti}, D., {Di Nino}, D., {Kennicutt},
  R.~C., {Regan}, M., \& {Schinnerer}, E. 2011, \apj, 727, 88

\bibitem[{{Chang} \& {Chakrabarti}(2011)}]{Chang2011}
{Chang}, P. \& {Chakrabarti}, S. 2011, \mnras, 416, 618

\bibitem[{{Churchwell} {et~al.}(2009){Churchwell}, {Babler}, {Meade},
  {Whitney}, {Benjamin}, {Indebetouw}, {Cyganowski}, {Robitaille}, {Povich},
  {Watson}, \& {Bracker}}]{Churchwell2009}
{Churchwell}, E., {Babler}, B.~L., {Meade}, M.~R., {Whitney}, B.~A.,
  {Benjamin}, R., {Indebetouw}, R., {Cyganowski}, C., {Robitaille}, T.~P.,
  {Povich}, M., {Watson}, C., \& {Bracker}, S. 2009, \pasp, 121, 213

\bibitem[{{Colombo} {et~al.}(2014){Colombo}, {Hughes}, {Schinnerer}, {Meidt},
  {Leroy}, {Pety}, {Dobbs}, {Garc{\'{\i}}a-Burillo}, {Dumas}, {Thompson},
  {Schuster}, \& {Kramer}}]{Colombo2014}
{Colombo}, D., {Hughes}, A., {Schinnerer}, E., {Meidt}, S.~E., {Leroy}, A.~K.,
  {Pety}, J., {Dobbs}, C.~L., {Garc{\'{\i}}a-Burillo}, S., {Dumas}, G.,
  {Thompson}, T.~A., {Schuster}, K.~F., \& {Kramer}, C. 2014, \apj, 784, 3

\bibitem[{{Combes} \& {Gerin}(1985)}]{CombesGerin1985}
{Combes}, F. \& {Gerin}, M. 1985, A\&A, 150, 327

\bibitem[{{Considere} \& {Athanassoula}(1988)}]{ConsidereAthanassoula1988}
{Considere}, S. \& {Athanassoula}, E. 1988, A\&As, 76, 365

\bibitem[{{Contopoulos} \& {Grosbol}(1986)}]{Contopoulos1986}
{Contopoulos}, G. \& {Grosbol}, P. 1986, \aap, 155, 11

\bibitem[{{Contopoulos} \& {Grosbol}(1988)}]{Contopoulos1988}
---. 1988, \aap, 197, 83

\bibitem[{{Cowie}(1981)}]{Cowie1981}
{Cowie}, L.~L. 1981, \apj, 245, 66

\bibitem[{{Dame} \& {Thaddeus}(2011)}]{Dame2011}
{Dame}, T.~M. \& {Thaddeus}, P. 2011, \apjl, 734, L24

\bibitem[{{Dawson} {et~al.}(2011){Dawson}, {McClure-Griffiths}, {Kawamura},
  {Mizuno}, {Onishi}, {Mizuno}, \& {Fukui}}]{Dawson2011}
{Dawson}, J.~R., {McClure-Griffiths}, N.~M., {Kawamura}, A., {Mizuno}, N.,
  {Onishi}, T., {Mizuno}, A., \& {Fukui}, Y. 2011, \apj, 728, 127

\bibitem[{{Dawson} {et~al.}(2013){Dawson}, {McClure-Griffiths}, {Wong},
  {Dickey}, {Hughes}, {Fukui}, \& {Kawamura}}]{Dawson2013}
{Dawson}, J.~R., {McClure-Griffiths}, N.~M., {Wong}, T., {Dickey}, J.~M.,
  {Hughes}, A., {Fukui}, Y., \& {Kawamura}, A. 2013, \apj, 763, 56

\bibitem[{{de Vaucouleurs}(1959)}]{deVaucouleurs1959}
{de Vaucouleurs}, G. 1959, Handbuch der Physik, 53, 275

\bibitem[{{Dobbs}(2008)}]{Dobbs2008}
{Dobbs}, C.~L. 2008, MNRAS, 391, 844

\bibitem[{{Dobbs} \& {Bonnell}(2006)}]{DobbsBonnell2006}
{Dobbs}, C.~L. \& {Bonnell}, I.~A. 2006, MNRAS, 367, 873

\bibitem[{{Dobbs} \& {Bonnell}(2008)}]{DobbsBonnell2008}
---. 2008, MNRAS, 385, 1893

\bibitem[{{Dobbs} \& {Burkert}(2012)}]{DobbsBurkert2012}
{Dobbs}, C.~L. \& {Burkert}, A. 2012, \mnras, 421, 2940

\bibitem[{{Dobbs} {et~al.}(2011){Dobbs}, {Burkert}, \& {Pringle}}]{Dobbs2011}
{Dobbs}, C.~L., {Burkert}, A., \& {Pringle}, J.~E. 2011, \mnras, 417, 1318

\bibitem[{{Dobbs} {et~al.}(2008){Dobbs}, {Glover}, {Clark}, \&
  {Klessen}}]{Dobbs+2008}
{Dobbs}, C.~L., {Glover}, S.~C.~O., {Clark}, P.~C., \& {Klessen}, R.~S. 2008,
  \mnras, 389, 1097

\bibitem[{{Dobbs} \& {Pringle}(2010)}]{DobbsPringle2010}
{Dobbs}, C.~L. \& {Pringle}, J.~E. 2010, MNRAS, 409, 396

\bibitem[{{Dobbs} \& {Pringle}(2013)}]{DobbsPringle2013}
---. 2013, \mnras, 432, 653

\bibitem[{{Dobbs} {et~al.}(2014){Dobbs}, {Pringle}, \& {Naylor}}]{Dobbs2014}
{Dobbs}, C.~L., {Pringle}, J.~E., \& {Naylor}, T. 2014, \mnras, 437, L31

\bibitem[{{Dobbs} {et~al.}(2010){Dobbs}, {Theis}, {Pringle}, \&
  {Bate}}]{Dobbs+2010}
{Dobbs}, C.~L., {Theis}, C., {Pringle}, J.~E., \& {Bate}, M.~R. 2010, MNRAS,
  403, 625

\bibitem[{{D'Onghia} {et~al.}(2013){D'Onghia}, {Vogelsberger}, \&
  {Hernquist}}]{D'Onghia+2013}
{D'Onghia}, E., {Vogelsberger}, M., \& {Hernquist}, L. 2013, ApJ, 766, 34

\bibitem[{{Donner} {et~al.}(1991){Donner}, {Engstrom}, \&
  {Sundelius}}]{Donner1991}
{Donner}, K.~J., {Engstrom}, S., \& {Sundelius}, B. 1991, \aap, 252, 571

\bibitem[{{Donner} \& {Thomasson}(1994)}]{DonnerThomasson1994}
{Donner}, K.~J. \& {Thomasson}, M. 1994, A\&A, 290, 785

\bibitem[{{Dopita} {et~al.}(1985){Dopita}, {Mathewson}, \& {Ford}}]{Dopita1985}
{Dopita}, M.~A., {Mathewson}, D.~S., \& {Ford}, V.~L. 1985, \apj, 297, 599

\bibitem[{{Douglas} {et~al.}(2010){Douglas}, {Acreman}, {Dobbs}, \&
  {Brunt}}]{Douglas2010}
{Douglas}, K.~A., {Acreman}, D.~M., {Dobbs}, C.~L., \& {Brunt}, C.~M. 2010,
  \mnras, 407, 405

\bibitem[{{Drimmel}(2000)}]{Drimmel2000}
{Drimmel}, R. 2000, \aap, 358, L13

\bibitem[{{Dubinski} {et~al.}(2009){Dubinski}, {Berentzen}, \&
  {Shlosman}}]{Dubinski+2009}
{Dubinski}, J., {Berentzen}, I., \& {Shlosman}, I. 2009, ApJ, 697, 293

\bibitem[{{Dubinski} {et~al.}(2008){Dubinski}, {Gauthier}, {Widrow}, \&
  {Nickerson}}]{Dubinski2008}
{Dubinski}, J., {Gauthier}, J.-R., {Widrow}, L., \& {Nickerson}, S. 2008, in
  Astronomical Society of the Pacific Conference Series, Vol. 396, Formation
  and Evolution of Galaxy Disks, ed. J.~G. {Funes} \& E.~M. {Corsini}, 321

\bibitem[{{Durbala} {et~al.}(2009){Durbala}, {Buta}, {Sulentic}, \&
  {Verdes-Montenegro}}]{Durbala+2009}
{Durbala}, A., {Buta}, R., {Sulentic}, J.~W., \& {Verdes-Montenegro}, L. 2009,
  MNRAS, 397, 1756

\bibitem[{{Dwarkadas} \& {Balbus}(1996)}]{Dwarkadus1996}
{Dwarkadas}, V.~V. \& {Balbus}, S.~A. 1996, \apj, 467, 87

\bibitem[{{Earn} \& {Sellwood}(1995)}]{EarnSellwood1995}
{Earn}, D.~J.~D. \& {Sellwood}, J.~A. 1995, ApJ, 451, 533

\bibitem[{{Eden} {et~al.}(2012){Eden}, {Moore}, {Plume}, \&
  {Morgan}}]{Eden2012}
{Eden}, D.~J., {Moore}, T.~J.~T., {Plume}, R., \& {Morgan}, L.~K. 2012, \mnras,
  422, 3178

\bibitem[{{Efremov} \& {Ivanov}(1982)}]{Efremov1982}
{Efremov}, I.~N. \& {Ivanov}, G.~R. 1982, \apss, 86, 117

\bibitem[{{Efremov}(2011)}]{Efremov2011}
{Efremov}, Y.~N. 2011, Astronomy Reports, 55, 108

\bibitem[{{Egusa} {et~al.}(2009){Egusa}, {Kohno}, {Sofue}, {Nakanishi}, \&
  {Komugi}}]{Egusa+2009}
{Egusa}, F., {Kohno}, K., {Sofue}, Y., {Nakanishi}, H., \& {Komugi}, S. 2009,
  ApJ, 697, 1870

\bibitem[{{Egusa} {et~al.}(2004){Egusa}, {Sofue}, \& {Nakanishi}}]{Egusa2004}
{Egusa}, F., {Sofue}, Y., \& {Nakanishi}, H. 2004, \pasj, 56, L45

\bibitem[{{Elmegreen}(1979)}]{Elmegreen1979}
{Elmegreen}, B.~G. 1979, \apj, 231, 372

\bibitem[{{Elmegreen}(1982)}]{Elmegreen1982}
---. 1982, \apj, 253, 655

\bibitem[{{Elmegreen}(1989)}]{Elmegreen1989}
---. 1989, \apj, 344, 306

\bibitem[{{Elmegreen}(1990)}]{Elmegreen1990}
---. 1990, Annals of the New York Academy of Sciences, 596, 40

\bibitem[{{Elmegreen}(1994)}]{Elmegreen1994}
---. 1994, \apj, 433, 39

\bibitem[{{Elmegreen}(1995)}]{Elmegreen1995}
{Elmegreen}, B.~G. 1995, in Astronomical Society of the Pacific Conference
  Series, Vol.~80, The Physics of the Interstellar Medium and Intergalactic
  Medium, ed. A.~{Ferrara}, C.~F. {McKee}, C.~{Heiles}, \& P.~R. {Shapiro}, 218

\bibitem[{{Elmegreen}(2007)}]{Elmegreen2007}
---. 2007, \apj, 668, 1064

\bibitem[{{Elmegreen}(2011)}]{Elmegreen2011}
---. 2011, ApJ, 737, 10

\bibitem[{{Elmegreen} \&
  {Elmegreen}(1983{\natexlab{a}})}]{ElmegreenElmegreen1983}
{Elmegreen}, B.~G. \& {Elmegreen}, D.~M. 1983{\natexlab{a}}, \apj, 267, 31

\bibitem[{{Elmegreen} \& {Elmegreen}(1983{\natexlab{b}})}]{Elmegreen1983}
---. 1983{\natexlab{b}}, \mnras, 203, 31

\bibitem[{{Elmegreen} \& {Elmegreen}(1986)}]{Elmegreen1986}
---. 1986, \apj, 311, 554

\bibitem[{{Elmegreen} \& {Elmegreen}(1990)}]{ElmegreenElmegreen1990}
---. 1990, \apj, 355, 52

\bibitem[{{Elmegreen} {et~al.}(2003){Elmegreen}, {Elmegreen}, \&
  {Leitner}}]{Elmegreen+2003a}
{Elmegreen}, B.~G., {Elmegreen}, D.~M., \& {Leitner}, S.~N. 2003, ApJ, 590, 271

\bibitem[{{Elmegreen} {et~al.}(1989){Elmegreen}, {Seiden}, \&
  {Elmegreen}}]{Elmegreen+1989}
{Elmegreen}, B.~G., {Seiden}, P.~E., \& {Elmegreen}, D.~M. 1989, \apj, 343, 602

\bibitem[{{Elmegreen} \& {Thomasson}(1993)}]{ElmegreenThomasson1993}
{Elmegreen}, B.~G. \& {Thomasson}, M. 1993, A\&A, 272, 37

\bibitem[{{Elmegreen} {et~al.}(1998){Elmegreen}, {Wilcots}, \&
  {Pisano}}]{Elmegreen1998}
{Elmegreen}, B.~G., {Wilcots}, E., \& {Pisano}, D.~J. 1998, \apjl, 494, L37

\bibitem[{{Elmegreen} {et~al.}(1999){Elmegreen}, {Chromey}, {Bissell}, \&
  {Corrado}}]{Elmegreen1999}
{Elmegreen}, D.~M., {Chromey}, F.~R., {Bissell}, B.~A., \& {Corrado}, K. 1999,
  \aj, 118, 2618

\bibitem[{{Elmegreen} \& {Elmegreen}(1982)}]{ElmegreenElmegreen1982}
{Elmegreen}, D.~M. \& {Elmegreen}, B.~G. 1982, MNRAS, 201, 1021

\bibitem[{{Elmegreen} \& {Elmegreen}(1987)}]{Elmegreen1987}
---. 1987, \apj, 314, 3

\bibitem[{{Elmegreen} {et~al.}(2011){Elmegreen}, {Elmegreen}, {Yau},
  {Athanassoula}, {Bosma}, {Buta}, {Helou}, {Ho}, {Gadotti}, {Knapen},
  {Laurikainen}, {Madore}, {Masters}, {Meidt}, {Men{\'e}ndez-Delmestre},
  {Regan}, {Salo}, {Sheth}, {Zaritsky}, {Aravena}, {Skibba}, {Hinz}, {Laine},
  {Gil de Paz}, {Mu{\~n}oz-Mateos}, {Seibert}, {Mizusawa}, {Kim}, \& {Erroz
  Ferrer}}]{Elmegreen+2011}
{Elmegreen}, D.~M., {Elmegreen}, B.~G., {Yau}, A., {Athanassoula}, E., {Bosma},
  A., {Buta}, R.~J., {Helou}, G., {Ho}, L.~C., {Gadotti}, D.~A., {Knapen},
  J.~H., {Laurikainen}, E., {Madore}, B.~F., {Masters}, K.~L., {Meidt}, S.~E.,
  {Men{\'e}ndez-Delmestre}, K., {Regan}, M.~W., {Salo}, H., {Sheth}, K.,
  {Zaritsky}, D., {Aravena}, M., {Skibba}, R., {Hinz}, J.~L., {Laine}, J., {Gil
  de Paz}, A., {Mu{\~n}oz-Mateos}, J.-C., {Seibert}, M., {Mizusawa}, T., {Kim},
  T., \& {Erroz Ferrer}, S. 2011, ApJ, 737, 32

\bibitem[{{Eneev} {et~al.}(1973){Eneev}, {Kozlov}, \& {Sunyaev}}]{Eneev1973}
{Eneev}, T.~M., {Kozlov}, N.~N., \& {Sunyaev}, R.~A. 1973, \aap, 22, 41

\bibitem[{{Englmaier} \& {Gerhard}(1999)}]{EnglmaierGerhard1999}
{Englmaier}, P. \& {Gerhard}, O. 1999, MNRAS, 304, 512

\bibitem[{{Englmaier} \& {Shlosman}(2000)}]{Englmaier2000}
{Englmaier}, P. \& {Shlosman}, I. 2000, \apj, 528, 677

\bibitem[{{Feitzinger} {et~al.}(1981){Feitzinger}, {Glassgold}, {Gerola}, \&
  {Seiden}}]{Feitzinger1981}
{Feitzinger}, J.~V., {Glassgold}, A.~E., {Gerola}, H., \& {Seiden}, P.~E. 1981,
  \aap, 98, 371

\bibitem[{{Ferreras} {et~al.}(2012){Ferreras}, {Cropper}, {Kawata}, {Page}, \&
  {Hoversten}}]{Ferreras+2012}
{Ferreras}, I., {Cropper}, M., {Kawata}, D., {Page}, M., \& {Hoversten}, E.~A.
  2012, MNRAS, 424, 1636

\bibitem[{{Font} {et~al.}(2011){Font}, {Beckman}, {Epinat}, {Fathi},
  {Guti{\'e}rrez}, \& {Hernandez}}]{Font+2011}
{Font}, J., {Beckman}, J.~E., {Epinat}, B., {Fathi}, K., {Guti{\'e}rrez}, L.,
  \& {Hernandez}, O. 2011, \apjl, 741, L14

\bibitem[{{Foyle} {et~al.}(2011){Foyle}, {Rix}, {Dobbs}, {Leroy}, \&
  {Walter}}]{Foyle+2011}
{Foyle}, K., {Rix}, H.-W., {Dobbs}, C.~L., {Leroy}, A.~K., \& {Walter}, F.
  2011, ApJ, 735, 101

\bibitem[{{Fuchs}(2001)}]{Fuchs2001a}
{Fuchs}, B. 2001, A\&A, 368, 107

\bibitem[{{Fuchs} {et~al.}(2005){Fuchs}, {Dettbarn}, \&
  {Tsuchiya}}]{Fuchs+2005}
{Fuchs}, B., {Dettbarn}, C., \& {Tsuchiya}, T. 2005, A\&A, 444, 1

\bibitem[{{Fuchs} \& {M{\"o}llenhoff}(1999)}]{FuchsMollenhoff1999}
{Fuchs}, B. \& {M{\"o}llenhoff}, C. 1999, A\&A, 352, L36

\bibitem[{{Fujii} {et~al.}(2011){Fujii}, {Baba}, {Saitoh}, {Makino}, {Kokubo},
  \& {Wada}}]{Fujii+2011}
{Fujii}, M.~S., {Baba}, J., {Saitoh}, T.~R., {Makino}, J., {Kokubo}, E., \&
  {Wada}, K. 2011, ApJ, 730, 109

\bibitem[{{Fujimoto}(1968)}]{Fujimoto1968}
{Fujimoto}, M. 1968, in IAU Symposium, Vol.~29, IAU Symposium, 453--+

\bibitem[{{Fukui} {et~al.}(2006){Fukui}, {Yamamoto}, {Fujishita}, {Kudo},
  {Torii}, {Nozawa}, {Takahashi}, {Matsumoto}, {Machida}, {Kawamura},
  {Yonekura}, {Mizuno}, {Onishi}, \& {Mizuno}}]{Fukui2006}
{Fukui}, Y., {Yamamoto}, H., {Fujishita}, M., {Kudo}, N., {Torii}, K.,
  {Nozawa}, S., {Takahashi}, K., {Matsumoto}, R., {Machida}, M., {Kawamura},
  A., {Yonekura}, Y., {Mizuno}, N., {Onishi}, T., \& {Mizuno}, A. 2006,
  Science, 314, 106

\bibitem[{{Fux}(1997)}]{Fux1997}
{Fux}, R. 1997, A\&A, 327, 983

\bibitem[{{Fux}(1999)}]{Fux1999}
---. 1999, \aap, 345, 787

\bibitem[{{Garcia Gomez} \& {Athanassoula}(1993)}]{Garcia1993}
{Garcia Gomez}, C. \& {Athanassoula}, E. 1993, \aaps, 100, 431

\bibitem[{{Gerber} \& {Lamb}(1994)}]{Gerber1994}
{Gerber}, R.~A. \& {Lamb}, S.~A. 1994, \apj, 431, 604

\bibitem[{{Gerola} \& {Seiden}(1978)}]{Gerola1978}
{Gerola}, H. \& {Seiden}, P.~E. 1978, \apj, 223, 129

\bibitem[{{Gittins} \& {Clarke}(2004)}]{Gittins2004}
{Gittins}, D.~M. \& {Clarke}, C.~J. 2004, \mnras, 349, 909

\bibitem[{{Goldreich} \&
  {Lynden-Bell}(1965{\natexlab{a}})}]{GoldreichLynden-Bell1965a}
{Goldreich}, P. \& {Lynden-Bell}, D. 1965{\natexlab{a}}, MNRAS, 130, 97

\bibitem[{{Goldreich} \&
  {Lynden-Bell}(1965{\natexlab{b}})}]{GoldreichLynden-Bell1965}
---. 1965{\natexlab{b}}, MNRAS, 130, 125

\bibitem[{{Goldreich} \& {Tremaine}(1978)}]{GoldreichTremaine1978}
{Goldreich}, P. \& {Tremaine}, S. 1978, ApJ, 222, 850

\bibitem[{{Goldreich} \& {Tremaine}(1979)}]{GoldreichTremaine1979}
---. 1979, ApJ, 233, 857

\bibitem[{{Gonzalez} \& {Graham}(1996)}]{Gonzalez1996}
{Gonzalez}, R.~A. \& {Graham}, J.~R. 1996, \apj, 460, 651

\bibitem[{{Gordon}(1978)}]{Gordon1978}
{Gordon}, M.~A. 1978, \apj, 222, 100

\bibitem[{{Gouda}(2012)}]{Gouda2012}
{Gouda}, N. 2012, in Astronomical Society of the Pacific Conference Series,
  Vol. 458, Galactic Archaeology: Near-Field Cosmology and the Formation of the
  Milky Way, ed. W.~{Aoki}, M.~{Ishigaki}, T.~{Suda}, T.~{Tsujimoto}, \&
  N.~{Arimoto}, 417

\bibitem[{{Grand} {et~al.}(2012{\natexlab{a}}){Grand}, {Kawata}, \&
  {Cropper}}]{Grand+2012b}
{Grand}, R.~J.~J., {Kawata}, D., \& {Cropper}, M. 2012{\natexlab{a}}, MNRAS,
  426, 167

\bibitem[{{Grand} {et~al.}(2012{\natexlab{b}}){Grand}, {Kawata}, \&
  {Cropper}}]{Grand+2012a}
---. 2012{\natexlab{b}}, MNRAS, 421, 1529

\bibitem[{{Grand} {et~al.}(2013){Grand}, {Kawata}, \& {Cropper}}]{Grand+2013}
---. 2013, A\&A, 553, A77

\bibitem[{{Grand} {et~al.}(2014){Grand}, {Kawata}, \& {Cropper}}]{Grand+2014}
---. 2014, \mnras, 439, 623

\bibitem[{{Grosb{\o}l}(1994)}]{Grosbol1994}
{Grosb{\o}l}, P. 1994, in Lecture Notes in Physics, Berlin Springer Verlag,
  Vol. 433, Galactic Dynamics and N-Body Simulations, ed. G.~{Contopoulos},
  N.~K. {Spyrou}, \& L.~{Vlahos}, 101--141

\bibitem[{{Grosb{\o}l} {et~al.}(2004){Grosb{\o}l}, {Patsis}, \&
  {Pompei}}]{Grosbol+2004}
{Grosb{\o}l}, P., {Patsis}, P.~A., \& {Pompei}, E. 2004, A\&A, 423, 849

\bibitem[{{Grosbol} \& {Patsis}(1998)}]{GrosbolPatsis1998}
{Grosbol}, P.~J. \& {Patsis}, P.~A. 1998, A\&A, 336, 840

\bibitem[{{Harsoula} {et~al.}(2011){Harsoula}, {Kalapotharakos}, \&
  {Contopoulos}}]{Harsoula+2011}
{Harsoula}, M., {Kalapotharakos}, C., \& {Contopoulos}, G. 2011, \mnras, 411,
  1111

\bibitem[{{Hernquist}(1990)}]{Hernquist1990b}
{Hernquist}, L. 1990, {Dynamical status of M51.}, ed. R.~{Wielen}, 108--117

\bibitem[{{Hodge} \& {Merchant}(1966)}]{Hodge1966}
{Hodge}, P.~W. \& {Merchant}, A.~E. 1966, \apj, 144, 875

\bibitem[{{Hohl}(1971)}]{Hohl1971}
{Hohl}, F. 1971, ApJ, 168, 343

\bibitem[{{Hohl}(1976)}]{Hohl1976}
---. 1976, \aj, 81, 30

\bibitem[{{Holmberg}(1941)}]{Holmberg1941}
{Holmberg}, E. 1941, \apj, 94, 385

\bibitem[{{Honma}(2013)}]{Honma2013}
{Honma}, M. 2013, in Astronomical Society of the Pacific Conference Series,
  Vol. 476, Astronomical Society of the Pacific Conference Series, ed.
  R.~{Kawabe}, N.~{Kuno}, \& S.~{Yamamoto}, 81

\bibitem[{{Hopkins} {et~al.}(2011){Hopkins}, {Quataert}, \&
  {Murray}}]{Hopkins2011}
{Hopkins}, P.~F., {Quataert}, E., \& {Murray}, N. 2011, \mnras, 417, 950

\bibitem[{{Howard} {et~al.}(1993){Howard}, {Keel}, {Byrd}, \&
  {Burkey}}]{Howard1993}
{Howard}, S., {Keel}, W.~C., {Byrd}, G., \& {Burkey}, J. 1993, \apj, 417, 502

\bibitem[{{Hubble}(1943)}]{Hubble1943}
{Hubble}, E. 1943, \apj, 97, 112

\bibitem[{{Hubble}(1926{\natexlab{a}})}]{Hubble1926b}
{Hubble}, E.~P. 1926{\natexlab{a}}, \apj, 63, 236

\bibitem[{{Hubble}(1926{\natexlab{b}})}]{Hubble1926}
---. 1926{\natexlab{b}}, \apj, 64, 321

\bibitem[{{Hubble}(1929)}]{Hubble1929}
---. 1929, \apj, 69, 103

\bibitem[{{Hunter}(1965)}]{Hunter1965}
{Hunter}, C. 1965, MNRAS, 129, 321

\bibitem[{{Hunter} \& {Gallagher}(1985)}]{Hunter1985}
{Hunter}, D.~A. \& {Gallagher}, III, J.~S. 1985, \apjs, 58, 533

\bibitem[{{Iye}(1978)}]{Iye1978}
{Iye}, M. 1978, PASJ, 30, 223

\bibitem[{{Iye} {et~al.}(1983){Iye}, {Aoki}, {Ueda}, \& {Noguchi}}]{Iye+1983}
{Iye}, M., {Aoki}, S., {Ueda}, T., \& {Noguchi}, M. 1983, Ap\&SS, 89, 363

\bibitem[{{Jalali} \& {Hunter}(2005)}]{JalaliHunter2005}
{Jalali}, M.~A. \& {Hunter}, C. 2005, ApJ, 630, 804

\bibitem[{{Jog} \& {Solomon}(1984)}]{JogSolomon1984}
{Jog}, C.~J. \& {Solomon}, P.~M. 1984, ApJ, 276, 114

\bibitem[{{Julian} \& {Toomre}(1966)}]{JulianToomre1966}
{Julian}, W.~H. \& {Toomre}, A. 1966, ApJ, 146, 810

\bibitem[{{Jungwiert} \& {Palous}(1994)}]{Jungwiert1994}
{Jungwiert}, B. \& {Palous}, J. 1994, \aap, 287, 55

\bibitem[{{Kaleida} \& {Scowen}(2010)}]{Kaleida2010}
{Kaleida}, C. \& {Scowen}, P.~A. 2010, \aj, 140, 379

\bibitem[{{Kalnajs}(1965)}]{Kalnajs1965}
{Kalnajs}, A.~J. 1965, PhD thesis, HARVARD UNIVERSITY.

\bibitem[{{Kalnajs}(1971)}]{Kalnajs1971}
---. 1971, ApJ, 166, 275

\bibitem[{{Kalnajs}(1972)}]{Kalnajs1972}
---. 1972, Astrophysical Letters, 11, 41

\bibitem[{{Kalnajs}(1973)}]{Kalnajs1973}
---. 1973, Proceedings of the Astronomical Society of Australia, 2, 174

\bibitem[{{Kalnajs} \& {Athanassoula-Georgala}(1974)}]{Kalnajs1974}
{Kalnajs}, A.~J. \& {Athanassoula-Georgala}, E. 1974, \mnras, 168, 287

\bibitem[{{Kamaya}(1998)}]{Kamaya1998}
{Kamaya}, H. 1998, \aj, 116, 1719

\bibitem[{{Kazantzidis} {et~al.}(2008){Kazantzidis}, {Bullock}, {Zentner},
  {Kravtsov}, \& {Moustakas}}]{Kazantzidis2008}
{Kazantzidis}, S., {Bullock}, J.~S., {Zentner}, A.~R., {Kravtsov}, A.~V., \&
  {Moustakas}, L.~A. 2008, \apj, 688, 254

\bibitem[{{Kendall} {et~al.}(2011){Kendall}, {Kennicutt}, \&
  {Clarke}}]{Kendall+2011}
{Kendall}, S., {Kennicutt}, R.~C., \& {Clarke}, C. 2011, MNRAS, 414, 538

\bibitem[{{Kennicutt} \& {Hodge}(1982)}]{KennicuttHodge1982}
{Kennicutt}, Jr., R. \& {Hodge}, P. 1982, ApJ, 253, 101

\bibitem[{{Kennicutt}(1981)}]{Kennicutt1981}
{Kennicutt}, Jr., R.~C. 1981, \aj, 86, 1847

\bibitem[{{Kennicutt}(1998)}]{Kennicutt1998}
---. 1998, \araa, 36, 189

\bibitem[{{Khoperskov} {et~al.}(2013){Khoperskov}, {Khoperskov}, {Zasov},
  {Bizyaev}, \& {Khrapov}}]{Khoperskov2013}
{Khoperskov}, A.~V., {Khoperskov}, S.~A., {Zasov}, A.~V., {Bizyaev}, D.~V., \&
  {Khrapov}, S.~S. 2013, \mnras, 431, 1230

\bibitem[{{Kim} {et~al.}(2001){Kim}, {Ryu}, \& {Jones}}]{Kim2001}
{Kim}, J., {Ryu}, D., \& {Jones}, T.~W. 2001, \apj, 557, 464

\bibitem[{{Kim} {et~al.}(2014){Kim}, {Kim}, \& {Kim}}]{Kim+2014}
{Kim}, W.-T., {Kim}, Y., \& {Kim}, J.-G. 2014, ArXiv e-prints

\bibitem[{{Kim} \& {Ostriker}(2006)}]{KimOstriker2006}
{Kim}, W.-T. \& {Ostriker}, E.~C. 2006, \apj, 646, 213

\bibitem[{{Kim} {et~al.}(2002){Kim}, {Ostriker}, \& {Stone}}]{Kim2002}
{Kim}, W.-T., {Ostriker}, E.~C., \& {Stone}, J.~M. 2002, \apj, 581, 1080

\bibitem[{{Knapen} {et~al.}(1992){Knapen}, {Beckman}, {Cepa}, {van der Hulst},
  \& {Rand}}]{Knapen1992}
{Knapen}, J.~H., {Beckman}, J.~E., {Cepa}, J., {van der Hulst}, T., \& {Rand},
  R.~J. 1992, \apjl, 385, L37

\bibitem[{{Kormendy} \& {Norman}(1979)}]{KormendyNorman1979}
{Kormendy}, J. \& {Norman}, C.~A. 1979, ApJ, 233, 539

\bibitem[{{Kwan} \& {Valdes}(1987)}]{Kwan1987}
{Kwan}, J. \& {Valdes}, F. 1987, \apj, 315, 92

\bibitem[{{La Vigne} {et~al.}(2006){La Vigne}, {Vogel}, \&
  {Ostriker}}]{LaVigne2006}
{La Vigne}, M.~A., {Vogel}, S.~N., \& {Ostriker}, E.~C. 2006, \apj, 650, 818

\bibitem[{{Lau} \& {Bertin}(1978)}]{LauBertin1978}
{Lau}, Y.~Y. \& {Bertin}, G. 1978, ApJ, 226, 508

\bibitem[{{Lau} {et~al.}(1976){Lau}, {Lin}, \& {Mark}}]{Lau+1976}
{Lau}, Y.~Y., {Lin}, C.~C., \& {Mark}, J.~W.-K. 1976, Proceedings of the
  National Academy of Science, 73, 1379

\bibitem[{{Lee} \& {Shu}(2012)}]{Lee2012}
{Lee}, W.-K. \& {Shu}, F.~H. 2012, \apj, 756, 45

\bibitem[{{Li} {et~al.}(2005){Li}, {Mac Low}, \& {Klessen}}]{Li2005}
{Li}, Y., {Mac Low}, M.-M., \& {Klessen}, R.~S. 2005, \apj, 626, 823

\bibitem[{{Lin} \& {Bertin}(1985)}]{LinBertin1985}
{Lin}, C.~C. \& {Bertin}, G. 1985, in IAU Symposium, Vol. 106, The Milky Way
  Galaxy, ed. {H.~van Woerden, R.~J.~Allen, \& W.~B.~Burton}, 513--530

\bibitem[{{Lin} \& {Shu}(1964)}]{LinShu1964}
{Lin}, C.~C. \& {Shu}, F.~H. 1964, ApJ, 140, 646

\bibitem[{{Lin} \& {Shu}(1966)}]{LinShu1966}
---. 1966, Proceedings of the National Academy of Science, 55, 229

\bibitem[{{Lin} \& {Shu}(1967)}]{LinShu1967}
{Lin}, C.~C. \& {Shu}, F.~H. 1967, in IAU Symposium, Vol.~31, Radio Astronomy
  and the Galactic System, ed. H.~{van Woerden}, 313

\bibitem[{{Lin} \& {Thurstans}(1984)}]{LinThurstans1984}
{Lin}, C.~C. \& {Thurstans}, R.~P. 1984, in ESA Special Publication, Vol. 207,
  Plasma Astrophysics, ed. T.~D. {Guyenne} \& J.~J. {Hunt}, 121--130

\bibitem[{{Lindblad}(1927)}]{Lindblad1927}
{Lindblad}, B. 1927, \mnras, 87, 420

\bibitem[{{Lindblad}(1935)}]{Lindblad1935}
---. 1935, \mnras, 95, 663

\bibitem[{{Lindblad}(1940)}]{Lindblad1940}
---. 1940, \apj, 92, 1

\bibitem[{{Lindblad}(1963)}]{Lindblad1963}
---. 1963, Stockholms Observatoriums Annaler, 22, 5

\bibitem[{{Lindblad}(1960)}]{Lindblad1960}
{Lindblad}, P.~O. 1960, Stockholms Observatoriums Annaler, 21, 4

\bibitem[{{Lindblad} \& {Lindblad}(1994)}]{LindbladLindblad1994}
{Lindblad}, P.~O. \& {Lindblad}, P.~A.~B. 1994, in Astronomical Society of the
  Pacific Conference Series, Vol.~66, Physics of the Gaseous and Stellar Disks
  of the Galaxy, ed. I.~R. {King}, 29

\bibitem[{{Lintott} {et~al.}(2011){Lintott}, {Schawinski}, {Bamford}, {Slosar},
  {Land}, {Thomas}, {Edmondson}, {Masters}, {Nichol}, {Raddick}, {Szalay},
  {Andreescu}, {Murray}, \& {Vandenberg}}]{Lintott2011}
{Lintott}, C., {Schawinski}, K., {Bamford}, S., {Slosar}, A., {Land}, K.,
  {Thomas}, D., {Edmondson}, E., {Masters}, K., {Nichol}, R.~C., {Raddick},
  M.~J., {Szalay}, A., {Andreescu}, D., {Murray}, P., \& {Vandenberg}, J. 2011,
  \mnras, 410, 166

\bibitem[{{Lubow} {et~al.}(1986){Lubow}, {Cowie}, \& {Balbus}}]{Lubow1986}
{Lubow}, S.~H., {Cowie}, L.~L., \& {Balbus}, S.~A. 1986, \apj, 309, 496

\bibitem[{{Lynden-Bell} \& {Kalnajs}(1972)}]{Lynden-BellKalnajs1972}
{Lynden-Bell}, D. \& {Kalnajs}, A.~J. 1972, MNRAS, 157, 1

\bibitem[{{Lynden-Bell} \& {Ostriker}(1967)}]{Lynden-BellOstriker1967}
{Lynden-Bell}, D. \& {Ostriker}, J.~P. 1967, MNRAS, 136, 293

\bibitem[{{Lynds}(1970)}]{Lynds1970}
{Lynds}, B.~T. 1970, in IAU Symposium, Vol.~38, The Spiral Structure of our
  Galaxy, ed. W.~{Becker} \& G.~I. {Kontopoulos}, 26

\bibitem[{{Ma}(2002)}]{Ma2002}
{Ma}, J. 2002, \aap, 388, 389

\bibitem[{{Mark}(1974)}]{Mark1974}
{Mark}, J.~W.-K. 1974, ApJ, 193, 539

\bibitem[{{Mark}(1976)}]{Mark1976b}
{Mark}, J.~W.~K. 1976, ApJ, 205, 363

\bibitem[{{Mart{\'{\i}}nez-Garc{\'{\i}}a}(2012)}]{Martinez-Garcia2012}
{Mart{\'{\i}}nez-Garc{\'{\i}}a}, E.~E. 2012, \apj, 744, 92

\bibitem[{{Mart{\'{\i}}nez-Garc{\'{\i}}a}
  {et~al.}(2009){Mart{\'{\i}}nez-Garc{\'{\i}}a}, {Gonz{\'a}lez-L{\'o}pezlira},
  \& {Bruzual-A}}]{Martinez2009}
{Mart{\'{\i}}nez-Garc{\'{\i}}a}, E.~E., {Gonz{\'a}lez-L{\'o}pezlira}, R.~A., \&
  {Bruzual-A}, G. 2009, \apj, 694, 512

\bibitem[{{Martinez-Valpuesta} \&
  {Gerhard}(2011)}]{Martinez-ValpuestaGerhard2011}
{Martinez-Valpuesta}, I. \& {Gerhard}, O. 2011, ApJL, 734, L20

\bibitem[{{Martos} {et~al.}(2004){Martos}, {Hernandez}, {Y{\'a}{\~n}ez},
  {Moreno}, \& {Pichardo}}]{Martos2004}
{Martos}, M., {Hernandez}, X., {Y{\'a}{\~n}ez}, M., {Moreno}, E., \&
  {Pichardo}, B. 2004, \mnras, 350, L47

\bibitem[{{Masset} \& {Tagger}(1997)}]{MassetTagger1997}
{Masset}, F. \& {Tagger}, M. 1997, A\&A, 322, 442

\bibitem[{{Masters} {et~al.}(2010){Masters}, {Mosleh}, {Romer}, {Nichol},
  {Bamford}, {Schawinski}, {Lintott}, {Andreescu}, {Campbell}, {Crowcroft},
  {Doyle}, {Edmondson}, {Murray}, {Raddick}, {Slosar}, {Szalay}, \&
  {Vandenberg}}]{Masters2010}
{Masters}, K.~L., {Mosleh}, M., {Romer}, A.~K., {Nichol}, R.~C., {Bamford},
  S.~P., {Schawinski}, K., {Lintott}, C.~J., {Andreescu}, D., {Campbell},
  H.~C., {Crowcroft}, B., {Doyle}, I., {Edmondson}, E.~M., {Murray}, P.,
  {Raddick}, M.~J., {Slosar}, A., {Szalay}, A.~S., \& {Vandenberg}, J. 2010,
  \mnras, 405, 783

\bibitem[{{McCray} \& {Kafatos}(1987)}]{McCray1987}
{McCray}, R. \& {Kafatos}, M. 1987, \apj, 317, 190

\bibitem[{{Meidt} {et~al.}(2009){Meidt}, {Rand}, \& {Merrifield}}]{Meidt+2009}
{Meidt}, S.~E., {Rand}, R.~J., \& {Merrifield}, M.~R. 2009, ApJ, 702, 277

\bibitem[{{Meidt} {et~al.}(2008{\natexlab{a}}){Meidt}, {Rand}, {Merrifield},
  {Debattista}, \& {Shen}}]{Meidt2008b}
{Meidt}, S.~E., {Rand}, R.~J., {Merrifield}, M.~R., {Debattista}, V.~P., \&
  {Shen}, J. 2008{\natexlab{a}}, \apj, 676, 899

\bibitem[{{Meidt} {et~al.}(2008{\natexlab{b}}){Meidt}, {Rand}, {Merrifield},
  {Shetty}, \& {Vogel}}]{Meidt+2008}
{Meidt}, S.~E., {Rand}, R.~J., {Merrifield}, M.~R., {Shetty}, R., \& {Vogel},
  S.~N. 2008{\natexlab{b}}, ApJ, 688, 224

\bibitem[{{Meidt} {et~al.}(2013){Meidt}, {Schinnerer}, {Garcia-Burillo},
  {Hughes}, {Colombo}, {Pety}, {Dobbs}, {Schuster}, {Kramer}, {Leroy}, {Dumas},
  \& {Thompson}}]{Meidt2013}
{Meidt}, S.~E., {Schinnerer}, E., {Garcia-Burillo}, S., {Hughes}, A.,
  {Colombo}, D., {Pety}, J., {Dobbs}, C.~L., {Schuster}, K.~F., {Kramer}, C.,
  {Leroy}, A.~K., {Dumas}, G., \& {Thompson}, T.~A. 2013, ArXiv e-prints

\bibitem[{{Merrifield} {et~al.}(2006){Merrifield}, {Rand}, \&
  {Meidt}}]{Merrifield+2006}
{Merrifield}, M.~R., {Rand}, R.~J., \& {Meidt}, S.~E. 2006, MNRAS, 366, L17

\bibitem[{{Miller} {et~al.}(1970){Miller}, {Prendergast}, \&
  {Quirk}}]{Miller1970}
{Miller}, R.~H., {Prendergast}, K.~H., \& {Quirk}, W.~J. 1970, \apj, 161, 903

\bibitem[{{Minchev} {et~al.}(2012){Minchev}, {Famaey}, {Quillen}, {Di Matteo},
  {Combes}, {Vlaji{\'c}}, {Erwin}, \& {Bland-Hawthorn}}]{Minchev+2012}
{Minchev}, I., {Famaey}, B., {Quillen}, A.~C., {Di Matteo}, P., {Combes}, F.,
  {Vlaji{\'c}}, M., {Erwin}, P., \& {Bland-Hawthorn}, J. 2012, A\&A, 548, A126

\bibitem[{{Mishurov} \& {Suchkov}(1975)}]{Mishurov1975}
{Mishurov}, I.~N. \& {Suchkov}, A.~A. 1975, \apss, 35, 285

\bibitem[{{Mo} {et~al.}(2010){Mo}, {van den Bosch}, \& {White}}]{Mo+2010}
{Mo}, H., {van den Bosch}, F.~C., \& {White}, S. 2010, {Galaxy Formation and
  Evolution}

\bibitem[{{Mouschovias} {et~al.}(2009){Mouschovias}, {Kunz}, \&
  {Christie}}]{Mouschovias2009}
{Mouschovias}, T.~C., {Kunz}, M.~W., \& {Christie}, D.~A. 2009, \mnras, 397, 14

\bibitem[{{Mouschovias} {et~al.}(1974){Mouschovias}, {Shu}, \&
  {Woodward}}]{Mouschovias1974}
{Mouschovias}, T.~C., {Shu}, F.~H., \& {Woodward}, P.~R. 1974, \aap, 33, 73

\bibitem[{{Mueller} \& {Arnett}(1976)}]{Mueller1976}
{Mueller}, M.~W. \& {Arnett}, W.~D. 1976, \apj, 210, 670

\bibitem[{{Nelson} \& {Matsuda}(1977)}]{Nelson1977}
{Nelson}, A.~H. \& {Matsuda}, T. 1977, \mnras, 179, 663

\bibitem[{{Nomura} \& {Kamaya}(2001)}]{Nomura2001}
{Nomura}, H. \& {Kamaya}, H. 2001, \aj, 121, 1024

\bibitem[{{Oh} {et~al.}(2008){Oh}, {Kim}, {Lee}, \& {Kim}}]{Oh+2008}
{Oh}, S.~H., {Kim}, W.-T., {Lee}, H.~M., \& {Kim}, J. 2008, ApJ, 683, 94

\bibitem[{{Ostriker} \& {Peebles}(1973)}]{OstrikerPeebles1973}
{Ostriker}, J.~P. \& {Peebles}, P.~J.~E. 1973, ApJ, 186, 467

\bibitem[{{Pasha}(2004{\natexlab{a}})}]{Pasha2004a}
{Pasha}, I.~I. 2004{\natexlab{a}}, ArXiv Astrophysics e-prints

\bibitem[{{Pasha}(2004{\natexlab{b}})}]{Pasha2004b}
---. 2004{\natexlab{b}}, ArXiv Astrophysics e-prints

\bibitem[{{Patsis} {et~al.}(1997){Patsis}, {Grosbol}, \&
  {Hiotelis}}]{Patsis1997}
{Patsis}, P.~A., {Grosbol}, P., \& {Hiotelis}, N. 1997, \aap, 323, 762

\bibitem[{{Patsis} {et~al.}(1994){Patsis}, {Hiotelis}, {Contopoulos}, \&
  {Grosbol}}]{Patsis1994}
{Patsis}, P.~A., {Hiotelis}, N., {Contopoulos}, G., \& {Grosbol}, P. 1994,
  \aap, 286, 46

\bibitem[{{Perryman} {et~al.}(2001){Perryman}, {de Boer}, {Gilmore}, {H{\o}g},
  {Lattanzi}, {Lindegren}, {Luri}, {Mignard}, {Pace}, \& {de
  Zeeuw}}]{Perryman+2001}
{Perryman}, M.~A.~C., {de Boer}, K.~S., {Gilmore}, G., {H{\o}g}, E.,
  {Lattanzi}, M.~G., {Lindegren}, L., {Luri}, X., {Mignard}, F., {Pace}, O., \&
  {de Zeeuw}, P.~T. 2001, A\&A, 369, 339

\bibitem[{{Pettitt} {et~al.}(2014){Pettitt}, {Dobbs}, {Acreman}, \&
  {Price}}]{Pettitt+2014}
{Pettitt}, A.~R., {Dobbs}, C.~L., {Acreman}, D.~M., \& {Price}, D.~J. 2014, in
  IAU Symposium, Vol. 298, IAU Symposium, ed. S.~{Feltzing}, G.~{Zhao}, N.~A.
  {Walton}, \& P.~{Whitelock}, 246--252

\bibitem[{{Pfleiderer}(1963)}]{Pfleiderer1963}
{Pfleiderer}, J. 1963, \zap, 58, 12

\bibitem[{{Pfleiderer} \& {Siedentopf}(1961)}]{Pfleiderer1961}
{Pfleiderer}, J. \& {Siedentopf}, H. 1961, \zap, 51, 201

\bibitem[{{Pichon} \& {Cannon}(1997)}]{PichonCannon1997}
{Pichon}, C. \& {Cannon}, R.~C. 1997, MNRAS, 291, 616

\bibitem[{{Polyachenko}(2004)}]{Polyachenko2004}
{Polyachenko}, E.~V. 2004, MNRAS, 348, 345

\bibitem[{{Polyachenko}(2005)}]{Polyachenko2005}
---. 2005, MNRAS, 357, 559

\bibitem[{{Pringle} \& {King}(2007)}]{Pringle2007}
{Pringle}, J.~E. \& {King}, A. 2007, {Astrophysical Flows}

\bibitem[{{Puerari} \& {Dottori}(1992)}]{PuerariDottori1992}
{Puerari}, I. \& {Dottori}, H.~A. 1992, A\&A, 93, 469

\bibitem[{{Purcell} {et~al.}(2011){Purcell}, {Bullock}, {Tollerud}, {Rocha}, \&
  {Chakrabarti}}]{Purcell2011}
{Purcell}, C.~W., {Bullock}, J.~S., {Tollerud}, E.~J., {Rocha}, M., \&
  {Chakrabarti}, S. 2011, \nat, 477, 301

\bibitem[{{Quillen} {et~al.}(2011){Quillen}, {Dougherty}, {Bagley}, {Minchev},
  \& {Comparetta}}]{Quillen+2011}
{Quillen}, A.~C., {Dougherty}, J., {Bagley}, M.~B., {Minchev}, I., \&
  {Comparetta}, J. 2011, MNRAS, 417, 762

\bibitem[{{Quirk} \& {Crutcher}(1973)}]{Quirk1973}
{Quirk}, W.~J. \& {Crutcher}, R.~M. 1973, \apj, 181, 359

\bibitem[{{Rafikov}(2001)}]{Rafikov2001}
{Rafikov}, R.~R. 2001, \mnras, 323, 445

\bibitem[{{Rand} \& {Wallin}(2004)}]{Rand2004}
{Rand}, R.~J. \& {Wallin}, J.~F. 2004, \apj, 614, 142

\bibitem[{{Rautiainen} \& {Salo}(1999)}]{RautiainenSalo1999}
{Rautiainen}, P. \& {Salo}, H. 1999, A\&A, 348, 737

\bibitem[{{Rautiainen} \& {Salo}(2000)}]{RautiainenSalo2000}
---. 2000, A\&A, 362, 465

\bibitem[{{Regan} \& {Wilson}(1993)}]{Regan1993}
{Regan}, M.~W. \& {Wilson}, C.~D. 1993, \aj, 105, 499

\bibitem[{{Reynolds}(1927)}]{Reynolds1927}
{Reynolds}, J.~H. 1927, The Observatory, 50, 185

\bibitem[{{Roberts}(1969)}]{Roberts1969}
{Roberts}, W.~W. 1969, ApJ, 158, 123

\bibitem[{{Roberts} {et~al.}(1975){Roberts}, {Roberts}, \&
  {Shu}}]{Roberts+1975}
{Roberts}, Jr., W.~W., {Roberts}, M.~S., \& {Shu}, F.~H. 1975, ApJ, 196, 381

\bibitem[{{Roberts} \& {Shu}(1972)}]{RobertsShu1972}
{Roberts}, Jr., W.~W. \& {Shu}, F.~H. 1972, Astrophysical Letters, 12, 49

\bibitem[{{Roberts} \& {Stewart}(1987)}]{RobertsStewart1987}
{Roberts}, Jr., W.~W. \& {Stewart}, G.~R. 1987, \apj, 314, 10

\bibitem[{{Roberts} \& {Yuan}(1970)}]{RobertsYuan1970}
{Roberts}, Jr., W.~W. \& {Yuan}, C. 1970, ApJ, 161, 887

\bibitem[{{Robertson} \& {Kravtsov}(2008)}]{RobertsonKravtsov2008}
{Robertson}, B.~E. \& {Kravtsov}, A.~V. 2008, ApJ, 680, 1083

\bibitem[{{Roca-F{\`a}brega} {et~al.}(2013){Roca-F{\`a}brega}, {Valenzuela},
  {Figueras}, {Romero-G{\'o}mez}, {Vel{\'a}zquez}, {Antoja}, \&
  {Pichardo}}]{Roca-Fabrega+2013}
{Roca-F{\`a}brega}, S., {Valenzuela}, O., {Figueras}, F., {Romero-G{\'o}mez},
  M., {Vel{\'a}zquez}, H., {Antoja}, T., \& {Pichardo}, B. 2013, MNRAS, 432,
  2878

\bibitem[{{Rodriguez-Fernandez} \&
  {Combes}(2008)}]{Rodriguez-FernandezCombes2008}
{Rodriguez-Fernandez}, N.~J. \& {Combes}, F. 2008, A\&A, 489, 115

\bibitem[{{Romeo}(1992)}]{Romeo1992}
{Romeo}, A.~B. 1992, \mnras, 256, 307

\bibitem[{{Romeo} {et~al.}(2010){Romeo}, {Burkert}, \& {Agertz}}]{Romeo2010}
{Romeo}, A.~B., {Burkert}, A., \& {Agertz}, O. 2010, \mnras, 407, 1223

\bibitem[{{Romeo} \& {Falstad}(2013)}]{Romeo2013}
{Romeo}, A.~B. \& {Falstad}, N. 2013, \mnras, 433, 1389

\bibitem[{{Romeo} \& {Wiegert}(2011)}]{Romeo2011}
{Romeo}, A.~B. \& {Wiegert}, J. 2011, \mnras, 416, 1191

\bibitem[{{Romero-G{\'o}mez} {et~al.}(2007){Romero-G{\'o}mez}, {Athanassoula},
  {Masdemont}, \& {Garc{\'{\i}}a-G{\'o}mez}}]{Romero-Gomez+2007}
{Romero-G{\'o}mez}, M., {Athanassoula}, E., {Masdemont}, J.~J., \&
  {Garc{\'{\i}}a-G{\'o}mez}, C. 2007, A\&A, 472, 63

\bibitem[{{Romero-G{\'o}mez} {et~al.}(2006){Romero-G{\'o}mez}, {Masdemont},
  {Athanassoula}, \& {Garc{\'{\i}}a-G{\'o}mez}}]{Romero-Gomez+2006}
{Romero-G{\'o}mez}, M., {Masdemont}, J.~J., {Athanassoula}, E., \&
  {Garc{\'{\i}}a-G{\'o}mez}, C. 2006, A\&A, 453, 39

\bibitem[{{Rosse}(1850)}]{Rosse1850}
{Rosse}, E.~o. 1850, Philosophical Transactions of the Royal Society, 140, 499

\bibitem[{{Rots}(1975)}]{Rots1975b}
{Rots}, A.~H. 1975, \aap, 45, 43

\bibitem[{{Rots} \& {Shane}(1975)}]{Rots1975}
{Rots}, A.~H. \& {Shane}, W.~W. 1975, \aap, 45, 25

\bibitem[{{Ro{\v s}kar} {et~al.}(2012){Ro{\v s}kar}, {Debattista}, {Quinn}, \&
  {Wadsley}}]{Roskar+2012}
{Ro{\v s}kar}, R., {Debattista}, V.~P., {Quinn}, T.~R., \& {Wadsley}, J. 2012,
  MNRAS, 426, 2089

\bibitem[{{Rubin} \& {Ford}(1970)}]{Rubin1970}
{Rubin}, V.~C. \& {Ford}, Jr., W.~K. 1970, \apj, 159, 379

\bibitem[{{Safronov}(1960)}]{Safranov1960}
{Safronov}, V.~S. 1960, Annales d'Astrophysique, 23, 979

\bibitem[{{Sakamoto} {et~al.}(1999){Sakamoto}, {Okumura}, {Ishizuki}, \&
  {Scoville}}]{Sakamoto+1999}
{Sakamoto}, K., {Okumura}, S.~K., {Ishizuki}, S., \& {Scoville}, N.~Z. 1999,
  ApJS, 124, 403

\bibitem[{{Salo} \& {Laurikainen}(2000{\natexlab{a}})}]{Salo2000b}
{Salo}, H. \& {Laurikainen}, E. 2000{\natexlab{a}}, \mnras, 319, 377

\bibitem[{{Salo} \& {Laurikainen}(2000{\natexlab{b}})}]{Salo2000}
---. 2000{\natexlab{b}}, \mnras, 319, 393

\bibitem[{{Salo} {et~al.}(2010){Salo}, {Laurikainen}, {Buta}, \&
  {Knapen}}]{Salo2010}
{Salo}, H., {Laurikainen}, E., {Buta}, R., \& {Knapen}, J.~H. 2010, \apjl, 715,
  L56

\bibitem[{{S{\'a}nchez-Gil} {et~al.}(2011){S{\'a}nchez-Gil}, {Jones},
  {P{\'e}rez}, {Bland-Hawthorn}, {Alfaro}, \& {O'Byrne}}]{Sanchez2011}
{S{\'a}nchez-Gil}, M.~C., {Jones}, D.~H., {P{\'e}rez}, E., {Bland-Hawthorn},
  J., {Alfaro}, E.~J., \& {O'Byrne}, J. 2011, \mnras, 415, 753

\bibitem[{{Sandage}(1961)}]{Sandage1961}
{Sandage}, A. 1961, {The Hubble atlas of galaxies}

\bibitem[{{Sandage}(1986)}]{Sandage1986}
---. 1986, \aap, 161, 89

\bibitem[{{Sanders} \& {Huntley}(1976)}]{SandersHuntley1976}
{Sanders}, R.~H. \& {Huntley}, J.~M. 1976, ApJ, 209, 53

\bibitem[{{Santill{\'a}n} {et~al.}(2000){Santill{\'a}n}, {Kim}, {Franco},
  {Martos}, {Hong}, \& {Ryu}}]{Santillan2000}
{Santill{\'a}n}, A., {Kim}, J., {Franco}, J., {Martos}, M., {Hong}, S.~S., \&
  {Ryu}, D. 2000, \apj, 545, 353

\bibitem[{{Schwarz}(1981)}]{Schwarz1981}
{Schwarz}, M.~P. 1981, ApJ, 247, 77

\bibitem[{{Seigar} {et~al.}(2005){Seigar}, {Block}, {Puerari}, {Chorney}, \&
  {James}}]{Seigar+2005}
{Seigar}, M.~S., {Block}, D.~L., {Puerari}, I., {Chorney}, N.~E., \& {James},
  P.~A. 2005, MNRAS, 359, 1065

\bibitem[{{Seigar} {et~al.}(2006){Seigar}, {Bullock}, {Barth}, \&
  {Ho}}]{Seigar+2006}
{Seigar}, M.~S., {Bullock}, J.~S., {Barth}, A.~J., \& {Ho}, L.~C. 2006, ApJ,
  645, 1012

\bibitem[{{Seigar} {et~al.}(2003){Seigar}, {Chorney}, \& {James}}]{Seigar+2003}
{Seigar}, M.~S., {Chorney}, N.~E., \& {James}, P.~A. 2003, MNRAS, 342, 1

\bibitem[{{Seigar} \& {James}(1998)}]{SeigarJames1998}
{Seigar}, M.~S. \& {James}, P.~A. 1998, MNRAS, 299, 685

\bibitem[{{Seigar} \& {James}(2002)}]{SeigarJames2002}
---. 2002, \mnras, 337, 1113

\bibitem[{{Sellwood}(1981)}]{Sellwood1981}
{Sellwood}, J.~A. 1981, A\&A, 99, 362

\bibitem[{{Sellwood}(1989)}]{Sellwood1989}
{Sellwood}, J.~A. 1989, in Dynamics of Astrophysical Discs, ed.
  {J.~A.~Sellwood}, 155--171

\bibitem[{{Sellwood}(1994)}]{Sellwood1994}
{Sellwood}, J.~A. 1994, in Galactic and Solar System Optical Astrometry, ed.
  L.~V. {Morrison} \& G.~F. {Gilmore}, 156

\bibitem[{{Sellwood}(2000)}]{Sellwood2000}
---. 2000, Ap\&SS, 272, 31

\bibitem[{{Sellwood}(2010{\natexlab{a}})}]{Sellwood2010}
---. 2010{\natexlab{a}}, MNRAS, 409, 145

\bibitem[{{Sellwood}(2010{\natexlab{b}})}]{Sellwood2010review}
---. 2010{\natexlab{b}}, ArXiv e-prints

\bibitem[{{Sellwood}(2011)}]{Sellwood2011}
---. 2011, MNRAS, 410, 1637

\bibitem[{{Sellwood}(2012)}]{Sellwood2012}
---. 2012, \apj, 751, 44

\bibitem[{{Sellwood} \& {Athanassoula}(1986)}]{SellwoodAthanassoula1986}
{Sellwood}, J.~A. \& {Athanassoula}, E. 1986, MNRAS, 221, 195

\bibitem[{{Sellwood} \& {Binney}(2002)}]{SellwoodBinney2002}
{Sellwood}, J.~A. \& {Binney}, J.~J. 2002, MNRAS, 336, 785

\bibitem[{{Sellwood} \& {Carlberg}(1984)}]{SellwoodCarlberg1984}
{Sellwood}, J.~A. \& {Carlberg}, R.~G. 1984, ApJ, 282, 61

\bibitem[{{Sellwood} \& {Carlberg}(2014)}]{Sellwood2014}
---. 2014, \apj, 785, 137

\bibitem[{{Sellwood} \& {Kahn}(1991)}]{SellwoodKahn1991}
{Sellwood}, J.~A. \& {Kahn}, F.~D. 1991, MNRAS, 250, 278

\bibitem[{{Sellwood} \& {Lin}(1989)}]{SellwoodLin1989}
{Sellwood}, J.~A. \& {Lin}, D.~N.~C. 1989, MNRAS, 240, 991

\bibitem[{{Sellwood} \& {Sparke}(1988)}]{SellwoodSparke1988}
{Sellwood}, J.~A. \& {Sparke}, L.~S. 1988, MNRAS, 231, 25P

\bibitem[{{Sempere} {et~al.}(1995){Sempere}, {Garcia-Burillo}, {Combes}, \&
  {Knapen}}]{Sempere1995}
{Sempere}, M.~J., {Garcia-Burillo}, S., {Combes}, F., \& {Knapen}, J.~H. 1995,
  \aap, 296, 45

\bibitem[{{Shetty} \& {Ostriker}(2006)}]{ShettyOstriker2006}
{Shetty}, R. \& {Ostriker}, E.~C. 2006, ApJ, 647, 997

\bibitem[{{Shetty} \& {Ostriker}(2008)}]{Shetty2008}
---. 2008, \apj, 684

\bibitem[{{Shetty} {et~al.}(2007){Shetty}, {Vogel}, {Ostriker}, \&
  {Teuben}}]{Shetty+2007}
{Shetty}, R., {Vogel}, S.~N., {Ostriker}, E.~C., \& {Teuben}, P.~J. 2007, ApJ,
  665, 1138

\bibitem[{{Shu}(1992)}]{Shu1992b}
{Shu}, F.~H. 1992, {Physics of Astrophysics, Vol. II} (University Science
  Books)

\bibitem[{{Shu} {et~al.}(1972){Shu}, {Milione}, {Gebel}, {Yuan}, {Goldsmith},
  \& {Roberts}}]{Shu+1972}
{Shu}, F.~H., {Milione}, V., {Gebel}, W., {Yuan}, C., {Goldsmith}, D.~W., \&
  {Roberts}, W.~W. 1972, \apj, 173, 557

\bibitem[{{Shu} {et~al.}(1973){Shu}, {Milione}, \& {Roberts}}]{Shu+1973}
{Shu}, F.~H., {Milione}, V., \& {Roberts}, Jr., W.~W. 1973, ApJ, 183, 819

\bibitem[{{Shu} {et~al.}(1971){Shu}, {Stachnik}, \& {Yost}}]{Shu+1971}
{Shu}, F.~H., {Stachnik}, R.~V., \& {Yost}, J.~C. 1971, \apj, 166, 465

\bibitem[{{Sleath} \& {Alexander}(1995)}]{Sleath1995}
{Sleath}, J.~P. \& {Alexander}, P. 1995, \mnras, 275, 507

\bibitem[{{Speights} \& {Westpfahl}(2011)}]{SpeightsWestpfahl2011}
{Speights}, J.~C. \& {Westpfahl}, D.~J. 2011, ApJ, 736, 70

\bibitem[{{Speights} \& {Westpfahl}(2012)}]{SpeightsWestpfahl2012}
---. 2012, ApJ, 752, 52

\bibitem[{{Stark} {et~al.}(1987){Stark}, {Elmegreen}, \& {Chance}}]{Stark1987}
{Stark}, A.~A., {Elmegreen}, B.~G., \& {Chance}, D. 1987, \apj, 322, 64

\bibitem[{{Struck} {et~al.}(2011){Struck}, {Dobbs}, \& {Hwang}}]{Struck+2011}
{Struck}, C., {Dobbs}, C.~L., \& {Hwang}, J.-S. 2011, MNRAS, 414, 2498

\bibitem[{{Struck-Marcell}(1990)}]{Struck-Marcell1990}
{Struck-Marcell}, C. 1990, \aj, 99, 71

\bibitem[{{Sundelius} {et~al.}(1987){Sundelius}, {Thomasson}, {Valtonen}, \&
  {Byrd}}]{Sundelius1987}
{Sundelius}, B., {Thomasson}, M., {Valtonen}, M.~J., \& {Byrd}, G.~G. 1987,
  \aap, 174, 67

\bibitem[{{Sygnet} {et~al.}(1988){Sygnet}, {Tagger}, {Athanassoula}, \&
  {Pellat}}]{Sygnet+1988}
{Sygnet}, J.~F., {Tagger}, M., {Athanassoula}, E., \& {Pellat}, R. 1988, MNRAS,
  232, 733

\bibitem[{{Tagger} {et~al.}(1987){Tagger}, {Sygnet}, {Athanassoula}, \&
  {Pellat}}]{Tagger+1987}
{Tagger}, M., {Sygnet}, J.~F., {Athanassoula}, E., \& {Pellat}, R. 1987, ApJl,
  318, L43

\bibitem[{{Takahara}(1978)}]{Takahara1978}
{Takahara}, F. 1978, PASJ, 30, 253

\bibitem[{{Tamburro} {et~al.}(2008){Tamburro}, {Rix}, {Walter}, {Brinks}, {de
  Blok}, {Kennicutt}, \& {Mac Low}}]{Tamburro+2008}
{Tamburro}, D., {Rix}, H.-W., {Walter}, F., {Brinks}, E., {de Blok}, W.~J.~G.,
  {Kennicutt}, R.~C., \& {Mac Low}, M.-M. 2008, AJ, 136, 2872

\bibitem[{{Tashpulatov}(1970)}]{Tashpulatov1970}
{Tashpulatov}, N. 1970, \sovast, 14, 227

\bibitem[{{Tasker} \& {Tan}(2009)}]{Tasker2009}
{Tasker}, E.~J. \& {Tan}, J.~C. 2009, \apj, 700, 358

\bibitem[{{Taylor} \& {Cordes}(1993)}]{Taylor1993}
{Taylor}, J.~H. \& {Cordes}, J.~M. 1993, \apj, 411, 674

\bibitem[{{Theis} \& {Spinneker}(2003)}]{Theis2003}
{Theis}, C. \& {Spinneker}, C. 2003, \apss, 284, 495

\bibitem[{{Thomasson} {et~al.}(1989){Thomasson}, {Donner}, {Sundelius}, {Byrd},
  {Huang}, \& {Valtonen}}]{Thomasson1989}
{Thomasson}, M., {Donner}, K.~J., {Sundelius}, B., {Byrd}, G.~G., {Huang},
  T.-Y., \& {Valtonen}, M.~J. 1989, \aap, 211, 25

\bibitem[{{Thomasson} {et~al.}(1990){Thomasson}, {Elmegreen}, {Donner}, \&
  {Sundelius}}]{Thomasson+1990}
{Thomasson}, M., {Elmegreen}, B.~G., {Donner}, K.~J., \& {Sundelius}, B. 1990,
  ApJl, 356, L9

\bibitem[{{Thornley}(1996)}]{Thornley1996}
{Thornley}, M.~D. 1996, ApJl, 469, L45+

\bibitem[{{Thornley} \& {Mundy}(1997)}]{ThornleyMundy1997a}
{Thornley}, M.~D. \& {Mundy}, L.~G. 1997, ApJ, 484, 202

\bibitem[{{Toomre}(1964)}]{Toomre1964}
{Toomre}, A. 1964, ApJ, 139, 1217

\bibitem[{{Toomre}(1969)}]{Toomre1969}
---. 1969, ApJ, 158, 899

\bibitem[{{Toomre}(1977)}]{Toomre1977}
---. 1977, ARAA, 15, 437

\bibitem[{{Toomre}(1981)}]{Toomre1981}
{Toomre}, A. 1981, in Structure and Evolution of Normal Galaxies, ed. S.~M.
  {Fall} \& D.~{Lynden-Bell}, 111--136

\bibitem[{{Toomre}(1990)}]{Toomre1990}
---. 1990, {Gas-hungry Sc spirals.}, ed. {Wielen, R.}, 292--303

\bibitem[{{Toomre} \& {Kalnajs}(1991)}]{ToomreKalnajs1991}
{Toomre}, A. \& {Kalnajs}, A.~J. 1991, in Dynamics of Disc Galaxies, ed.
  {B.~Sundelius}, 341--+

\bibitem[{{Toomre} \& {Toomre}(1972)}]{Toomre1972}
{Toomre}, A. \& {Toomre}, J. 1972, \apj, 178, 623

\bibitem[{{Tremaine} \& {Weinberg}(1984)}]{Tremaine1984}
{Tremaine}, S. \& {Weinberg}, M.~D. 1984, \apjl, 282, L5

\bibitem[{{Tsoutsis} {et~al.}(2008){Tsoutsis}, {Efthymiopoulos}, \&
  {Voglis}}]{Tsoutsis+2008}
{Tsoutsis}, P., {Efthymiopoulos}, C., \& {Voglis}, N. 2008, \mnras, 387, 1264

\bibitem[{{Tsoutsis} {et~al.}(2009){Tsoutsis}, {Kalapotharakos},
  {Efthymiopoulos}, \& {Contopoulos}}]{Tsoutsis+2009}
{Tsoutsis}, P., {Kalapotharakos}, C., {Efthymiopoulos}, C., \& {Contopoulos},
  G. 2009, \aap, 495, 743

\bibitem[{{Tutukov} \& {Fedorova}(2006)}]{Tutukov2006}
{Tutukov}, A.~V. \& {Fedorova}, A.~V. 2006, Astronomy Reports, 50, 785

\bibitem[{{Vall{\'e}e}(2005)}]{Vallee2005}
{Vall{\'e}e}, J.~P. 2005, \aj, 130, 569

\bibitem[{{van den Bergh}(1959)}]{vandenBergh1959}
{van den Bergh}, S. 1959, \aj, 64, 347

\bibitem[{{Vauterin} \& {Dejonghe}(1996)}]{VauterinDejonghe1996}
{Vauterin}, P. \& {Dejonghe}, H. 1996, A\&A, 313, 465

\bibitem[{{Verley} {et~al.}(2007){Verley}, {Leon}, {Verdes-Montenegro},
  {Combes}, {Sabater}, {Sulentic}, {Bergond}, {Espada}, {Garc{\'{\i}}a},
  {Lisenfeld}, \& {Odewahn}}]{Verley2007}
{Verley}, S., {Leon}, S., {Verdes-Montenegro}, L., {Combes}, F., {Sabater}, J.,
  {Sulentic}, J., {Bergond}, G., {Espada}, D., {Garc{\'{\i}}a}, E.,
  {Lisenfeld}, U., \& {Odewahn}, S.~C. 2007, \aap, 472, 121

\bibitem[{{Vogel} {et~al.}(1988){Vogel}, {Kulkarni}, \& {Scoville}}]{Vogel1988}
{Vogel}, S.~N., {Kulkarni}, S.~R., \& {Scoville}, N.~Z. 1988, \nat, 334, 402

\bibitem[{{Voglis} {et~al.}(2006{\natexlab{a}}){Voglis}, {Stavropoulos}, \&
  {Kalapotharakos}}]{Voglis+2006a}
{Voglis}, N., {Stavropoulos}, I., \& {Kalapotharakos}, C. 2006{\natexlab{a}},
  \mnras, 372, 901

\bibitem[{{Voglis} {et~al.}(2006{\natexlab{b}}){Voglis}, {Tsoutsis}, \&
  {Efthymiopoulos}}]{Voglis+2006b}
{Voglis}, N., {Tsoutsis}, P., \& {Efthymiopoulos}, C. 2006{\natexlab{b}},
  \mnras, 373, 280

\bibitem[{{Vorontsov-Velyaminov}(1959)}]{Vorontsov-Velyaminov1959}
{Vorontsov-Velyaminov}, B.~A. 1959, in Atlas and catalog of interacting
  galaxies (1959), 0

\bibitem[{{Wada}(1994)}]{Wada1994}
{Wada}, K. 1994, PASJ, 46, 165

\bibitem[{{Wada}(2008)}]{Wada2008}
---. 2008, ApJ, 675, 188

\bibitem[{{Wada} {et~al.}(2011){Wada}, {Baba}, \& {Saitoh}}]{Wada+2011}
{Wada}, K., {Baba}, J., \& {Saitoh}, T.~R. 2011, ApJ, 735, 1

\bibitem[{{Wada} \& {Koda}(2004)}]{WadaKoda2004}
{Wada}, K. \& {Koda}, J. 2004, MNRAS, 349, 270

\bibitem[{{Wang} \& {Silk}(1994)}]{Wang1994}
{Wang}, B. \& {Silk}, J. 1994, \apj, 427, 759

\bibitem[{{Westerlund} \& {Mathewson}(1966)}]{Westerlund1966}
{Westerlund}, B.~E. \& {Mathewson}, D.~S. 1966, \mnras, 131, 371

\bibitem[{{Westpfahl}(1998)}]{Westpfahl1998}
{Westpfahl}, D.~J. 1998, \apjs, 115, 203

\bibitem[{{Willett} {et~al.}(2013){Willett}, {Lintott}, {Bamford}, {Masters},
  {Simmons}, {Casteels}, {Edmondson}, {Fortson}, {Kaviraj}, {Keel}, {Melvin},
  {Nichol}, {Raddick}, {Schawinski}, {Simpson}, {Skibba}, {Smith}, \&
  {Thomas}}]{Willett2013}
{Willett}, K.~W., {Lintott}, C.~J., {Bamford}, S.~P., {Masters}, K.~L.,
  {Simmons}, B.~D., {Casteels}, K.~R.~V., {Edmondson}, E.~M., {Fortson}, L.~F.,
  {Kaviraj}, S., {Keel}, W.~C., {Melvin}, T., {Nichol}, R.~C., {Raddick},
  M.~J., {Schawinski}, K., {Simpson}, R.~J., {Skibba}, R.~A., {Smith}, A.~M.,
  \& {Thomas}, D. 2013, \mnras, 435, 2835

\bibitem[{{Woodward}(1975)}]{Woodward1975}
{Woodward}, P.~R. 1975, ApJ, 195, 61

\bibitem[{{Woodward}(1976)}]{Woodward1976}
---. 1976, \apj, 207, 484

\bibitem[{{Zhang}(1996)}]{Zhang1996}
{Zhang}, X. 1996, ApJ, 457, 125

\bibitem[{{Zimmer} {et~al.}(2004){Zimmer}, {Rand}, \& {McGraw}}]{Zimmer2004}
{Zimmer}, P., {Rand}, R.~J., \& {McGraw}, J.~T. 2004, \apj, 607, 285

\end{thebibliography}

\end{document}